\documentclass[twocolumn]{aastex631}

\usepackage{amssymb,amsmath}
\usepackage{todonotes}
\usepackage{xspace}
\usepackage{subfigure}
\usepackage{tabularx}
\usepackage{makecell}

\usepackage{forest, tikz}
\usetikzlibrary{decorations.pathreplacing}

\def\dssj{\textsc{DSSJ}\xspace}
\def\knnsj{$k$\textsc{NNSJ}\xspace}
\def\knn{$k$\textsc{NN}\xspace}

\def\snaps{\textsc{SNAPS}\xspace}
\def\antares{\textsc{ANTARES}\xspace}
\def\snapshotA{\texttt{SNAPShot1}\xspace}
\def\lsstsynt{\texttt{LSST5M}\xspace}

\shorttitle{AASTeX v6.3.1 Sample article}
\shortauthors{Gowanlock et al.}

\begin{document}

\title{The Solar System Notification Alert Processing System (SNAPS): Asteroid Population Outlier Detection}

\correspondingauthor{Michael Gowanlock}
\email{michael.gowanlock@nau.edu}

\author[0000-0002-0826-6204]{Michael Gowanlock}
\affiliation{School of Informatics, Computing, and Cyber Systems \\
Northern Arizona University \\
P.O. Box 5693 \\
Flagstaff, AZ 86011, USA}
\affiliation{Department of Astronomy and Planetary Science \\
Northern Arizona University \\
P.O. Box 6010 \\
Flagstaff, AZ 86011, USA}

\author[0000-0003-4580-3790]{David E. Trilling}
\affiliation{Department of Astronomy and Planetary Science \\
Northern Arizona University \\
P.O. Box 6010 \\
Flagstaff, AZ 86011, USA}
\affiliation{School of Informatics, Computing, and Cyber Systems \\
Northern Arizona University \\
P.O. Box 5693 \\
Flagstaff, AZ 86011, USA}

\author[0000-0002-6676-1713]{Daniel Kramer}
\affiliation{School of Informatics, Computing, and Cyber Systems \\
Northern Arizona University \\
P.O. Box 5693 \\
Flagstaff, AZ 86011, USA}
\affiliation{Department of Astronomy and Planetary Science \\
Northern Arizona University \\
P.O. Box 6010 \\
Flagstaff, AZ 86011, USA}

\author[0000-0002-6292-9056]{Maria Chernyavskaya}
\affiliation{Department of Astronomy and Planetary Science \\
Northern Arizona University \\
P.O. Box 6010 \\
Flagstaff, AZ 86011, USA}

\author{Andrew McNeill}
\affiliation{Department of Physics\\Lehigh University\\ 16 Memorial Drive East\\ Bethlehem, PA 18015, USA}



\begin{abstract}
The Solar System Notification Alert Processing System (SNAPS) is a ZTF and Rubin Observatory alert broker that will send alerts to the community regarding interesting events in the Solar System. SNAPS is actively monitoring Solar System objects and one of its functions is to compare objects (primarily main belt asteroids) to one another to find those that are outliers relative to the population. In this paper, we use the SNAPShot1 dataset which contains 31,693 objects from ZTF and derive outlier scores for each of these objects. SNAPS employs an unsupervised approach; consequently, to derive outlier rankings for each object, we propose four different outlier metrics such that we can explore variants of outlier scores and add confidence to outlier rankings. We also provide outlier scores for each object in each permutation of 15 feature spaces, between 2 and 15 features, which yields 32,752 total feature spaces. We show that we can derive population outlier rankings each month at Rubin Observatory scale using 4$\times$Nvidia A100 GPUs, and present several avenues of scientific investigation that can be explored using population outlier detection.
\end{abstract}


\section{Introduction}\label{sec:intro}
The Rubin Observatory's Legacy Survey of Space and Time~\citep[LSST;][]{LSST} will have an major impact on time-domain astronomy. To prepare for this large all-sky survey, many researchers have been designing data processing pipelines~\citep{van2021ztf,coughlin2021ztf} for precursor surveys, such as the Zwicky Transient Facility~\citep[ZTF;][]{bellm2018zwicky}. While ZTF generates roughly a tenth of the LSST data volume, it and other LSST precursor surveys such as the Asteroid Terrestrial-impact Last Alert System~\citep[ATLAS;][]{ATLAS-survey}, Automated Survey for SuperNovae~\citep[ASAS-SN;][]{ASAS-SN-survey}, and the Catalina Real-Time Transient Survey~\citep[CRTS;][]{CRTS-survey} are instrumental for ensuring that the necessary preparations have been made to maximize the scientific return of LSST.
Several alert brokers are currently being developed to prepare for LSST including Fink~\citep{Moller2020}, ALeRCE~\citep{Forster2021,Sanchez2021}, ANTARES~\citep{matheson2021antares}, Lasair~\citep{Smith2019}, among others\footnote{For an update-to-date list of LSST alert brokers, see the following URL: \url{https://www.lsst.org/scientists/alert-brokers}.}.

Recent LSST preparation efforts have focused on classifying astrophysical sources~\citep{soraisam2020classification}. In the context of Solar System science, the vast majority of objects are asteroids, and so classification is less important in this context (e.g., classifying an asteroid vs. comet is not very illuminating). Instead, a major focus of the Solar System community is to detect interesting transient events in addition to detecting asteroids that are different relative to the population. 

In this paper, we present our population outlier detection software infrastructure for the Solar System Notification Alert Processing System (\snaps). \citet{trilling2023} presented the first \snaps data release (\snapshotA), which contains 31,693 asteroids from ZTF. Each asteroid has been observed at least 51 times and has been assigned a rotation period and light curve amplitude among other derived properties. Beyond Solar System science, ZTF has been similarly used for deriving the rotation periods of fully convective stars~\citep{lu2022bridging}. It is clear that the time-domain capabilities of new observatories are revolutionizing our understanding of the Solar System and beyond.

At present, \snaps is operating as a downstream broker from \antares~\citep{matheson2021antares}, and will continue to do so in the LSST era~\citep[more detail on \snaps can be found in][]{trilling2023}. Population outlier detection refers to detecting asteroids that have interesting properties relative to the other asteroids in the database. Because we have few examples of outlying objects in the Solar System, we use an unsupervised machine learning approach that detects interesting objects without the use of binary classification labels (i.e., inlier vs. outlier). Our approach ranks the objects by assigning them an outlier score, and this will allow users of \snaps to customize the number of objects that they may want to investigate further; for instance, a user may find an interesting object and then use telescopic follow-up observations for additional analysis. Furthermore, we use an ensemble of algorithms and outlier detection metrics to guide the search for outliers, as the methods may disagree on which objects have the highest outlier scores.

The paper is organized as follows. Section~\ref{sec:datasets} describes the datasets used throughout the paper. Section~\ref{sec:system_description} presents the population outlier detection system. Section~\ref{sec:results} highlights key results by deriving population outlier scores for objects in our ZTF catalog, in addition to showing system performance using this catalog, and expected performance at LSST scale. Section~\ref{sec:science} shows example avenues of scientific investigation that can be carried out using the system. Finally, Section~\ref{sec:conclusions} discusses population outlier detection in the time-domain era of astronomy and future work directions.

\section{Datasets}\label{sec:datasets}
In this section, we describe the datasets used in our evaluation, including \snapshotA~\citep{trilling2023}, which is a ZTF dataset. We also outline a synthetic dataset that is used to assess the performance of the system at LSST scale.

\subsection{SNAPShot1}\label{sec:data_snapshot1}
The first \snaps data release is \snapshotA, and a detailed description of this data release is outlined in \citet{trilling2023}. This data release contains $|D|=31,693$ ZTF asteroids that have at least 51 observations, such that properties can be derived, including light curve properties, such as rotation periods and amplitudes. When we describe the data normalization procedure in the next section, we will summarize the properties of each asteroid.

\snapshotA includes data from the NEOWISE catalog~\citep{neowise}, and we used the \texttt{NEOWISE\_DIAM} and \texttt{NEOWISE\_VALBEDO} features as those properties may be useful for detecting outliers. However, some of the ZTF objects were not included in the NEOWISE catalog; consequently, in cases where values were undefined (using NaN values), we replaced those values with the mean of the distribution. This ensures that those objects with undefined values are considered typical, and therefore, will not be ranked as an outlier based on the \texttt{NEOWISE\_DIAM} and \texttt{NEOWISE\_VALBEDO} properties.

\begin{deluxetable*}{l|l|l|l|r|r}
\tablecaption{Features used for outlier detection in \snapshotA. The table shows each feature and description. The text describes that \texttt{NEOWISE\_DIAM} and \texttt{NEOWISE\_VALBEDO} do not provide values for all ZTF objects; therefore, those objects that are missing values, are replaced by the mean of the distribution of those features. The mean and standard deviation ($\mu$, $\sigma$) of the features before normalization are shown as they are used to generate a synthetic dataset to assess population outlier detection at LSST scale.}\label{tab:population_outlier_properties}
\tablewidth{\columnwidth}
\tabletypesize{\footnotesize}
\tablehead{
\colhead{Feature}&\colhead{Description}&\colhead{NaN to Mean}&\colhead{Derived}&\colhead{$\mu$}&\colhead{$\sigma$}
}
\startdata
G\_BR             &The Bowell G value in the $r$ filter \citep{bowellApplicationPhotometricModels1989}&&\checkmark&0.226& 0.234\\
H\_BR             &The absolute magnitude of the object, in the Bowell HG system, in the $r$ filter&&\checkmark&14.108& 1.877\\
G\_BG             &The Bowell G value in the $g$ filter&&\checkmark&0.216& 0.255\\
H\_BG             &The absolute magnitude of the object, in the Bowell HG system, in the $g$ filter&&\checkmark&14.685& 1.902\\
LCAMP            &The amplitude of the light curve&&\checkmark&0.322& 0.300\\
ROTPER           &The rotation period (in hours) of the object&&\checkmark&101.150& 643.548\\
GRCOLOR          &The $g-r$ color of the object&&\checkmark&0.596& 0.146\\
SIGGRCOLOR       &The $1\sigma$ error of the $g-r$ color&&\checkmark&0.008& 0.006\\
PERIPOWER        &The normalized Lomb-Scargle Periodogram power of the rotation period&&\checkmark&0.514& 0.289\\
NEOWISE\_DIAM     &The diameter of the object calculated by the NEOWise Survey&\checkmark&&8.263& 15.675\\
NEOWISE\_VALBEDO  &The albedo in the V-band, calculated by the NEOWise Survey& \checkmark&&0.196& 0.185\\
MPC\_A            &The object's semi-major axis in AU&&&2.698& 1.837\\
MPC\_E            &The object's eccentricity&&&0.144& 0.100\\
MPC\_I            &The object's inclination in degrees&&&9.660& 8.917\\
HAVG             &The object's average absolute magnitude&&\checkmark&14.045& 1.830\\
\enddata
\end{deluxetable*}

\subsection{LSST Synthetic Dataset (\lsstsynt)}\label{sec:datasets_lsst_synthetic}
The LSST catalog will contain roughly 5 million asteroids at the end of the 10~year survey. To examine the performance of the system at LSST scale, we generate a synthetic dataset with 5 million objects with the same $d=15$ dimensions used for \snapshotA described in Table~\ref{tab:population_outlier_properties} and denote this dataset as \lsstsynt. We replicate the features of each object in \snapshotA  and then add noise to the values of each feature. The noise added to each feature is drawn from a Gaussian distribution using $\mu=0$ where the standard deviation, $\sigma$, is that of each feature in \snapshotA (as shown in Table~\ref{tab:population_outlier_properties}). While we do not expect that this dataset will be exceedingly representative of the LSST catalog, it is sufficient for testing the performance of population outlier detection with \snaps at LSST scale (see Section~\ref{sec:LSST_scale}).

\section{Finding Outlying Asteroids: An Unsupervised Approach}\label{sec:system_description}

The two major categories of machine learning are supervised and unsupervised learning~\citep{ZHOU2017350}. Supervised learning takes as input feature vectors with class labels and attempts to learn the function that maximizes classifying objects with the correct class label. In contrast, unsupervised learning does not use class labels and instead attempts to identify patterns in data. 

Other time-domain alert brokers will classify sources
outside of the Solar System, such as Fink~\citep{Moller2020}, ALeRCE~\citep{Forster2021,Sanchez2021}, and ANTARES~\citep{matheson2021antares}. Regarding the Solar System, the alert broker Fink~\citep{Moller2020} has been used to
identify new candidate Solar System objects~\citep{2023A&A...680A..17L}, and they also design a new phase function model
that corrects for the geometric properties of asteroids, including
spin coordinates, to better characterize sparse photometry from
ZTF~\citep{2024A&A...687A..38C}. Unlike much of the prior work, source classification is not the target of our alert broker; rather, we are interested in identifying objects exhibiting interesting behavior either relative to its prior observational record, or as compared to the greater population of objects. Here, we refer to the former as \emph{real-time} and the latter as \emph{population} outlier detection, respectively. Since there are very few known examples of small bodies exhibiting interesting/transient behavior, we are focused on unsupervised approaches to outlier detection, as we do not have sufficient examples of outlier activity needed to train supervised approaches. Furthermore, since LSST is expected to detect many new objects which may exhibit exotic properties for which we have no known examples, we do not want to inadvertently exclude these objects from being detected as outliers, which may occur if a supervised approach is used.

\subsection{Terminology}
Because this paper includes material from several fields of computer science and astronomy, we outline terminology that is used throughout the paper as follows. 
\begin{itemize}
\item \emph{Feature vector/point:} A feature vector is a set of numerical attributes that define an object, e.g., an asteroid (example properties include color, rotation period, and albedo). Because these feature vectors occupy a position in a $d$ dimensional feature space, they are also called \emph{points}. Throughout this paper, we refer to asteroid properties as feature vectors and points.
\item \emph{Dataset/database:}  A collection of feature vectors denoted as $D$. The cardinality of the set (number of objects/asteroids) is denoted as $|D|$.
\item \emph{Feature space:} The space encompassing all of the feature vectors (or points) in the dataset, $D$.
\item \emph{Dimensionality: } The number of features included in each feature vector ($d$). We examine $d\leq 15$.
\item \emph{Self-join:} Performing a search on all feature vectors within a dataset ($D$) as compared to each other.
\item \emph{Distance Similarity Search:} This is also known as a range query or a radius search, where a search is conducted around a query point using a radius $\epsilon$ and all neighbors that are found within $\epsilon$ are returned. A distance similarity self-join performs distance similarity searches on all points in the dataset, $D$.
\item \emph{$k$-Nearest Neighbor Search:} Finding the closest $k$ points within the feature space from a given query point. A \knn self-join performs $k$-nearest neighbor searches on all points in a dataset, $D$.
\item \emph{Multi-GPU Speedup:} A performance measure that describes the scalability of a GPU algorithm that uses multiple GPUs. It is the ratio of the time to compute on one GPU ($T_1$) to $T_{n_{GPU}}$, where $T_{n_{GPU}}$ is the time to compute on $n_{GPU}$ GPUs. In this paper we evaluate up to $n_{GPU}=4$ and so the maximum possible speedup is 4$\times$.
\item \emph{Multi-GPU Parallel Efficiency:} A performance measure that describes how well a resource is utilized (the GPU in this context). It is the multi-GPU speedup above divided by $n_{GPU}$, and the maximum value is 1.0 (or 100\%).
\end{itemize}	

\subsection{Background and GPU-Accelerated Outlier Detection Algorithms}

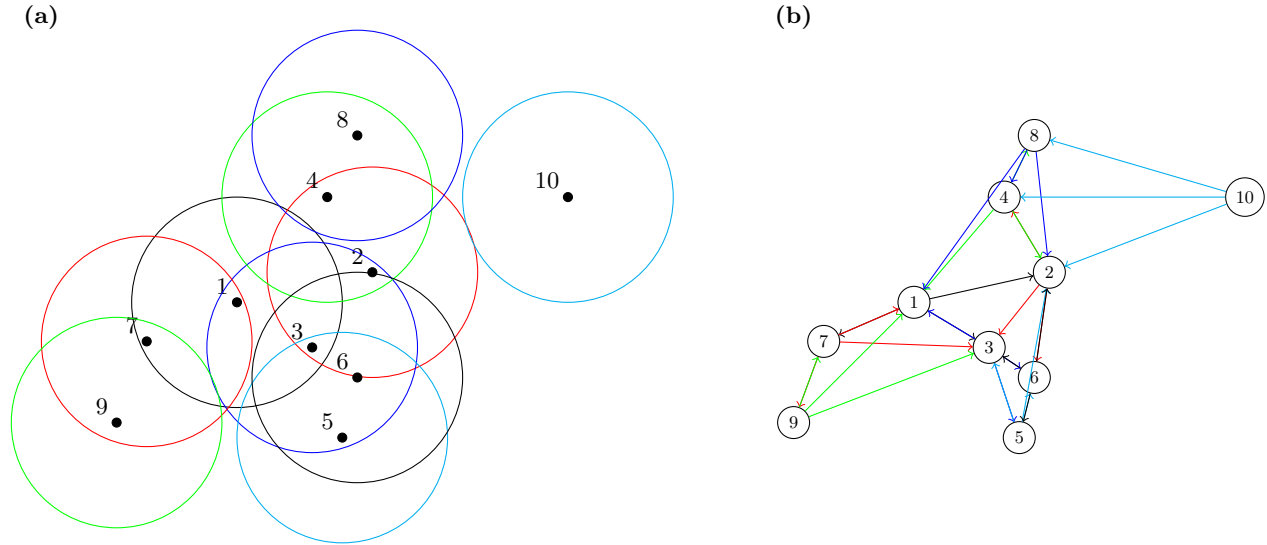
\begin{figure*}[!t]
\centering


\begin{tikzpicture}[scale=2]
\node[] at (-1,3) {\textbf{(a)}};
\node[] at (4,3) {\textbf{(b)}};
\foreach \x/\y/\z/\w in {0.3/1.1/1/black, 1.2/1.3/2/red, 0.8/0.8/3/blue, 0.9/1.8/4/green, 1/0.2/5/cyan, 1.1/0.6/6/black, -0.3/0.84/7/red, 1.1/2.21/8/blue, -0.5/0.3/9/green, 2.5/1.8/10/cyan} {
\filldraw [fill=black](\x,\y) circle (0.03cm) node[anchor=south east]{\z};	
\draw [color=\w](\x,\y) circle (0.7cm);
}
%
%
%
\def\offset{4.5}
\foreach \x/\y/\z/\w\label in {0.3/1.1/1/black/A, 1.2/1.3/2/red/B, 0.8/0.8/3/blue/C, 0.9/1.8/4/green/D, 1/0.2/5/cyan/E, 1.1/0.6/6/black/F, -0.3/0.84/7/red/G, 1.1/2.21/8/blue/H, -0.5/0.3/9/green/I, 2.5/1.8/10/cyan/J} {
\node[draw,circle,scale=0.75](\label) at (\x+\offset,\y){\z};
}
\draw [->, color=black] (A) edge (C);
\draw [->, color=black] (A) edge (G);
\draw [->, color=black] (A) edge (B);
\draw [->, color=red] (B) edge (D);
\draw [->, color=red] (B) edge (C);
\draw [->, color=red] (B) edge (F);
\draw [->, color=blue] (C) edge (F);
\draw [->, color=blue] (C) edge (A);
\draw [->, color=blue] (C) edge (E);
\draw [->, color=green] (D) edge (H);
\draw [->, color=green] (D) edge (B);
\draw [->, color=green] (D) edge (A);
\draw [->, color=cyan] (E) edge (F);
\draw [->, color=cyan] (E) edge (C);
\draw [->, color=cyan] (E) edge (B);
\draw [->, color=black] (F) edge (C);
\draw [->, color=black] (F) edge (E);
\draw [->, color=black] (F) edge (B);
\draw [->, color=red] (G) edge (I);
\draw [->, color=red] (G) edge (A);
\draw [->, color=red] (G) edge (C);
\draw [->, color=blue] (H) edge (D);
\draw [->, color=blue] (H) edge (B);
\draw [->, color=blue] (H) edge (A);
\draw [->, color=green] (I) edge (G);
\draw [->, color=green] (I) edge (A);
\draw [->, color=green] (I) edge (C);
\draw [->, color=cyan] (J) edge (B);
\draw [->, color=cyan] (J) edge (H);
\draw [->, color=cyan] (J) edge (D);
\end{tikzpicture}
    \caption{Illustrative example of a $d=2$ feature space comparing outlier detection approaches for (a) \dssj with $\epsilon=0.67$; (b) graph representation of \knnsj with $k=3$. (a) The points are shown as black dots, where a fixed radius is drawn around each point. (b) The \knn graph is shown where a directed edge between two nodes denotes the $k=3$ neighbors for each point. Since $k=3$ and there are 10 points, there are a total of 30 directed edges. Colors are used to improve readability in both subfigures.  The \dssj and \knnsj approaches will be described in greater detail in Sections~\ref{sec:DSSJ_description}--\ref{sec:summary_comparison_outlier_detection_approaches}.}
   \label{fig:comparison_outlier_approaches}
\end{figure*}

Many of the pioneering efforts on outlier detection use an intuitive notion of an outlier, which defines an outlier as having a large fraction of the dataset exterior to a search distance around a given point~\citep{knorr1997unified}. Subsequent efforts have focused on the notion of neighbors around a point, where the definition of an outlier is a function of the properties of its closest neighbors~\citep{campos2016evaluation}. This second class of algorithms that use $k$ nearest neighbors have been shown to outperform other unsupervised learning methods~\citep{campos2016evaluation}. Given these two major approaches to outlier detection, we select two methods as follows: (1) the Distance Similarity Self-Join (\dssj), and (2) the $k$-Nearest Neighbor Self-Join (\knnsj).

The \dssj operation performs distance similarity searches on all feature vectors (or points) within a dataset. Each distance similarity search returns all points within a search distance, $\epsilon$, of a point. Thus, the ``self-join'' refers to searching all points within a dataset as compared to each other.  The $k$-Nearest Neighbor Self-Join (\knnsj) finds the $k$ nearest neighbors of all points in a dataset. Compared to \dssj, which uses a fixed search radius $\epsilon$, the \knn algorithm employs varying search radii for each point in the dataset to find at least $k$ neighbors per point. For example, in a densely populated region of the data space, the search radius will be small such that it finds at least $k$ neighbors, whereas in a sparsely populated region, the search radius will need to be larger. An example of the \dssj and \knnsj approaches is outlined in Figure~\ref{fig:comparison_outlier_approaches}.

The typical approach for selecting parameters for an outlier detection algorithm (e.g., $k$ in the \knnsj approach) is to use the Receiver Operating Characteristic (ROC)~\citep{campos2016evaluation}. However, ROC requires a set of labels that define inliers and outliers. Because we do not have a labeled set of outliers, we elect to use a different approach. We select a search radius $\epsilon$ for \dssj and the number of neighbors $k$ for \knnsj by reaching a trade-off between overfitting to the local data density or underfitting to the global data density of each feature space. While this computation adds additional complexity to the detection of outliers across all of the feature spaces, it ensures that we are not arbitrarily selecting parameters $\epsilon$ or $k$.

\subsection{Population Outlier Detection and Accelerating Scientific Discovery}\label{sec:num_feature_spaces}

One facet of population outlier detection is the examination of multi-dimensional feature spaces. The features must be normalized to ensure that some properties are not given more weight than others. Otherwise, some features may dominate  the feature space, and obfuscate other features from contributing to the degree that an object is considered an outlier. Therefore, we normalized all features to be in the range [0, 1]. We used all of the \snapshotA features outlined in Table~\ref{tab:population_outlier_properties}, where the non-normalized mean and standard deviation values are reported.

To address a wide range of science cases, and to accelerate the scientific discovery process, we assign objects an outlier ranking across all permutations of the 15 feature spaces (Table~\ref{tab:population_outlier_properties} shows each feature), where the dimensionality $d\geq 2$. In this context, population outlier detection serves two purposes: (1) to filter those objects that are likely to be outliers from the typical population of asteroids; and, (2) to accelerate the scientific discovery process by providing additional information that may be of interest to a researcher. 

Regarding (2) above, the notion of accelerating the scientific discovery process by augmenting a human researcher with machine support has been described in other works which address providing multiple data variants to a user or guiding the scientific discovery process through data prioritization~\citep{wagstaff2013guiding,ComputerAidedDiscovery}. Population outlier detection across numerous multidimensional feature spaces provides both data prioritization and variant exploration.

To enable the permutation approach described above, \snaps will provide information regarding each permutation of the $d=15$ features/dimensions where $d\geq 2$. Therefore, the total number of feature spaces in \snapshotA is as follows: 

\begin{equation}
\sum_{i=2}^{15}\binom{15}{i}=32,752. 
\end{equation}

Below we describe the methods used for outlier detection and how we rank each object within each feature space.

\subsection{Outlier Detection Methods and Object Ranking}
We present two outlier detection methods each of which have two different metrics, where each outputs a ranked list of outlier objects. The ranked list method allows us to compare the rankings between outlier detection approaches. Because the different outlier detection methods are likely to return different rankings for each object, this allows us to have several outlier detection criteria that can be used to determine whether a given object should be investigated in greater detail. This may include telescopic follow-up to obtain additional information about an object. This approach allows \snaps users to determine their own thresholds for decision making.

We let the list of ranked objects for a given method be denoted as $R$ where the rank of object $i$ is denoted as $r_i \in R$ where $i=1,\ldots,|D|$. Object $i$ with the minimum ranking value  $r_i \in R$, {argmin}$(r_i \in R)$, has the greatest probability of being an outlier, whereas the object with the maximum value, {argmax}$(r_i \in R)$ is considered the most typical object in the feature space. We highlight that some of the outlier detection methods yield ties between objects, where two or more objects may be assigned the same outlier score, whereas other methods may not produce ties between object rankings.

\subsection{Distance Similarity Self-Join (DSSJ)}\label{sec:DSSJ_description}

The \dssj method finds all points within a search radius, $\epsilon$, around each point in a dataset. Intuitively, points having a significant number of points within $\epsilon$ are considered typical, and those with few points within $\epsilon$ will be considered outliers. 

\subsubsection{Ranking Metrics}\label{sec:ranking_metrics_dssj}
We outline two ranking metrics for \dssj as follows. One metric is \emph{distance-oblivious} where all points within the search radius, $\epsilon$, of a query point are considered equal, and thus their distances to the query point are not included in the ranking. The other metric is \emph{distance-aware} where the distances of points to the query point within the search radius $\epsilon$ are considered in the ranking calculation. These two metrics yield different rankings for the \dssj method.

\begin{enumerate}
\item \emph{Distance-oblivious:} The ranking function, $R$, is simply a ranking from the fewest to the greatest number of neighbors each point has within the search distance $\epsilon$.
\item \emph{Distance-aware:} The ranking function, $R$, is a ranking from the largest mean distance between an object and its points within $\epsilon$ and the smallest mean distance. Intuitively, if a given point has its neighbors at large distances, then this implies that it is an outlier.
\end{enumerate}

\subsubsection{Overfitting vs. Underfitting: On The Selection of The Search Radius $\epsilon$}\label{sec:dssj_under_overfit}
The selection of $\epsilon$ will directly impact the outlier ranking of each object in the dataset. Consider two extremes: (1) when $\epsilon\approx0$, there will be few (if any) neighbors found within $\epsilon$ around a given point/feature vector; and, (2) as $\epsilon\rightarrow\infty$ each feature vector will find all other feature vectors in the feature space. Both of these cases are impractical for finding outliers. The search radius $\epsilon$ must not be too small such that too few neighbors are found, while not being too large such that too many neighbors are found.

Considering the extreme cases above, a small value of $\epsilon$ implies that the local density of a feature vector determines whether it is an outlier, whereas a large value of $\epsilon$ implies that the global density makes this determination. Thus, in the former case, the data will be overfit to the local density, and in the latter case, the data will be underfit. Here, we provide a simple solution for selecting $\epsilon$, which considers the pragmatic nature of finding outlying objects in the context of large-scale astronomical surveys, by reaching a trade-off between underfitting and overfitting.

Consider two ranked lists $R^A$ and $R^B$, and $n$ which refers to some top fraction of the outliers in each list. In this paper we set $n=0.01|D|$ (1\% of the dataset). If the top-$n$ outlying objects in $R^A$ and $R^B$ are permuted and a given object in $R^A$ has a similar position in $R^B$, then the object has been found to be an outlier in both ranked lists. Likewise, objects that are not within the top-$n$ objects in both lists are inliers, and are thus typical objects in the dataset.

Given the above, we compare the similarity of two ranked lists of outliers, $R^A$ and $R^B$, by counting the total fraction, $f\in[0, 1]$, of feature vectors belonging in the top-$n$ in both lists, or more formally their intersection:

\begin{equation}  
f=|R^A \cap R^B|\cdot n^{-1},\label{eqn:similarity_sets}
\end{equation}
 where $r_i\in R^A < n$ and $r_i\in R^B < n$.

To reach a trade-off between underfitting and overfitting, we search a grid of evenly spaced $\epsilon$ values, $G$, where  each value of $\epsilon$ is denoted as $\epsilon_j$ where $j=1, 2, \ldots, |G|$, and $\epsilon_j\in[\epsilon_{min}, \epsilon_{max}]$. $\epsilon_{min}$ and  $\epsilon_{max}$ refer to the minimum and maximum values of $\epsilon$ searched respectively, and their selection will be described later in Section~\ref{sec:eps_search_grid}. 

For each value of $\epsilon_j$, we derive a corresponding ranking of outliers, denoted as $R_j$, where $R_1$ corresponds to the ranking given by $\epsilon_{min}$ and $R_{|G|}$ corresponds to the ranking given by $\epsilon_{max}$.

$\epsilon_{min}$ refers to a small search radius, and thus would produce a ranked list that overfits the data. Likewise, $\epsilon_{max}$ refers to a large search radius that would underfit the data. To find a good value of $\epsilon_j\in[\epsilon_{min}, \epsilon_{max}]$, we compute the similarity (using Equation~\ref{eqn:similarity_sets}) between $R_1$ and $R_j$, and the similarity between $R_{|G|}$ and $R_j$. These two sets of similarity values, corresponding to those generated as compared to $R_1$ and $R_{|G|}$ are denoted as $U^{min}$ and $U^{max}$, respectively, where $|U^{min}|=|U^{max}|=|G|$. 

Finally, the selected value of $\epsilon$ is given by the value of $j$ that minimizes the difference in similarity values, which is described below: 
\begin{equation}
\epsilon^{select}=\epsilon_{argmin(|U^{min}_j-U^{max}_j|)}.
\label{eqn:eps_select_minimization}
\end{equation}

\noindent\textbf{Illustrative Example:} To better illustrate these concepts, consider the following concrete example that computes $U^{min}$ and $U^{max}$ with $|G|=3$ (i.e., for illustrative purposes we only consider computing the fraction of points that match for 3 values of $\epsilon$). Here, $|G|=3$ implies that $j=1, 2, 3$, and that $\epsilon_{min}=\epsilon_1$ and $\epsilon_{max}=\epsilon_3$.

The set of top-$n$ outlying points found by $\epsilon_{1}$, $\epsilon_{2}$ and $\epsilon_{3}$ are as follows, where we set $n=10$. We color code the points that are common between sets.

$\epsilon_1=\epsilon_{min}=\{\textcolor{red}{1}, \textcolor{red}{2}, \textcolor{red}{3}, \textcolor{red}{4}, 5, 16, 17, 18, 19, 20\}$.

$\epsilon_2=\{\textcolor{red}{1}, \textcolor{red}{2}, \textcolor{red}{3}, \textcolor{red}{4}, \textcolor{blue}{8}, \textcolor{blue}{9}, \textcolor{blue}{10}, 100, 101, 102\}$.

$\epsilon_3=\epsilon_{max}=\{\textcolor{red}{1}, \textcolor{red}{2}, \textcolor{blue}{8}, \textcolor{blue}{9}, \textcolor{blue}{10}, 11, 12, 13, 14, 15\}$.

Using Equation~\ref{eqn:similarity_sets} we compute $U^{min}$ and $U^{max}$ which compares the similarity between the top-$n$ points found by $\epsilon_{min}$ and $\epsilon_{max}$, respectively. Specifically, $U^{min}$ is computed by comparing $\epsilon_1$ to those found by $\epsilon_1$ (itself), $\epsilon_2$, and $\epsilon_3$. $U^{max}$ is computed by comparing $\epsilon_3$ to $\epsilon_1$, $\epsilon_2$, and $\epsilon_3$ (itself). We highlight that because we compare the same set to itself, this guarantees that there is a 1.0 match fraction in $U^{min}$ and $U^{max}$.

Using the example values above and Equation~\ref{eqn:similarity_sets} we compute $U^{min}$ and $U^{max}$ as follows:

$U^{min}=\{10/10, 4/10, 2/10\}=\{1.0, 0.4, 0.2\}$.

$U^{max}=\{2/10, 5/10, 10/10\}=\{0.2, 0.5, 1.0\}$.

Observe that because we compare the smallest value of $\epsilon$ to increasingly larger values of $\epsilon$, the match fraction declines. And similarly because we  compare the largest value of $\epsilon$ to increasingly larger values of $\epsilon$, the match fraction increases. This produces an intermediate value of $\epsilon$ that will be selected that reaches a trade-off between the smallest and largest $\epsilon$ values.

\begin{figure}[!t]
\centering
\subfigure[]{
       \includegraphics[width=0.4\textwidth]{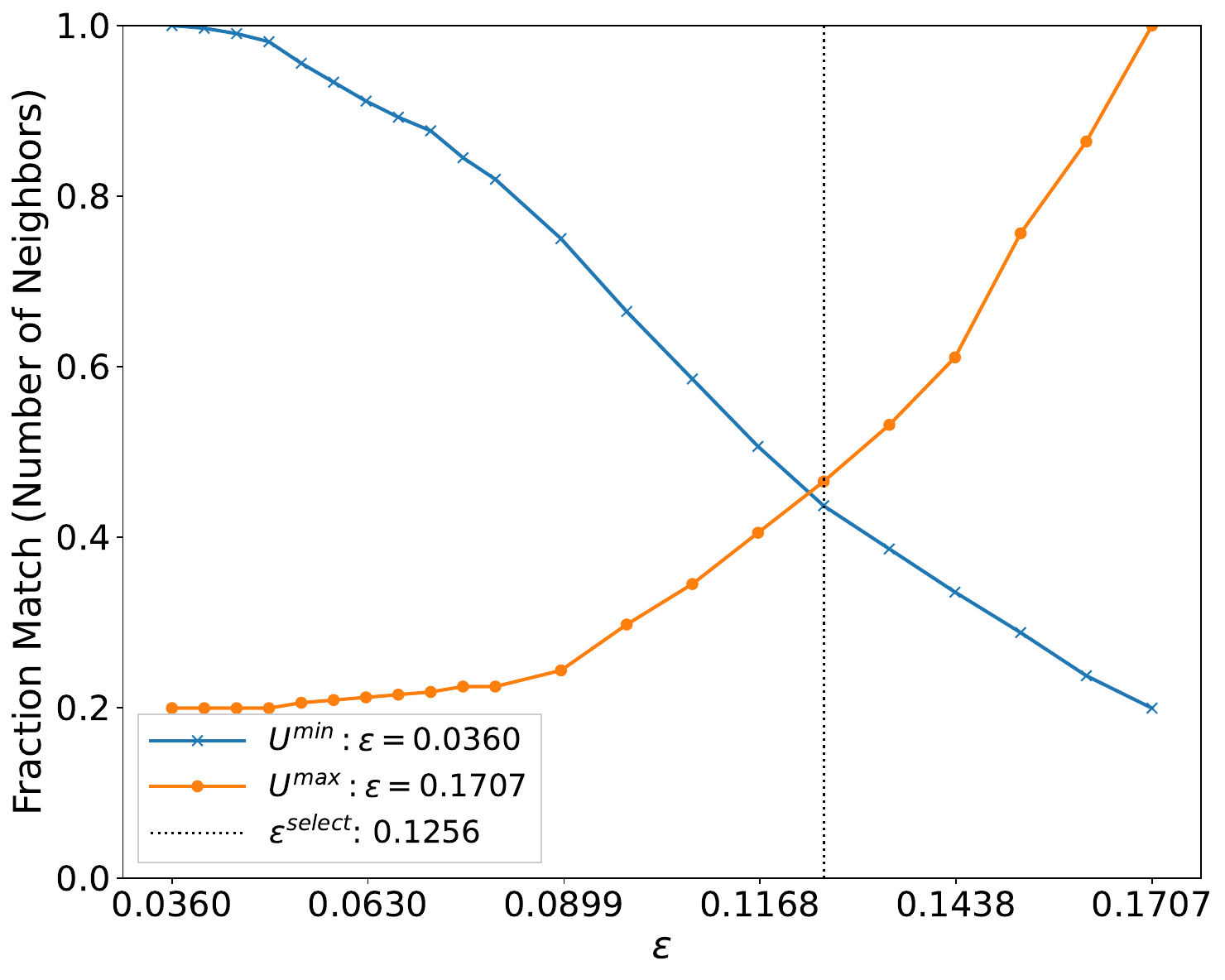}
}
\subfigure[]{
       \includegraphics[width=0.4\textwidth]{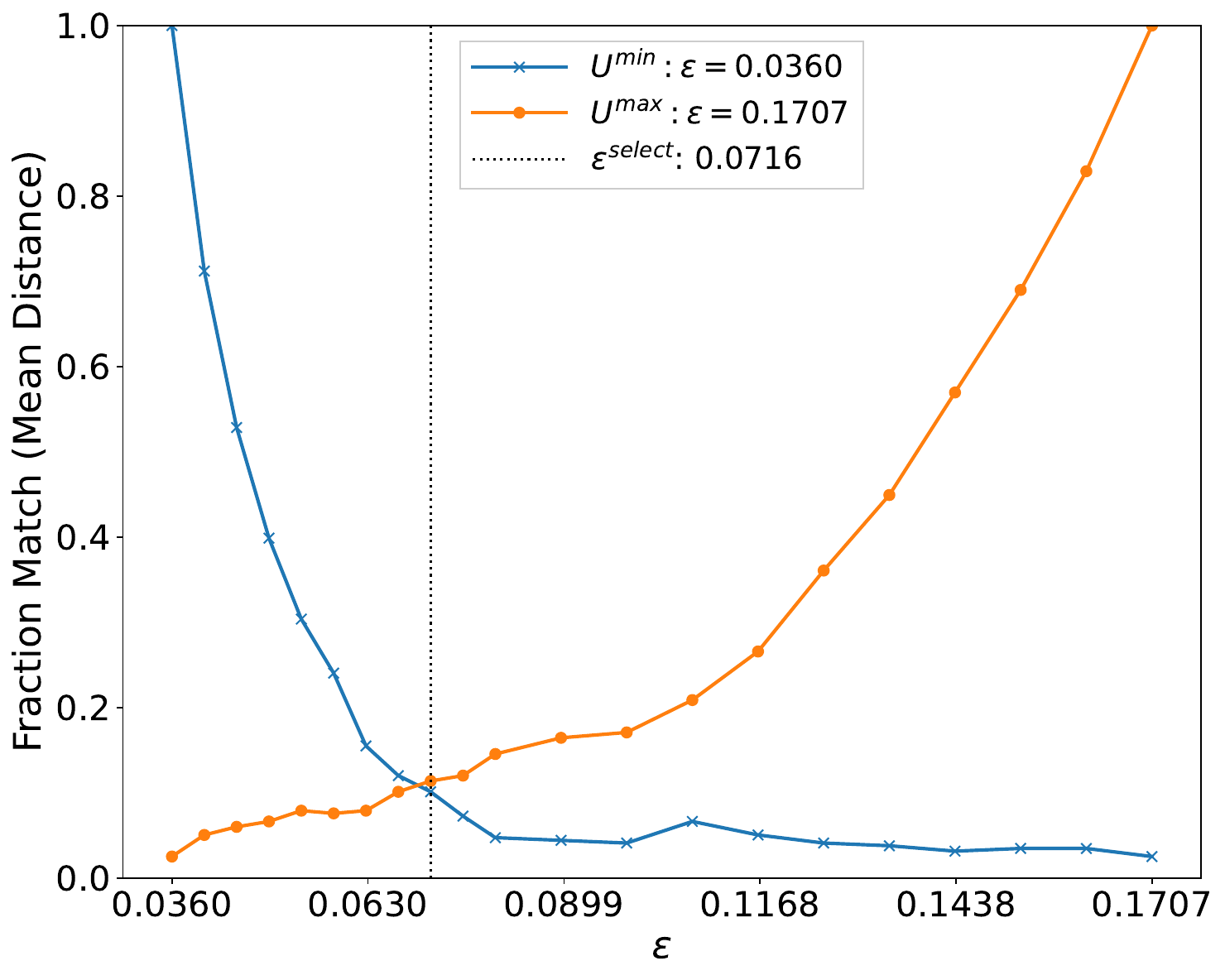}
}

    \caption{Example plots showing computing the best value of $\epsilon$ on an example $d=5$ feature space, which are the first five features in Table~\ref{tab:population_outlier_properties}. $U^{min}$ ($U^{max}$) refers to the fraction ($f$) of the top-$n$  objects that are in both ranked lists for $\epsilon=0.0360$ ($\epsilon=0.1707$) and the $\epsilon$ values on the horizontal axis. The value of $\epsilon^{select}$  is shown as the vertical dotted line for  (a) the number of neighbors metric (distance-oblivious), and (b) the mean distance metric (distance-aware).}
   \label{fig:epsilon_selection_example}
\end{figure}

Figure~\ref{fig:epsilon_selection_example} shows a plot of this selection procedure on one of the feature spaces, where (a) shows the number of neighbors metric, and (b) shows the mean distance metric (as defined in Section~\ref{sec:ranking_metrics_dssj}).  The plot shows how similar the rankings are between the $\epsilon$ value on the horizontal axis as compared to the smallest ($U^{min}$) and largest ($U^{max}$) $\epsilon$ values, which are those that are expected to overfit and underfit the data. Thus, as $\epsilon$ increases, the fraction match for $U^{min}$ decreases, whereas it increases for $U^{max}$. A trade-off for $\epsilon^{select}$ is reached where $\epsilon_{min} < \epsilon^{select} < \epsilon_{max}$. Note that $\epsilon$ is evenly sampled on the horizontal axis, which is somewhat misleading. While the axis shows \emph{linear} increases in $\epsilon$, this leads to a \emph{non-linear} increase in the \emph{search volume} which is a function of the data dimensionality. Thus, it is similar to a logarithmic sampling of the search volume on the horizontal axis. However, fine-grained sampling is unnecessary at high values of $\epsilon$, as a large value of $\epsilon^{select}$ is unlikely to minimize the difference between $U^{min}_j$ and $U^{max}_j$ (Equation~\ref{eqn:eps_select_minimization}). We will examine this in more detail when we show the distribution of $\epsilon^{select}$ values across all feature spaces in Section~\ref{sec:eval_eps_k_select}.

In this section, we omitted how $\epsilon_{min}$ and  $\epsilon_{max}$ are selected. In what follows, we describe how we select these values.

\subsubsection{Selection and Computation of the $\epsilon$ Search Grid}\label{sec:eps_search_grid}

A major challenge of using \dssj for outlier detection is that the values of $\epsilon_{min}$ and $\epsilon_{max}$ vary for each feature space. For instance, with lower dimensional data, the search radius needed to find an average number of neighbors per point is smaller than for high dimensional data, as the total volume of the feature space increases exponentially with the number of dimensions. Furthermore, since the data in the feature spaces are unlikely to follow any well-defined data distributions, the values of $\epsilon$ need to be found numerically as they cannot be found analytically.

To address this, we define selectivity which is the mean number of neighbors within $\epsilon$ found in a dataset for \dssj~\citep{gallet2021heterogeneous}. The definition is as follows: $s=(|A|-|D|)\cdot|D|^{-1}$ where $|D|$ is the number of feature vectors and $|A|$ is the total result set size (the total number of pairs of points within $\epsilon$ of each other) of the self-join operation. Thus, it computes the mean number of neighbors within $\epsilon$ in a dataset, excluding a point/feature vector finding itself. When processing \snapshotA, we set $s_{min}\approx0.001|D|$ and $s_{max}\approx0.15|D|$, corresponding to a minimum and maximum selectivity of 0.1\% and 15\% of the dataset. The results of \dssj using these two selectivity values will overfit and underfit the data respectively. 

To find the values of $\epsilon_{min}$ and $\epsilon_{max}$ that yield the minimum and maximum selectivity values described above, we use the method outlined in~\citet{Gowanlock2021KNNJPDC} which was utilized to select a search distance $\epsilon$ that finds on average at least $k$ neighbors for each point in a dataset for the $k$-nearest neighbor self-join algorithm. The method is lightweight as it samples the dataset to first estimate the mean distance between points in a dataset ($\epsilon_{mean}$), which is used as the upper bound on a practical search radius to find $\epsilon_{max}$ (or $\epsilon_{min}$). Then, the distances between a sample of points and points within the dataset are computed to create a histogram of distances between points. Using this histogram, a cumulative distance distribution histogram is computed which yields the relationship between a search distance and the average number of neighbors that will be found per point in the dataset. We use this histogram to select the search radii $\epsilon_{min}$ and $\epsilon_{max}$.   We refer the reader to~\citet{Gowanlock2021KNNJPDC} for further detail.

\begin{figure}[!t]
\centering


\begin{tikzpicture}[scale=2]
\filldraw[fill=red](1,1) circle (0.1cm) node[]{$q$};
\foreach \x\y in {1/1,2/1.6, 3/2.4} {
	\draw (1,1) circle (\x*0.5cm);
	\node at (1.12+0.5*\x,1) {$\epsilon_\x$};
}
\foreach \x/\y/\z in {0.3/1.1/1, 1.2/1.3/2, 0.8/0.8/3, 0.9/1.8/4, 1/0.2/5, 1.1/0.6/6, -0.3/0.84/7, 1.1/2.21/8, -0.5/0.3/9, 2.5/1.8/10} {
\filldraw [fill=black](\x,\y) circle (0.03cm) node[anchor=south east]{\z};	
}
\end{tikzpicture}
    \caption{Illustration of several result sets for query point $q$ in a $d=2$ feature space, where data points are shown as black dots and $q$ is shown as the red dot. Three evenly spaced search distances are shown ($\epsilon_1, \epsilon_2, \epsilon_3$), which are illustrated as circles centered on query point $q$. The result sets ($A$) are as follows: $A(\epsilon_1)=\{2, 3, 6\}$; $A(\epsilon_2)=\{1, 4, 5\} \cup A(\epsilon_1)$; $A(\epsilon_3)=\{7, 8\} \cup A(\epsilon_1) \cup A(\epsilon_2)$. Points 9 and 10 are not within any of the result sets. Thus, the result sets for $\epsilon_1$ and $\epsilon_2$ can be directly derived from that of $\epsilon_3$.}
   \label{fig:epsilon_max_example}
\end{figure}
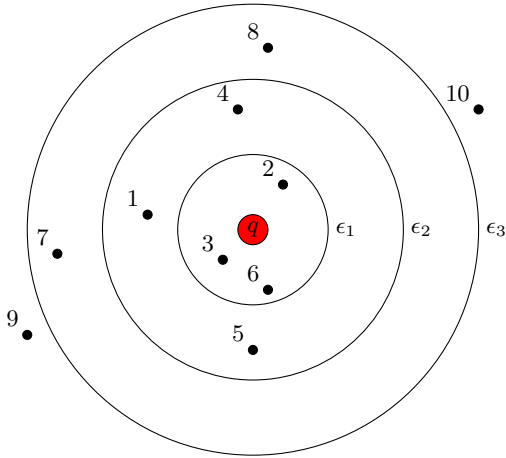

After $\epsilon_{min}$ and $\epsilon_{max}$ are found, the $\epsilon$ grid $G$ is computed, which is used to derive the rankings for all feature vectors in the feature space and a value of $\epsilon$ is selected that reaches a trade-off between overfitting and underfitting (Section~\ref{sec:dssj_under_overfit}). We use a computationally efficient approach to compute the result sets ($A$) for each  $\epsilon_j\in G$. Consider that the result set $A$ for $\epsilon_{max}$ contains all of the neighbors for lower values of $\epsilon$ as well. Thus, we execute \dssj once for $\epsilon_{max}$ and use this result set to derive the result sets for all $\epsilon_j \in G$. Figure~\ref{fig:epsilon_max_example} shows an example of three search radii ($\epsilon_1, \epsilon_2, \epsilon_3$), where $\epsilon_1=\epsilon_{min}$ and $\epsilon_3=\epsilon_{max}$. The figure shows that all result sets for $\epsilon_j<\epsilon_{max}$ can be derived from $\epsilon_{max}$. Thus, independently executing \dssj for all $\epsilon_j\in G$ is unnecessary (and would be very computationally expensive), as the result set for $\epsilon_{max}$ can be filtered to compute the result sets for smaller search distances.

\subsection{$k$-Nearest Neighbor Self-Join}\label{sec:KNNSJ_description}
Intuitively, the \knnsj algorithm can be employed for unsupervised outlier detection where an outlier is characterized as having a significant number of its \knn located at large distances.

\subsubsection{Ranking Metrics}\label{sec:ranking_metrics_knnsj}
Similarly to \dssj, the \knnsj algorithm has two ranking metrics outlined as follows, where one metric is distance-oblivious and one is distance-aware. This yields two different rankings for the \knnsj method.

\begin{enumerate}
\item \emph{Distance-oblivious:} The ranking function $R$ is simply an in-degree ranking of the \knn graph~\citep{hautamaki2004outlier}, which is also known as its reverse nearest neighbors. The in-degree refers to the number of occurrences that a point is found within another point's set of \knn. The fewer the number of occurrences, the more likely the point is an outlier.
\item \emph{Distance-aware:} The ranking function is a ranking of mean distances between a point and its \knn. Intuitively, if a given point has its $k$ neighbors at large distances, then this implies that it is an outlier.
\end{enumerate}

\subsubsection{On The Selection of $k$ and the \knn Search Grid}\label{sec:knnsj_under_overfit}

For brevity we briefly provide an overview of the procedure for \knnsj because the method is analogous to that of \dssj. We highlight similarities and key differences between outlier detection with \knnsj compared to \dssj. 

Similarly to \dssj, the detection of outliers with the \knnsj algorithm is dependent on the value of $k$. Unlike the \dssj algorithm, it is more straightforward to detect outliers with the \knnsj algorithm because we do not need to perform a search to find $\epsilon_{min}$ and $\epsilon_{max}$.  A low value of $k$ will overfit the data and a large value of $k$ will underfit the data. We reach a trade-off using an analogous method to \dssj which searches a grid of $k$ values where similarity is measured as compared to $k_{min}$ and $k_{max}$, and the value of $k$ which reaches a trade-off between under- and overfitting is selected. Likewise,  we use a computationally efficient approach to compute the grid of $k$ values, denoted as $k_j \in G$ where $j=1, 2, \ldots, |G|$. For each value of $k_j$, we compute the $k$ neighbors for $k_{max}$, which contain the neighbors for all $k_j<k_{max}$.  When processing \snapshotA, we set $k_{min}=2$ and $k_{max}=4096$.

\begin{figure}[!t]
\centering
\subfigure[]{
       \includegraphics[width=0.4\textwidth]{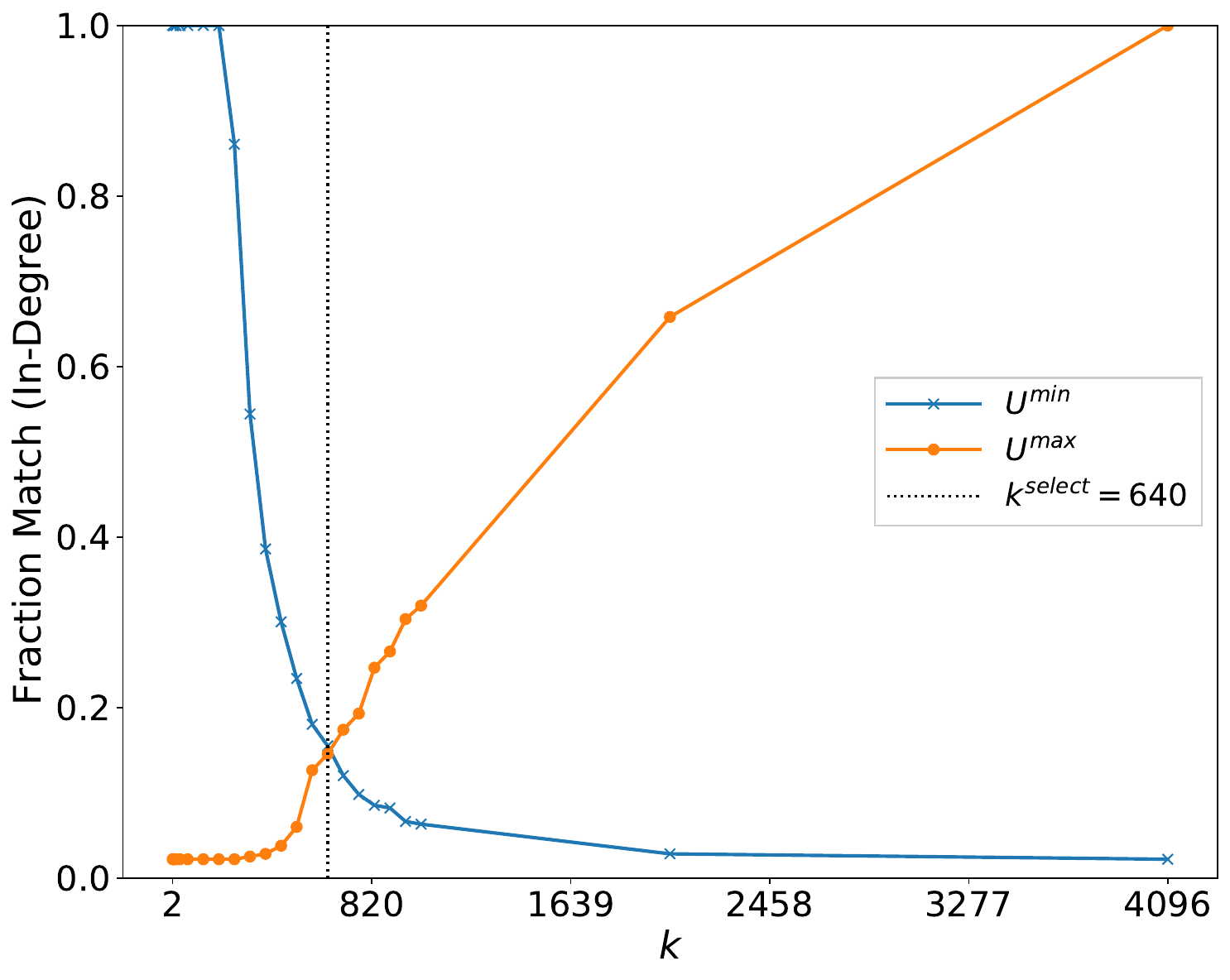}
}
\subfigure[]{
       \includegraphics[width=0.4\textwidth]{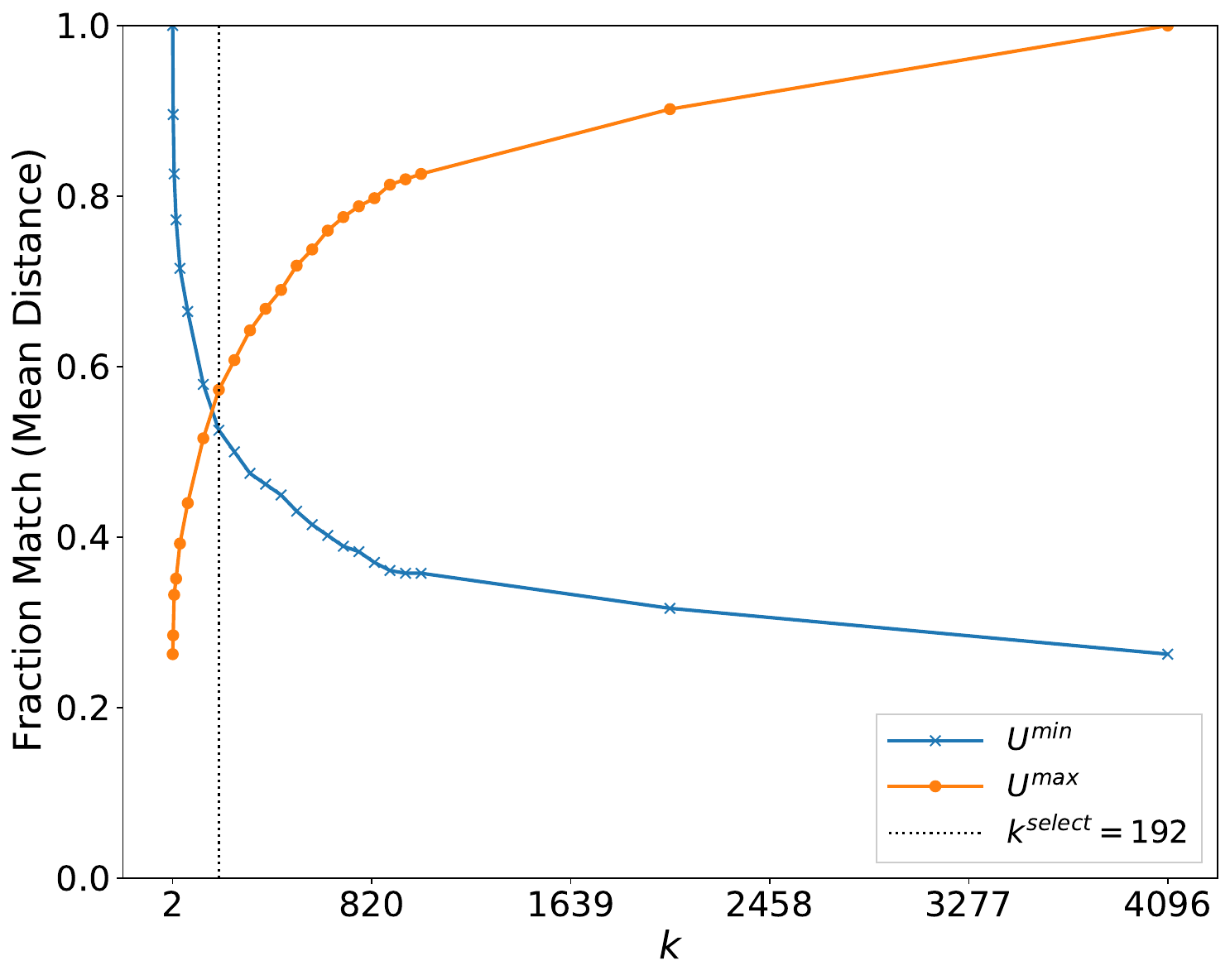}
}

    \caption{Example plots showing the computation of the best value of $k$ using $k_{min}=2$ and $k_{max}=4096$ on the same feature space shown in Figure~\ref{fig:epsilon_selection_example}. $U^{min}$ ($U^{max}$) refers to the fraction ($f$) of the top-$n$  objects that are in both ranked lists for $k=2$ ($k=4096$) and the $k$ values on the horizontal axis. The value of $k^{select}$  is shown as the vertical dotted line for  (a) the in-degree metric (distance-oblivious), and (b) the mean distance to each point's $k$ neighbors (distance-aware).}
   \label{fig:k_selection_example}
\end{figure}

Figure~\ref{fig:k_selection_example} shows a plot of this selection procedure on one of the feature spaces, where (a) shows the in-degree metric, and (b) shows the mean distance metric (as defined in Section~\ref{sec:ranking_metrics_knnsj}). The plot shows how similar the rankings are between the $k$ value on the horizontal axis as compared to the smallest and largest values of $k$ ($k_{min}=2$ and $k_{max}=4096$), which are those that are expected to overfit and underfit the data. Thus, as $k$ increases, the fraction match for $U^{min}$ decreases, whereas it increases for $U^{max}$. A trade-off for $k^{select}$ is reached where $k_{min} < k^{select} < k_{max}$. Note that $k$ is unevenly sampled on the horizontal axis. This is because it is unlikely that $k^{select}$ will be large, so we do not perform a fine-grained sampling at larger values of $k$. Furthermore, we find that $k^{select}$ is typically bound to a particular range of $k$ values for the in-degree and mean distance metrics, and so we finely sample those regions. We will show the distribution of $k^{select}$ across all feature spaces in Section~\ref{sec:eval_eps_k_select}.

\subsection{Summary: Comparison of Outlier Approaches}\label{sec:summary_comparison_outlier_detection_approaches}

The \dssj and \knnsj methods each have two metrics which yield four total outlier detection rankings for each object (Sections~\ref{sec:ranking_metrics_dssj}~and~\ref{sec:ranking_metrics_knnsj}). We briefly compare and contrast these methods; however, their characteristics will be better understood with an example on real-world data that we will show in Section~\ref{sec:results_visualization_comparison_outlier_methods}.

Figure~\ref{fig:comparison_outlier_approaches} shows an illustrative example of (a) \dssj and (b) \knnsj, where the positions of the points are identical in both subfigures. Table~\ref{tab:comparison_outlier_approaches} shows the corresponding metric values and the resulting rankings for each.  Observe that the rankings for each method vary. Only point 10 has the same rank (0) across the four methods, because it has no neighbors within $\epsilon$ and it has an in-degree value of 0. Furthermore, the \dssj number of neighbors metric and the \knnsj in-degree metric have numerous rankings that are the same between points because the values are non-negative integers. In contrast, the \dssj mean distance and \knn mean distance to $k$ neighbors metrics are real numbers so the probability that two values are the same is low, and therefore it is unlikely that two points will share the same rank. The only exception is when no neighbors are found within the search radius $\epsilon$ for the \dssj mean distance metric (e.g., point 10 in Figure~\ref{fig:comparison_outlier_approaches}), in which case it will be assigned a rank of 0. From this example it appears that there is little consensus among the outlier detection techniques; however, this is primarily because there are only 10 points in the example. On real-world data there is much more consensus between the four rankings, and this will be discussed in greater detail in Section~\ref{sec:results_visualization_comparison_outlier_methods}.

\begin{deluxetable*}{r|r|r|r|r}
\tablecaption{The metric values (rankings) for the example shown in Figure~\ref{fig:comparison_outlier_approaches}. A ranking of 0 indicates the most outlying object.}\label{tab:comparison_outlier_approaches}
\tablewidth{\columnwidth}
\tabletypesize{\footnotesize}
\tablehead{
\colhead{Point}&\colhead{\dssj Num. Neighbors}&\colhead{\dssj Mean Distance}& \colhead{\knn In-Degree} & \colhead{\knn Mean Distance to $k$ Neighbors}
}
\startdata
 1&2 (2)&0.6185 (1)&5 (4)&0.7197 (5)\\
 2&2 (2)&0.6117 (3)&6 (5)&0.6435 (7)\\
 3&4 (3)&0.5541 (5)&6 (5)&0.5254 (8)\\
 4&2 (2)&0.5196 (7)&3 (3)&0.6537 (6)\\
 5&2 (2)&0.5224 (6)&2 (2)&0.7209 (4)\\
 6&2 (2)&0.3864 (9)&3 (3)&0.4933 (9)\\
 7&2 (2)&0.6149 (2)&2 (2)&0.7768 (3)\\
 8&1 (1)&0.4562 (8)&2 (2)&0.9133 (2)\\
 9&1 (1)&0.5758 (4)&1 (1)&1.0334 (1)\\
10&0 (0)&0.0000 (0)&0 (0)&1.4839 (0)\\
\enddata
\end{deluxetable*}


 \subsection{High-Level Population Outlier Detection System Description and Optimizations}

 In the prior sections, we discussed the outlier detection methods. Here, we describe  the GPU-accelerated implementations that are used in the system, and the integration of the constituent components of the system. The implementation uses shared libraries written in C/C++/CUDA using Python (NumPy) interfaces.

\subsubsection{GPU-Accelerated Implementations}
We employ two state-of-the-art \dssj and \knnsj algorithms both of which use GPU acceleration. \citet{Gowanlock2019DaMoN} proposed a \dssj algorithm for moderate to high-dimensional data. To enable searching in high dimensional feature spaces and avoid the \emph{curse of dimensionality} problem\footnote{See \citet{zimek2012survey} for an overview of the curse of dimensionality problem.} where index searches become increasingly exhaustive, \citet{Gowanlock2019DaMoN} proposed indexing the data in $c$ dimensions, where $c<d$. Thus, a representation of the data in $c$ dimensions avoids exhaustive index searches in high dimensions. We configure our software to index in $c=d$ dimensions when $d\leq 6$, and $c=6$ when $d>6$. For example, when we process the feature space with the greatest dimensionality $(d=15)$, we only index $c=6$ dimensions.

Several optimizations were subsequently made to the  algorithm outlined in~\citet{Gowanlock2019DaMoN}, including exploiting instruction level parallelism and changing the order in which query points are processed to decrease load imbalance in the GPU kernel as outlined by~\citet{Gowanlock2023}. This optimized algorithm is employed in our software\footnote{\url{https://github.com/mgowanlock/gpu_self_join}}. 

\citet{Gowanlock2021KNNJPDC} describes a hybrid CPU+GPU \knnsj algorithm for low-dimensional data, where $d\lesssim8$. The algorithm distributes query points to either the CPU or GPU based on the expected amount of work required of each query point. In short, query points in low density regions are assigned to the CPU and those in high density regions are assigned to the GPU. Because query points in high density regions require a significant number of distance calculations to refine the candidate set (those points returned by an index but that need to be refined using distance calculations), it is preferable to assign these queries to the GPU instead of the CPU because the former architecture has much higher throughput for this operation. The other facet of distance calculation complexity is the data dimensionality, $d$. For moderate to high dimensional data, the GPU significantly outperforms the CPU, and so there are cases where the CPU+GPU algorithm yields an insignificant performance gain over a GPU-only approach. Consequently, because the data we are using has $d=15$ dimensions, it is preferable to use the GPU-only version of the algorithm proposed by~\citet{Gowanlock2021KNNJPDC}. While not shown in that paper, the same method described for the \dssj above is employed which indexes the data in $c<d$ dimensions to allow for \knn searches in high dimensions. We use the publicly available source code of this algorithm\footnote{\url{https://github.com/mgowanlock/hybrid_k_nearest_neighbor_self_join}}.

\subsubsection{Computational Optimizations}\label{sec:optimizations}
 We describe several optimizations that are used to improve the performance of the system. These methods are presented independently from the results, but they will be of particular interest when interpreting the results. Thus, the reader may wish to read Section~\ref{sec:results} and then return to this section later.

\noindent\textbf{Multiprocessing Each Feature Space:} 
As described in Section~\ref{sec:num_feature_spaces}, there are 32,752 total feature spaces across $d=2-15$ dimensions. We use the multiprocessing library in Python to offload the processing of each of these feature spaces. Thus, each feature space is computed independently and concurrently with the others. 

\noindent\textbf{Oversubscription of GPU Resources:}
It is well known that there is overhead when communicating between the host (which contains the CPU and main memory) and the GPU~\citep{capodieci2017sigamma}. For instance, the \knnsj and \dssj methods will need to transfer their respective result sets from the GPU back to main memory. During this time, the GPU will not have anything to compute and so the resource will be underutilized. Underutilizing the GPU will significantly increase the total time needed to compute all of the feature spaces. 

  To address the underutilization problem, we oversubscribe the GPU and assign a maximum of $o$ processes to the GPU at a given time. Therefore, when one processes finishes computing its task (a GPU kernel invocation) and starts transferring its results back to the host, another process can start executing its GPU kernel. The oversubscription value, $o$, cannot be too large, otherwise GPU global memory capacity will be exceeded, and on our platform the global memory of a single GPU is 40 GiB. This will be described in greater detail in Section~\ref{sec:all_performance_results}.

 \noindent\textbf{Multi-GPU Scalability:}
 Because the feature spaces are independent, it is straightforward to assign a feature space to a given GPU, where either the \dssj or \knnsj algorithms are executed on the feature space. Thus, our system exploits the four GPUs on our hardware platform (the platform is described in additional detail in Section~\ref{sec:exp_method}). Because the feature spaces are of varying dimensionality, the amount of work needed to compute each feature space varies. We schedule the computation of feature spaces to GPUs using dynamic scheduling, where the least loaded GPU is assigned the next feature space. This yields good load balancing across the four GPUs.  We describe multi-GPU scalability in greater detail Section~\ref{sec:all_performance_results}.

 \noindent\textbf{Nested Parallelism of Tasks Across the CPU and GPU:}
 Recall from Sections~\ref{sec:eps_search_grid}~and~\ref{sec:knnsj_under_overfit} that we require finding a good value of $\epsilon$ and $k$ for \dssj and \knnsj, respectively, such that we do not arbitrarily select a value for these parameters that underfits or overfits the data. We use the result set for $\epsilon_{max}$ and $k_{max}$ to derive all result sets for lower values of $\epsilon_j$ and $k_j$, respectively. The GPU computes these result sets, and transfers them back to the host, where we then compute the result sets for the smaller values of $\epsilon_j<\epsilon_{max}$ and $k_j<k_{max}$ in parallel on the CPU. This allows the CPU to contribute to the overall computation instead of simply orchestrating data transfers to/from the GPU and performing other minor host-oriented tasks that are required for GPU computation.

\begin{deluxetable*}{l|l}
\tablecaption{Parameters used throughout the evaluation. The values of the parameters are given in parentheses where applicable.}\label{tab:eval_parameters}
\tablewidth{\columnwidth}
\tabletypesize{\footnotesize}
\tablehead{
\colhead{Parameter}&\colhead{Description}
}
\startdata
$n$ ($0.01|D|$)&The number of points considered when computing the similarity between two ranked lists. \\
&1\% of the dataset is used as we are interested in selecting those objects that are outliers.\\
$G$ & The grid of $\epsilon$ or $k$ values that are searched for \dssj and \knnsj, respectively.\\
$s_{min} (\approx 0.001|D|)$&The minimum selectivity used when defining the $\epsilon$ grid.\\
$s_{max} (\approx 0.15|D|)$&The maximum selectivity used when defining the $\epsilon$ grid.\\
$\epsilon^{select}$&The selected value of $\epsilon$ that reaches a trade-off between underfitting and overfitting a feature space.\\
$k_{min} (2)$&The minimum value of $k$ used when defining the grid of $k$ values.\\
$k_{max} (4096)$&The maximum value of $k$ used when defining the grid of $k$ values.\\
$k^{select}$&The selected value of $k$ that reaches a trade-off between underfitting and overfitting a feature space.\\
$o$ & The oversubscription factor, which is the maximum number of processes that are assigned to one GPU at a time.\\
\enddata
\end{deluxetable*}

\section{Results}\label{sec:results}
\subsection{Experimental Methodology}\label{sec:exp_method}
Our platform consists of 2x Intel(R) Xeon(R) Platinum 8358 CPUs (64 total physical cores) with a base clock speed of 2.60 GHz and 512 GiB of main memory. The platform is equipped with 4$\times$ Nvidia A100 GPUs, each of which has 40 GiB of global memory. The source code uses the following programming languages: Python, C/C++, and CUDA. All host C/C++ code is compiled with the O3 compiler optimization flag, and all CUDA code was compiled with CUDA v11.7. The C/C++ and CUDA code uses 32-bit floating point precision. All performance related experiments use an average of three time trials; however, the variance in the time measurements is insignificant, and error bars on these plots would be too small to observe so we do not report this information.

The platform is located within our institution's computer cluster, Monsoon, and we ensure that our experiments are carried out without interference from other users running on the node. The population outlier detection pipeline requires storing output (i.e., writing the rankings for each feature space to disk and diagnostic plots), and this occurs over the network where files are written to a parallel filesystem. We include this and all other overheads in the 
 end-to-end response time such that we accurately capture the performance of the system. 

For convenience, Table~\ref{tab:eval_parameters} summarizes the parameters used throughout the evaluation, which were defined in prior sections. Some parameters will be varied in the sections that evaluate the performance of the algorithms.


\subsection{Distribution of the Selected Values of $\epsilon$ and $k$}\label{sec:eval_eps_k_select}

We begin by examining the selected values of $\epsilon$ for the \dssj algorithm. Figure~\ref{fig:epsilon_selection_example} showed selecting a good value of $\epsilon$ on a single feature space by reaching a trade-off between underfitting and overfitting the data. Here, we present the distribution of $\epsilon^{select}$ values across all 32,752 feature spaces for the number of neighbors and mean distance metrics. In other words, we show the resulting $\epsilon^{select}$ values from an analogous analysis to that shown in Figure~\ref{fig:epsilon_selection_example}, but summarized across all feature spaces.

Note that it is not possible to show the distribution using the $\epsilon^{select}$ values because the range [$\epsilon_{min}$, $\epsilon_{max}$] varies across each individual feature space. Thus, we show the index $j$ of the selected value of $\epsilon_j \in G$, denoted as arg($\epsilon^{select}$). For example, in Figure~\ref{fig:epsilon_selection_example}(a), instead of reporting $\epsilon^{select}=0.1256$, we report arg($\epsilon^{select}$)$=15$.

\begin{figure}[!t]
\centering

      \includegraphics[trim={1cm 1cm 1cm 1cm}, width=1.0\columnwidth]{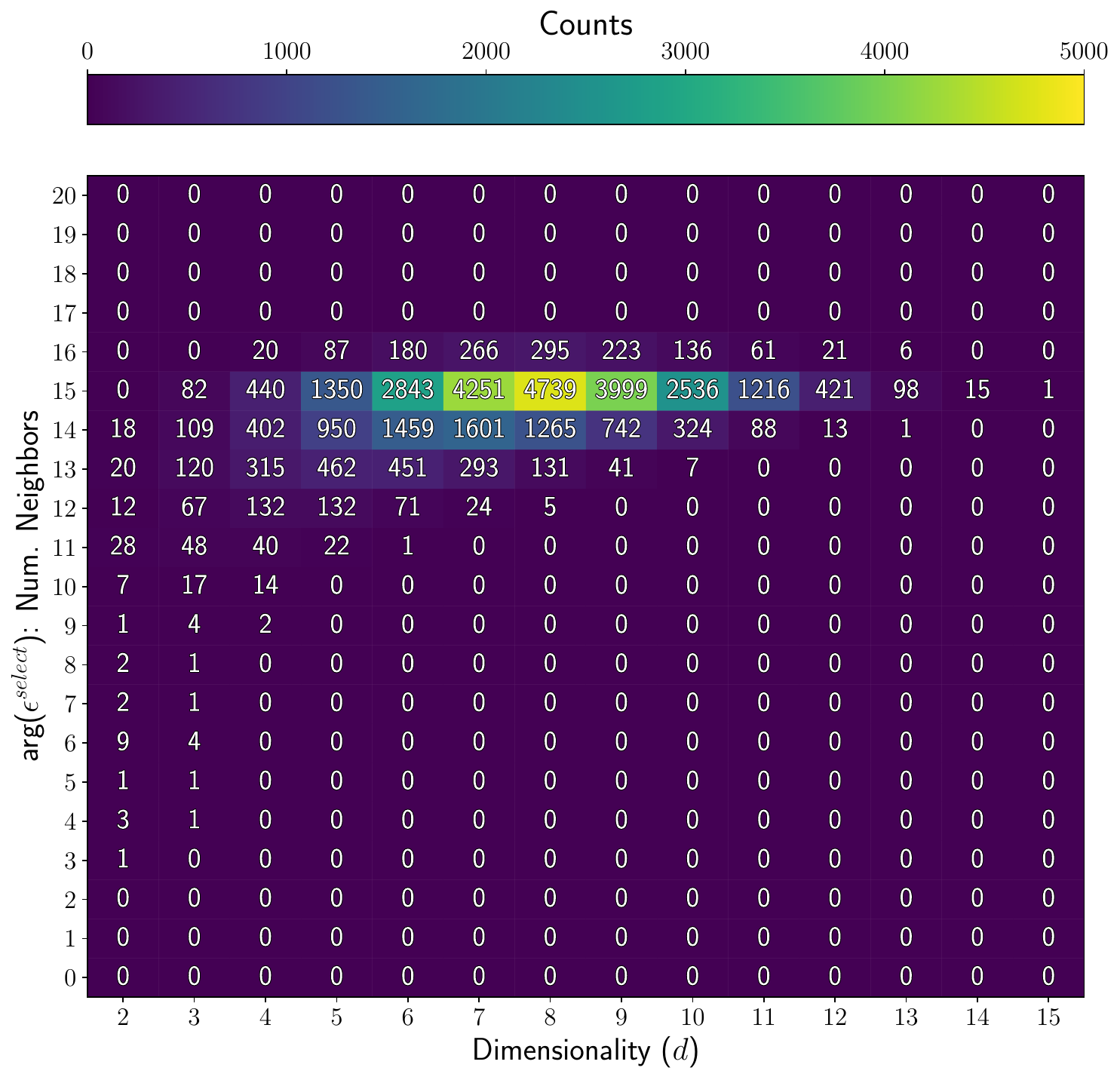}
    \caption{The distribution of $\epsilon^{select}$ values for the number of neighbors metric across all 32,752 feature spaces (Section~\ref{sec:num_feature_spaces}). Because the $\epsilon$ values can vary significantly between feature spaces, where $\epsilon$ tends to increase with dimensionality, we show the index $j$ of the selected value of $\epsilon_j \in G$, denoted as arg($\epsilon^{select}$).}
   \label{fig:heatmap_selected_eps_values_num_neighbors}
\end{figure}

\begin{figure}[!t]
\centering
       \includegraphics[trim={1cm 1cm 1cm 1cm}, width=1.0\columnwidth]{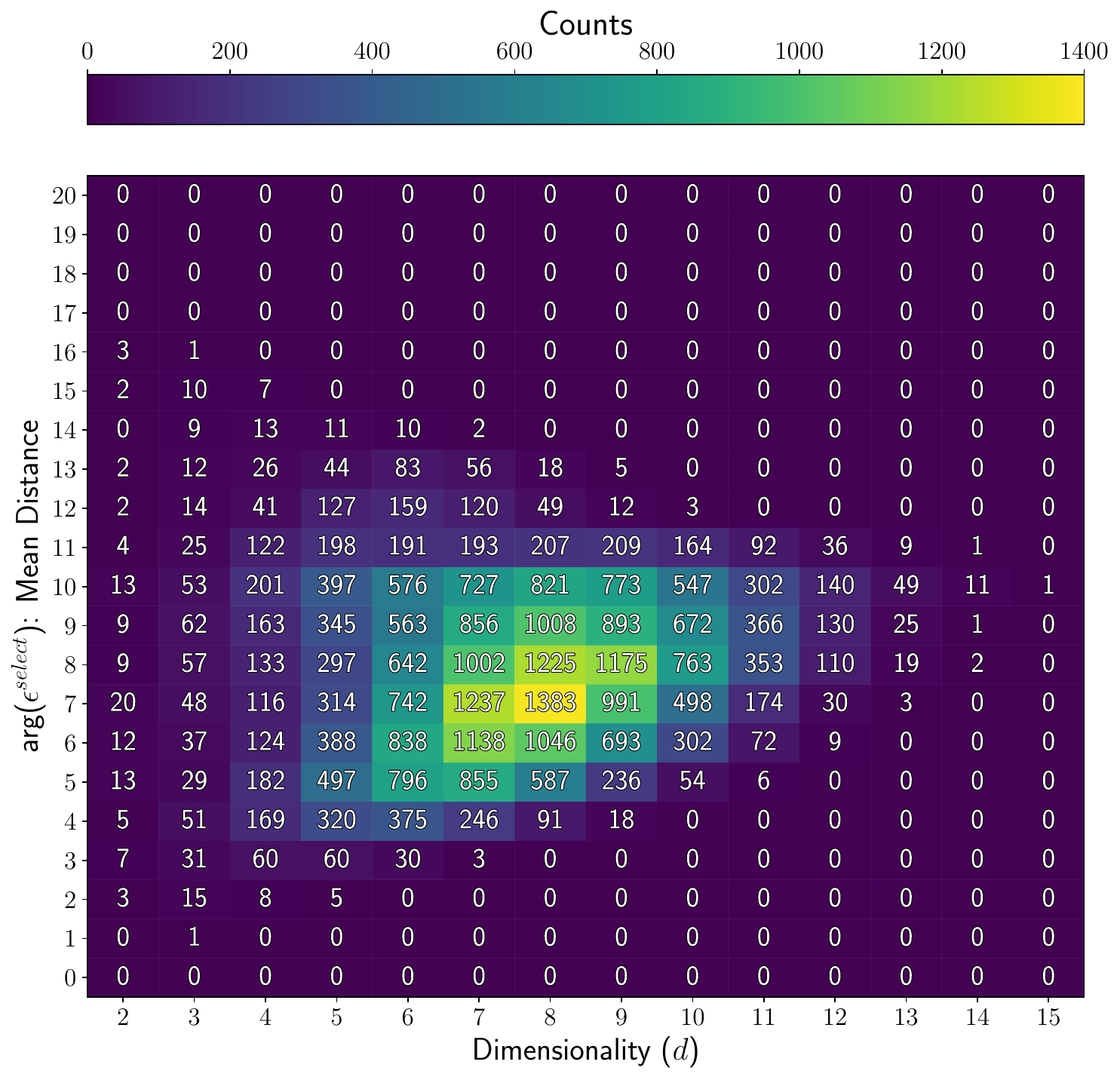}
    \caption{The same as Figure~\ref{fig:heatmap_selected_eps_values_num_neighbors}, but for the mean distance metric.}
   \label{fig:heatmap_selected_eps_values_mean_distance}
\end{figure}

Figures~\ref{fig:heatmap_selected_eps_values_num_neighbors}~and~\ref{fig:heatmap_selected_eps_values_mean_distance} plot arg($\epsilon^{select}$) as a function of the dimensionality of the feature spaces for the number of neighbors and mean distance metrics, respectively. Across all feature spaces, we find that $\epsilon^{select}$ is not found towards the extremes of the distribution of $\epsilon^{select}$ values for either metric, indicating that the range of selectivity parameters given by $s_{min}$ and $s_{max}$ do not need to be expanded to find a suitable $\epsilon^{select}$ for each feature space.  Comparing Figures~\ref{fig:heatmap_selected_eps_values_num_neighbors}~and~\ref{fig:heatmap_selected_eps_values_mean_distance}, on average, the former requires selecting a larger $\epsilon^{select}$ than the latter. We attribute this to the difference in ranking functions. The ranking function for the number of neighbors metric allows two or more points to share the same rank if they have the same number of neighbors within $\epsilon$. Thus a larger $\epsilon$ allows for the rankings to be more differentiated (i.e., fewer points have the same rank as $\epsilon$ increases), and so the under/overfitting trade-off is reached at a larger value of $\epsilon$. In contrast, the mean distance metric will not have two (or more) points with the same rank, as the probability of two points having the same mean distance to its neighbors is unlikely. This also explains why the distribution of arg($\epsilon^{select}$) is clustered in Figure~\ref{fig:heatmap_selected_eps_values_num_neighbors}, whereas the distribution is more diffuse in Figure~\ref{fig:heatmap_selected_eps_values_mean_distance}.


\begin{figure}[!t]
\centering
      \includegraphics[trim={1cm 1cm 1cm 1cm}, width=1.0\columnwidth]{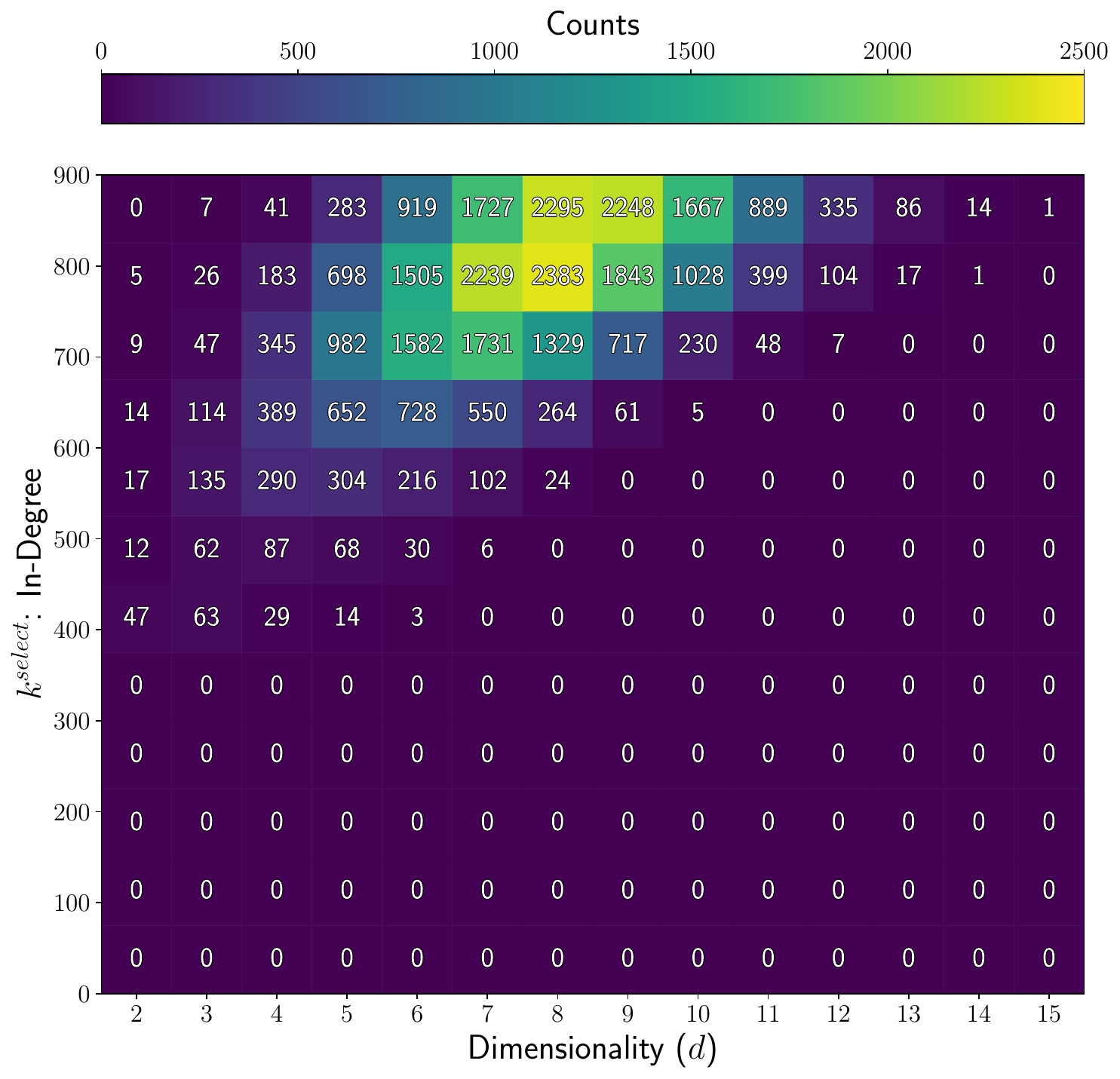}
    \caption{The distribution of $k^{select}$ values for the in-degree metric across all 32,752 feature spaces (Section~\ref{sec:num_feature_spaces}). To improve readability, the vertical axis is truncated and does not show all $k^{select}$ values which are mostly zeros. The plot shows 99.5\% of the $k^{select}$ values across all feature spaces.}
   \label{fig:heatmap_selected_k_values_in_degree}
\end{figure}

\begin{figure}[!t]
\centering
       \includegraphics[trim={1cm 1cm 1cm 1cm}, width=1.0\columnwidth]{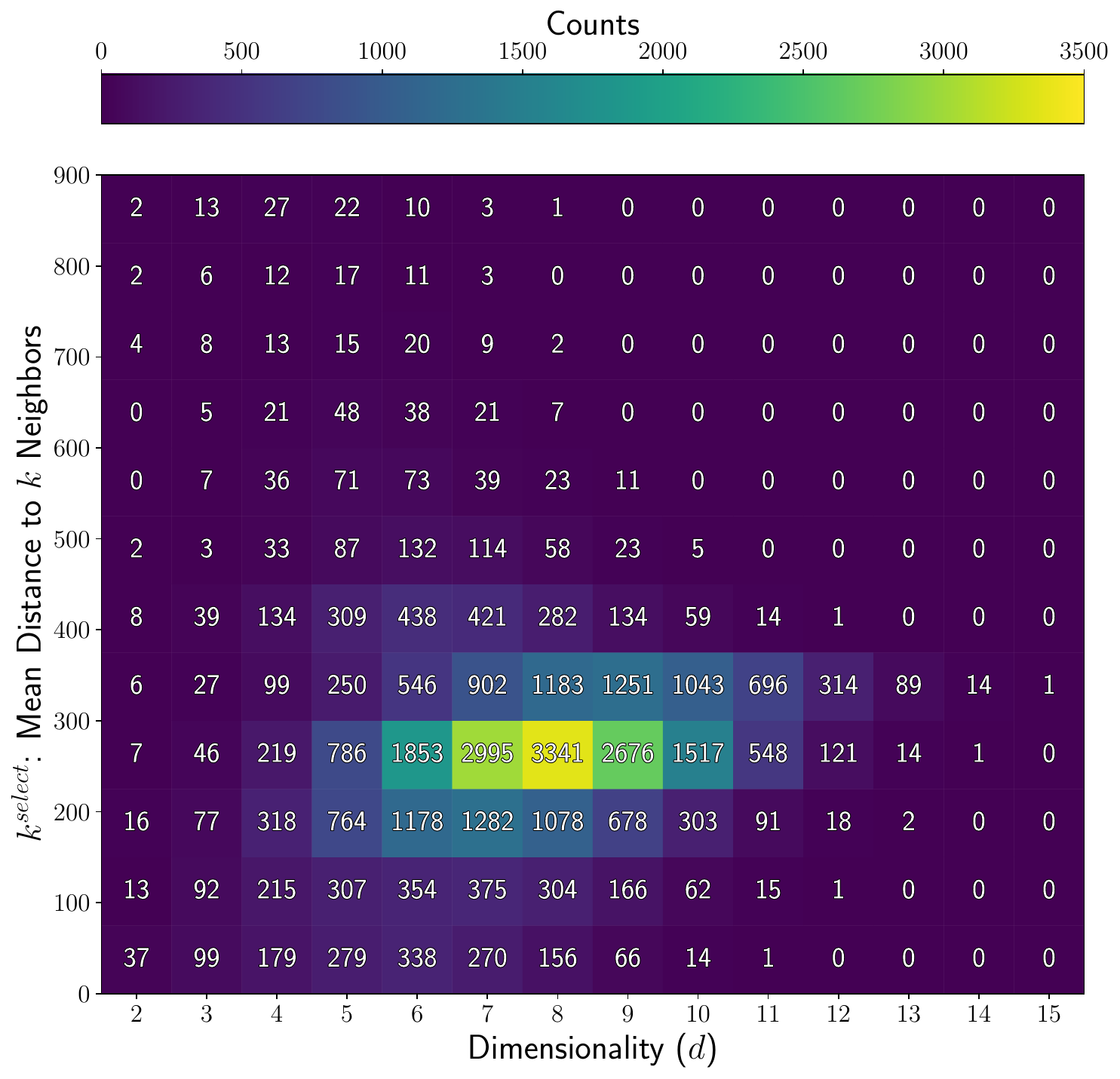}
    \caption{The same as Figure~\ref{fig:heatmap_selected_k_values_in_degree}, but for the mean distance to $k$ neighbors metric. The plot shows 98.5\% of the $k^{select}$ values across all feature spaces.}
   \label{fig:heatmap_selected_k_values_mean_distance}
\end{figure}

We examine the distribution of the selected values of $k$ for the \knnsj algorithm. Figures~\ref{fig:heatmap_selected_k_values_in_degree}~and~\ref{fig:heatmap_selected_k_values_mean_distance} plot $k^{select}$ as a function of the dimensionality of the feature spaces for the in-degree and mean distance metrics, respectively.  The range of $k$ values considered using the under/overfitting methodology is given by $k_{min}$ and $k_{max}$, which yields the range [2, 4096]. For both metrics, the selected values of $k$ are well below $k_{max}=4096$, and at most $k^{select}$ is found within the first $\sim25\%$ of the distribution at $k^{select}\lesssim 1000$. We observe similar trends between the in-degree metric (Figure~\ref{fig:heatmap_selected_k_values_in_degree}) and the number of neighbors metric for \dssj (Figure~\ref{fig:heatmap_selected_eps_values_num_neighbors}) where the distribution of $k^{select}$ is clustered, and this is due to points having the same rank if they have the same in-degree.

\subsection{Performance Results}\label{sec:all_performance_results}
We report the performance of the system by examining two metrics: (1) how performance varies as a function of oversubscription (Section~\ref{sec:optimizations}), and (2) how performance varies as a function of the number of GPUs (Section~\ref{sec:optimizations}). As outlined in Table~\ref{tab:eval_parameters}, we use $s_{max}\approx 0.15|D|$ for \dssj and $k_{max}=4096$ for \knnsj. These values largely dictate the amount of work computed, and these values constitute a large fraction of the size of the \snapshotA datasets, which only contains $|D|=31,693$ objects. Thus, we selected these parameters conservatively to show the worst-case performance of the system, as it is clear from Figures~\ref{fig:heatmap_selected_eps_values_num_neighbors}--\ref{fig:heatmap_selected_k_values_mean_distance} that $\epsilon^{select}$ and $k^{select}$ are found well below $\epsilon_{max}$ and $k_{max}$, respectively.

\begin{figure}[!t]
\centering
       \includegraphics[width=0.45\textwidth]{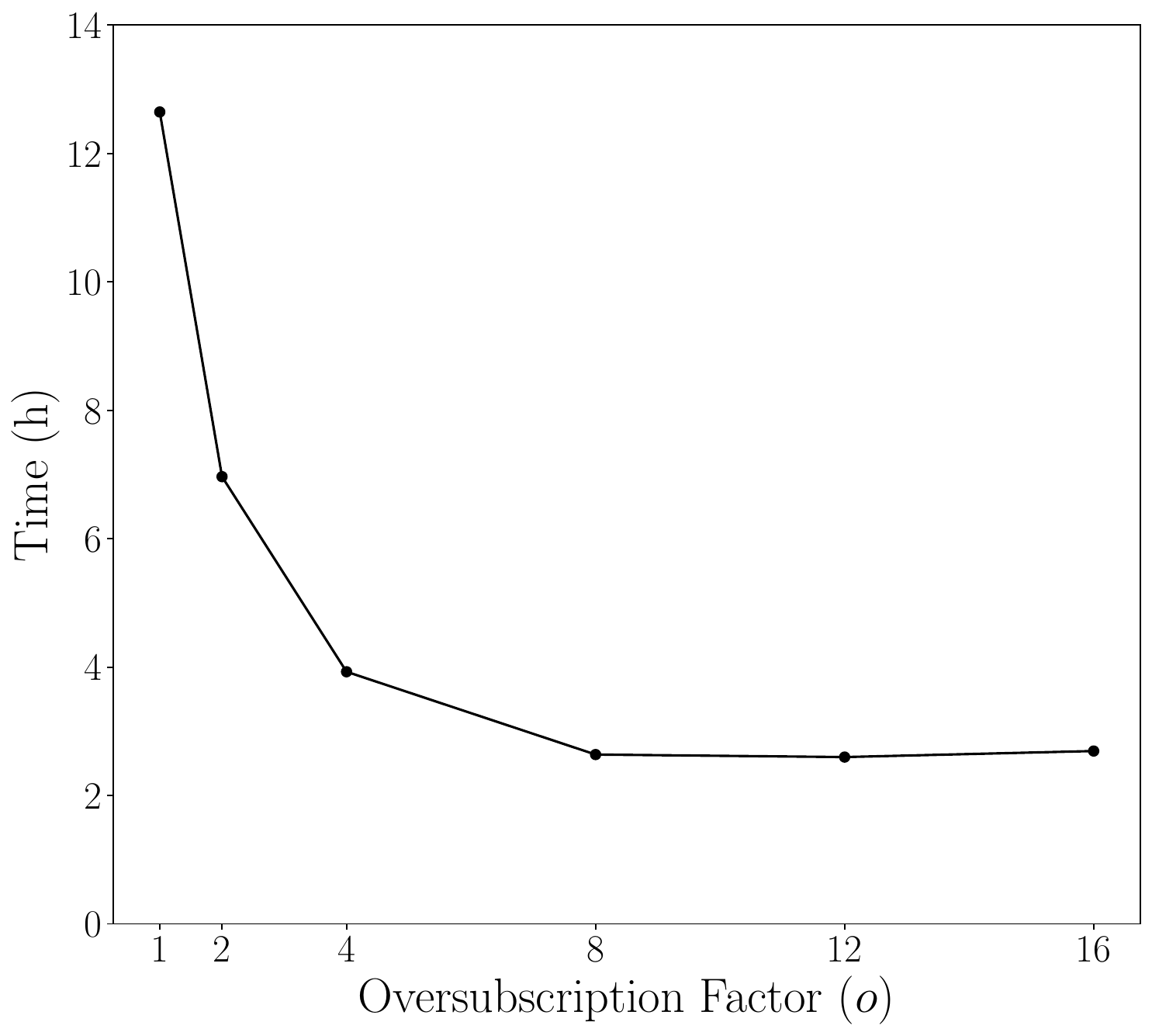}
    \caption{The response time (h) to compute $\epsilon^{select}$ and the corresponding rankings for both \dssj metrics as a function of the oversubscription ($o$) factor using 1 GPU. This includes the total time to find $\epsilon^{select}$ for both outlier detection metrics, and compute the rankings for all $|D|=32,752$ feature spaces.}
   \label{fig:performance_oversubscription_scalability_DSSJ}
\end{figure}

Recall that the oversubscription factor refers to the maximum number of processes that can be assigned to one GPU at a time. Because using the GPU requires data transfers between the host, and other host-side tasks be performed before and after the main GPU kernels are executed, it is beneficial to allow multiple processes concurrently access the GPU (and associated host side functionality). For instance, this allows PCIe data transfers to be overlapped with GPU computation, thus mitigating this data transfer cost. 

Figure~\ref{fig:performance_oversubscription_scalability_DSSJ} plots the response time (h) using 1 GPU as a function of the oversubcription factor ($o$) to find $\epsilon$ and compute the outlier rankings for \dssj.  From the figure we find that with $o=1$ (a maximum of a single process concurrently accessing the GPU), the response time is 12.65~h. However, if we use $o=8$ the response time is 2.64~h, which is 379\% faster than when no oversubscription is used ($o=1$). This demonstrates that the GPU is underutilized when $o=1$. Furthermore, we find that selecting $o=8-12$ achieves similar performance, but when $o=16$ the response time begins to increase. 

We find that using more than one GPU for \dssj does not improve performance. This is because there are significant host-side (CPU/main memory) tasks that are computed before the GPU kernel is executed and after the GPU returns the results to the host. Thus, the limiting factor for computing \dssj is not insufficient GPU resources, rather it is due to memory pressure and contention, which delays the execution of GPU kernels. This is largely a result of using Python as the ``glue'' programming language, where the bottleneck is not computation, but processing minor tasks in Python that require formatting the data needed for the C/C++ libraries and transferring results back to Python for further analysis.

We present the performance results for \knnsj. The \knnsj algorithm is more computationally expensive than \dssj, as the search radius needed to find at least $k$ neighbors will vary for each point in the dataset. For this reason, compared to \dssj the \knnsj algorithm achieves performance gains when multiple GPUs are used.

\begin{figure}[!t]
\centering
\subfigure[]{
       \includegraphics[width=0.45\textwidth]{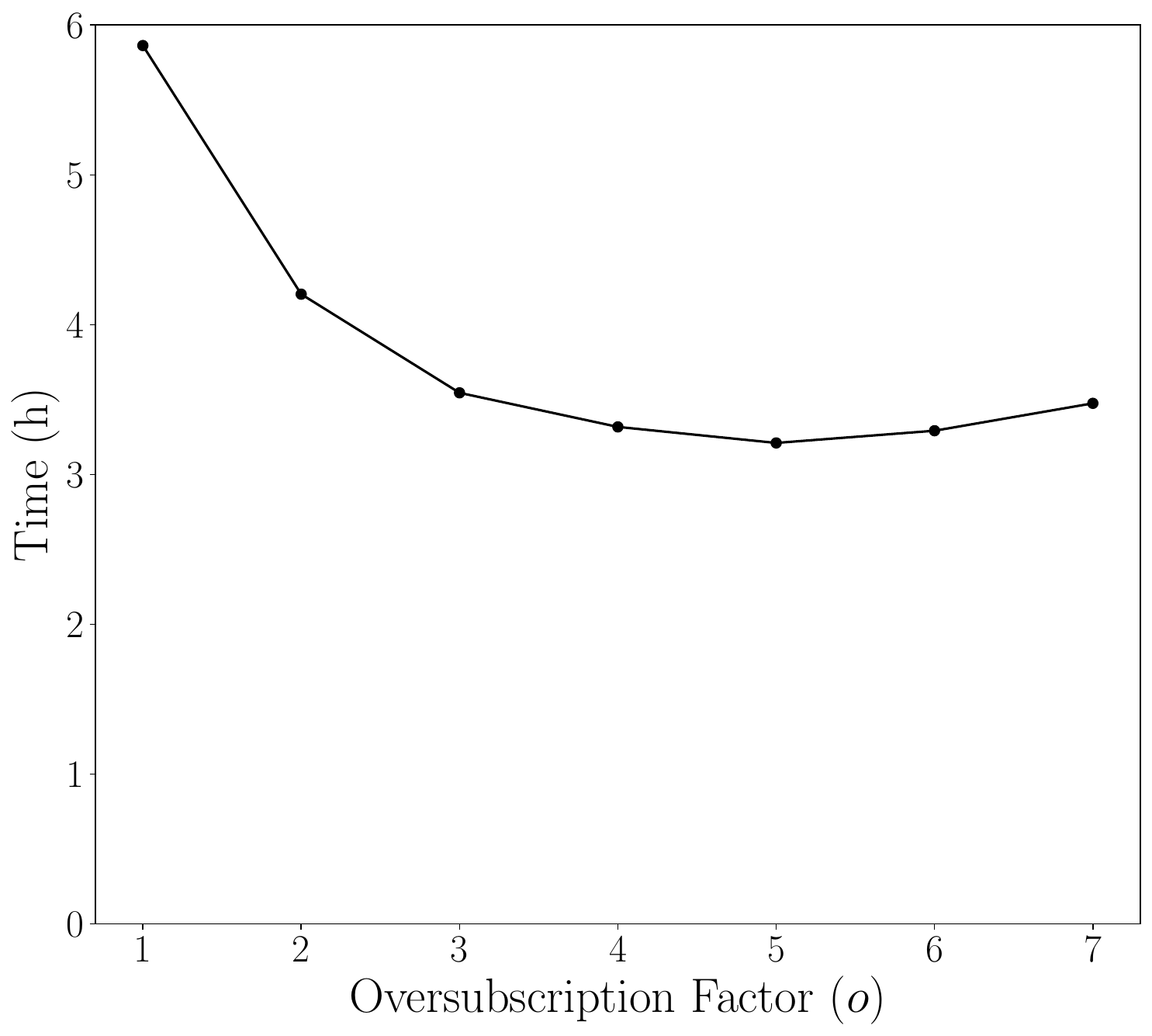}
       }
      \subfigure[]{
       \includegraphics[width=0.45\textwidth]{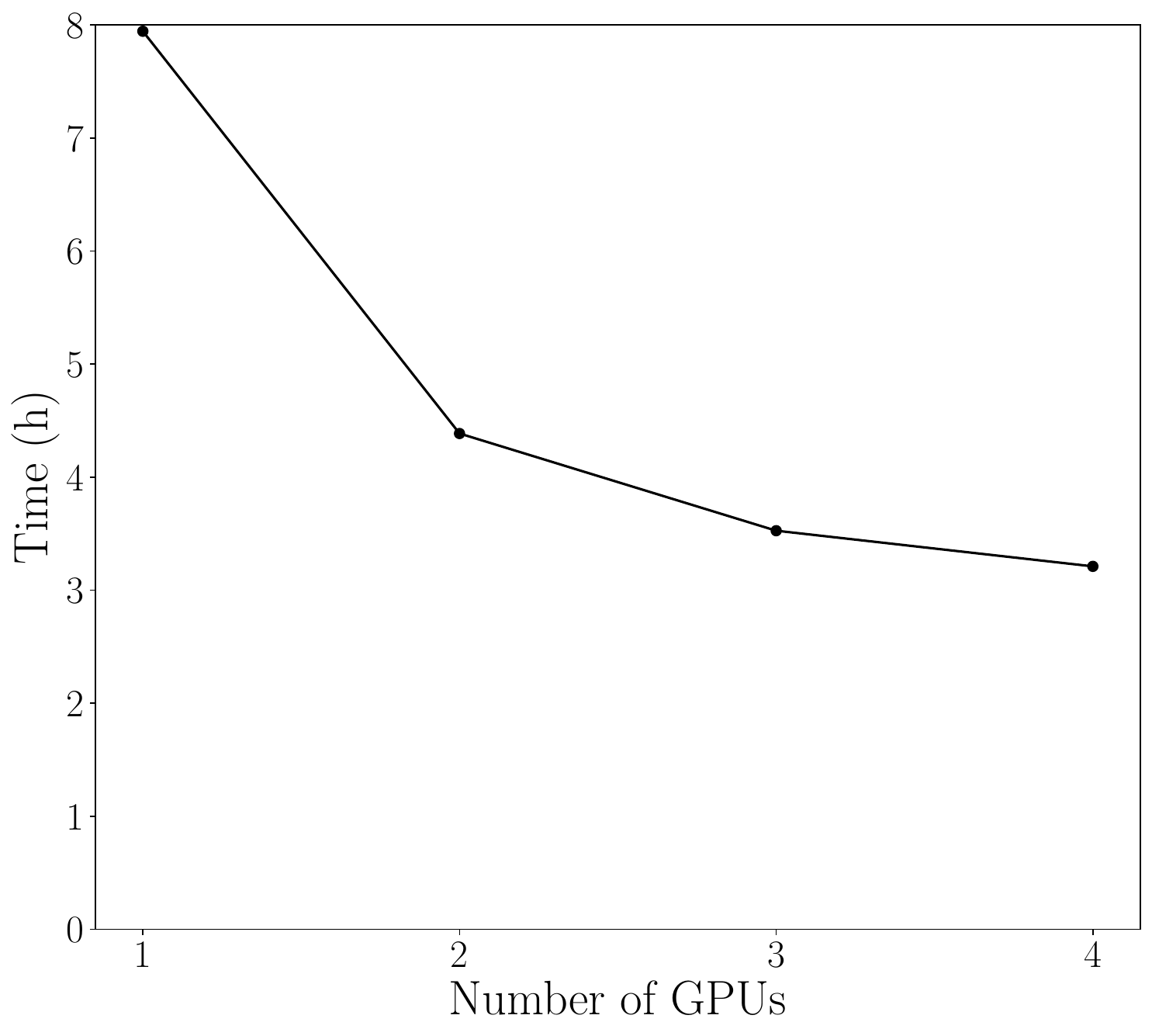}
       } 
    \caption{Performance results for \knnsj showing the total time to find $k^{select}$ for both outlier detection metrics, and compute the rankings for all $|D|=32,752$ feature spaces. (a) The response time (h) to compute $k^{select}$ and the corresponding rankings for both \knnsj metrics as a function of the oversubscription ($o$) factor with 4 GPUs. (b) The response time as a function of the number of GPUs, demonstrating the scalability when $o=5$.}
   \label{fig:performance_oversubscription_scalability_KNNSJ}
\end{figure}

Figure~\ref{fig:performance_oversubscription_scalability_KNNSJ}(a) plots the response time (h) for \knnsj  as a function of the oversubcription factor ($o$) using 4 GPUs. For the \knnsj algorithm, we find that oversubscription significantly improves performance over not using oversubscription where using $o=5$ is 83\% faster than not using oversubscription ($o=1$).

Figure~\ref{fig:performance_oversubscription_scalability_KNNSJ}(b) shows the response time (h) as a function of the number of GPUs where we set $o=5$ (the value of $o$ that achieved the best performance in panel (a)).  The speedup for 2--4 GPUs is as follows: 1.81$\times$, 2.25$\times$, and 2.47$\times$. We find that multi-GPU scalability is limited: a respectable performance gain is observed with 2 GPUs, but poor scalability is observed with 3 and 4 GPUs. This is due to the same reasons described above for \dssj where host-side tasks limit multi-GPU scalability.

As we will show in Section~\ref{sec:LSST_scale}, multi-GPU performance is much better when examining the much larger \lsstsynt dataset.

\subsection{Visualization and Comparison of Outlier Detection Methods}\label{sec:results_visualization_comparison_outlier_methods}
In this section, we show example visualizations of feature spaces and the detection of outliers. The visualizations illustrate the importance of using an ensemble of algorithms for outlier detection, as they all yield different outlier rankings which make some methods more or less suitable for a given scientific investigation. Because it is difficult to visualize feature spaces in more than 3 dimensions, we limit the visualizations to $d\leq3$.

\begin{figure*}[!t]
\centering
\subfigure[\dssj: Number of Neighbors Metric ]{
       \includegraphics[width=0.4\textwidth]{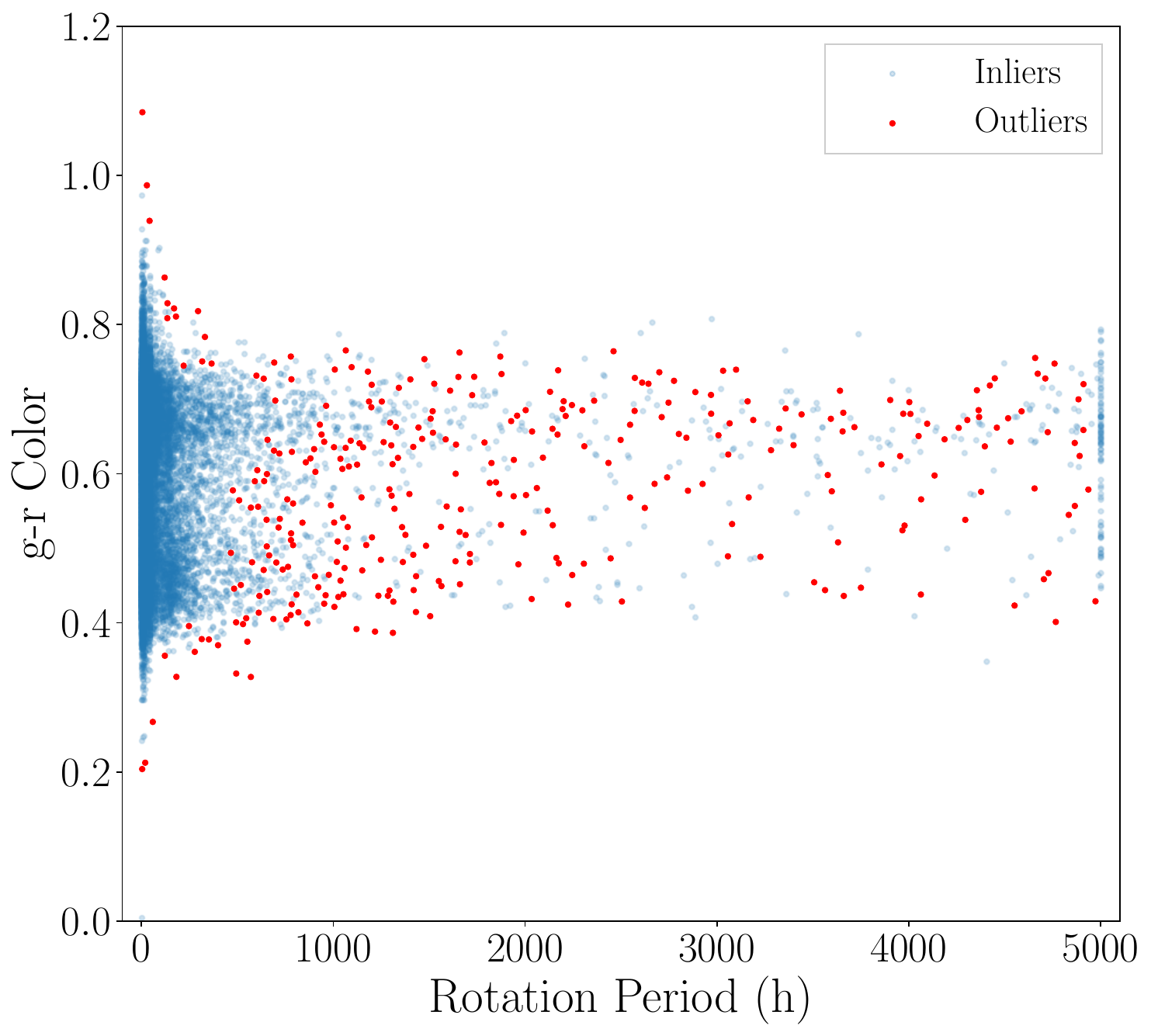}
}
\subfigure[\dssj: Mean Distance Metric]{
       \includegraphics[width=0.4\textwidth]{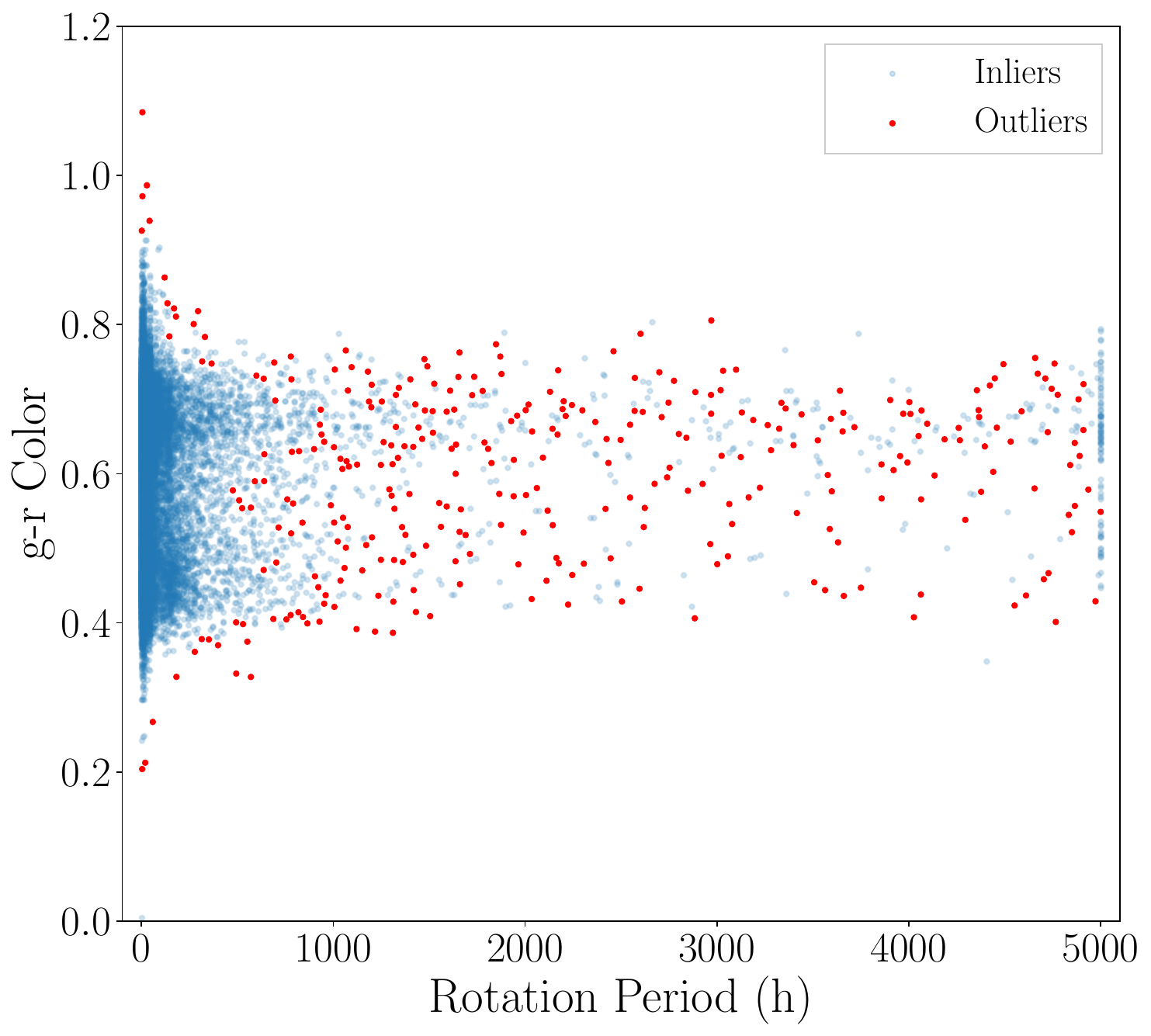}
}
\subfigure[\knnsj: In-Degree Metric ]{
       \includegraphics[width=0.4\textwidth]{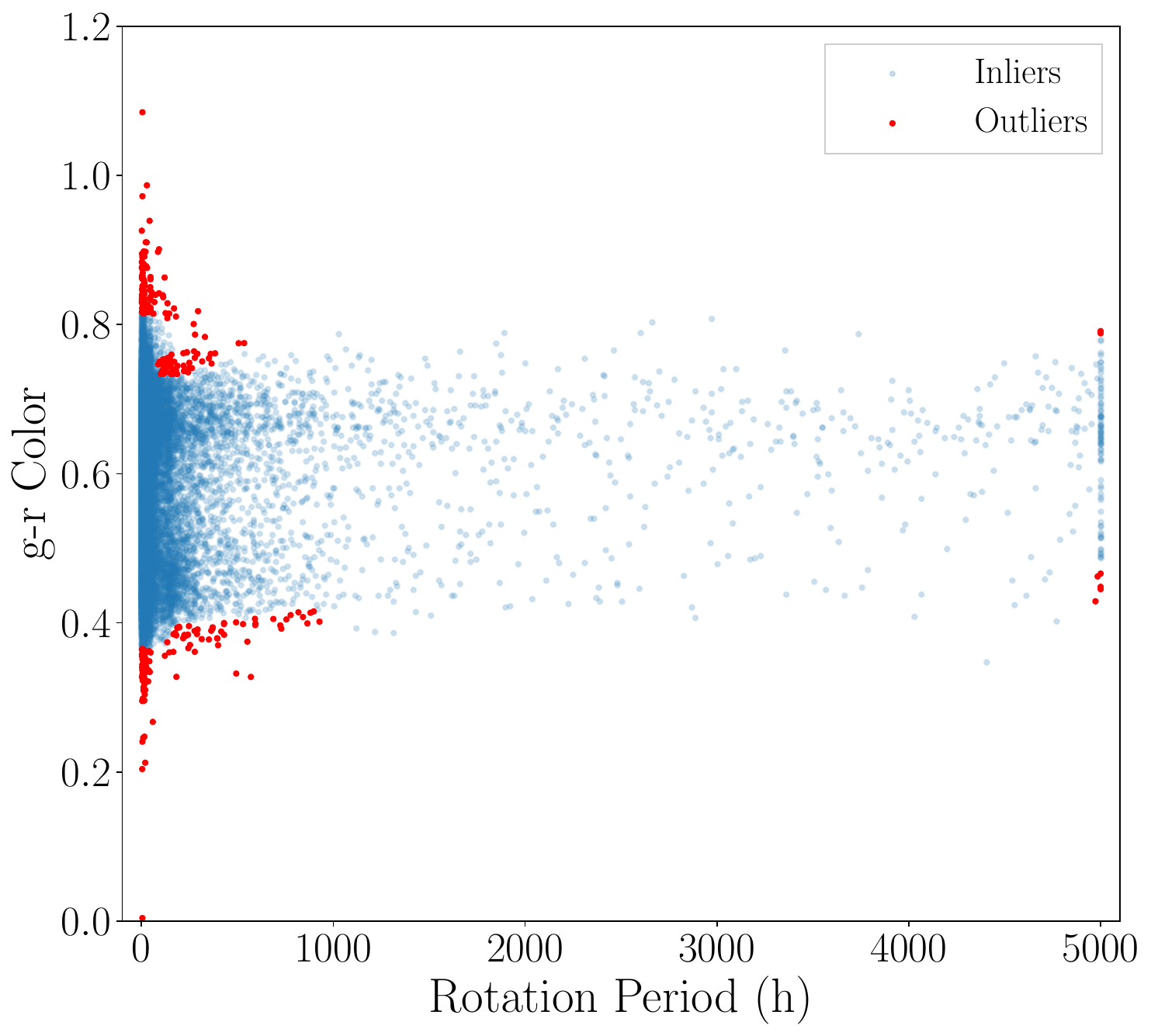}
}
\subfigure[\knnsj: Mean Distance to $k$ Neighbors Metric]{
       \includegraphics[width=0.4\textwidth]{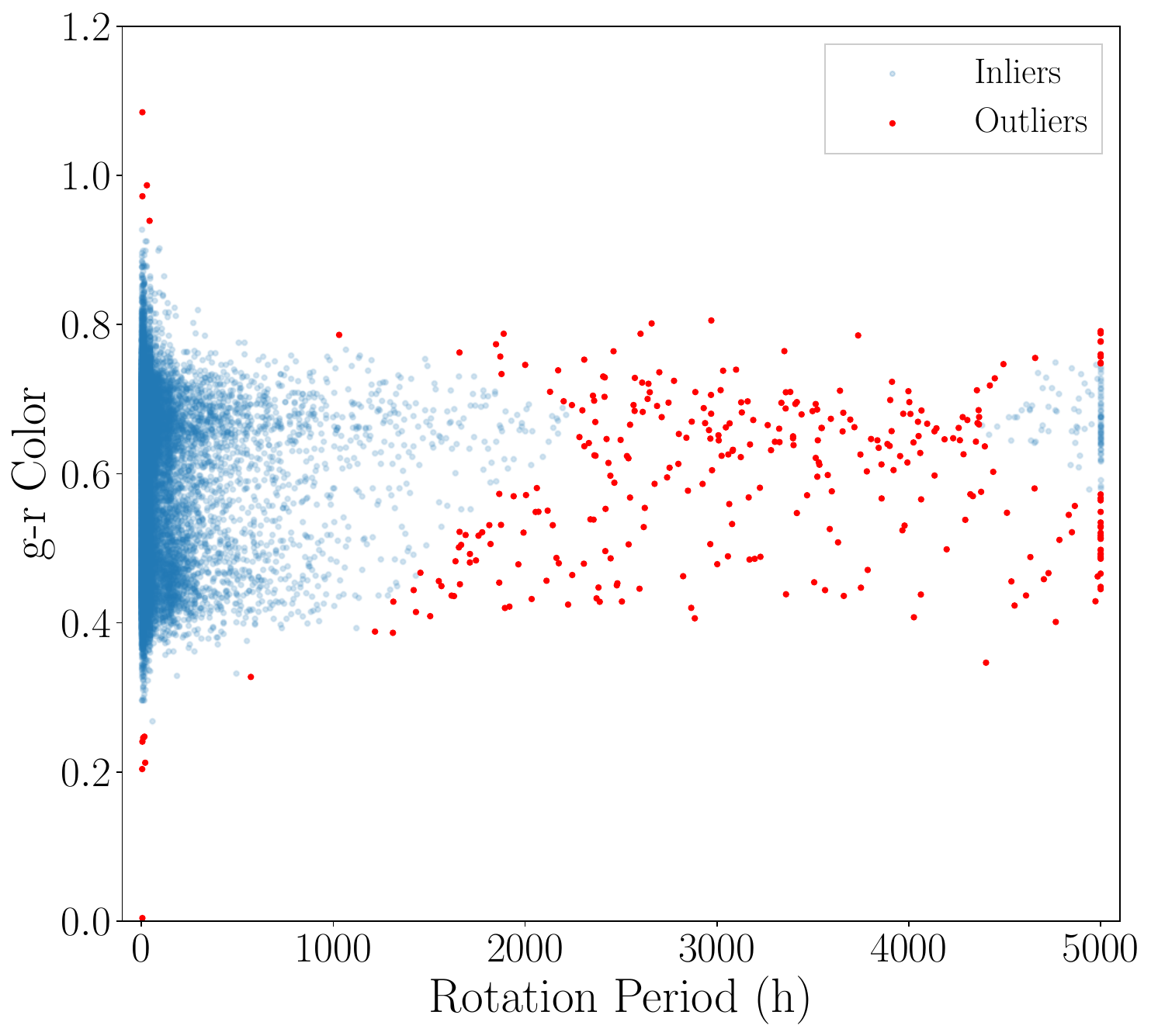}
}

    \caption{Comparison of the four outlier detection methods on an example $d=2$ feature space, showing the rotation period as a function of $g-r$ color. (a)--(b) show the two metrics for \dssj, and (c)--(d) show the two metrics for \knnsj. The points that fall within the top 1\% of the ranked list of outliers are shown as solid red points in each plot. The inliers (99\%) are shown as the translucent blue points.}
   \label{fig:DSSJ_KNNSJ_rotper_grcolor}
\end{figure*}

\begin{figure*}[!t]
\centering
\subfigure[\dssj: Number of Neighbors Metric ]{
       \includegraphics[width=0.4\textwidth]{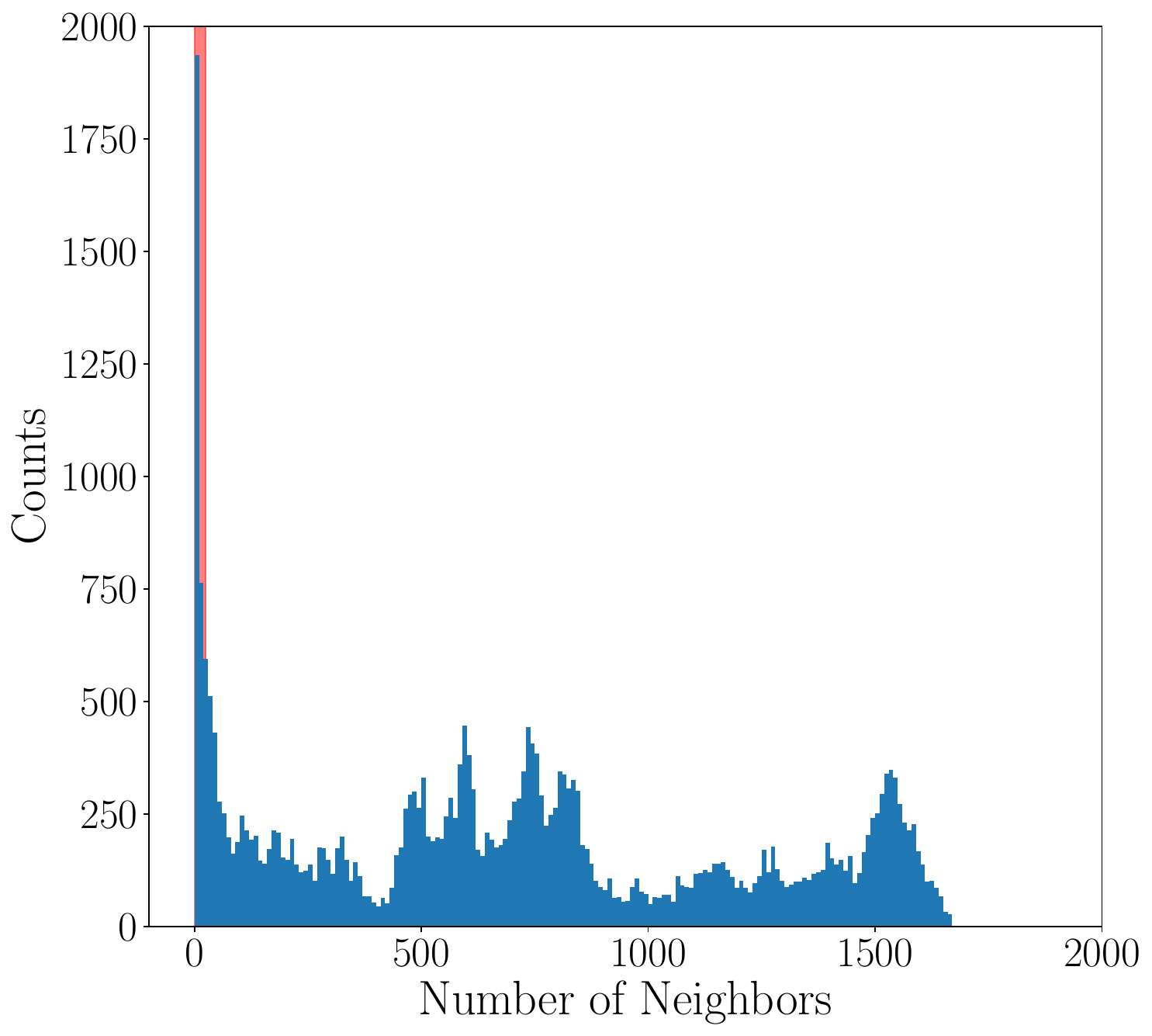}
}
\subfigure[\dssj: Mean Distance Metric]{
       \includegraphics[width=0.4\textwidth]{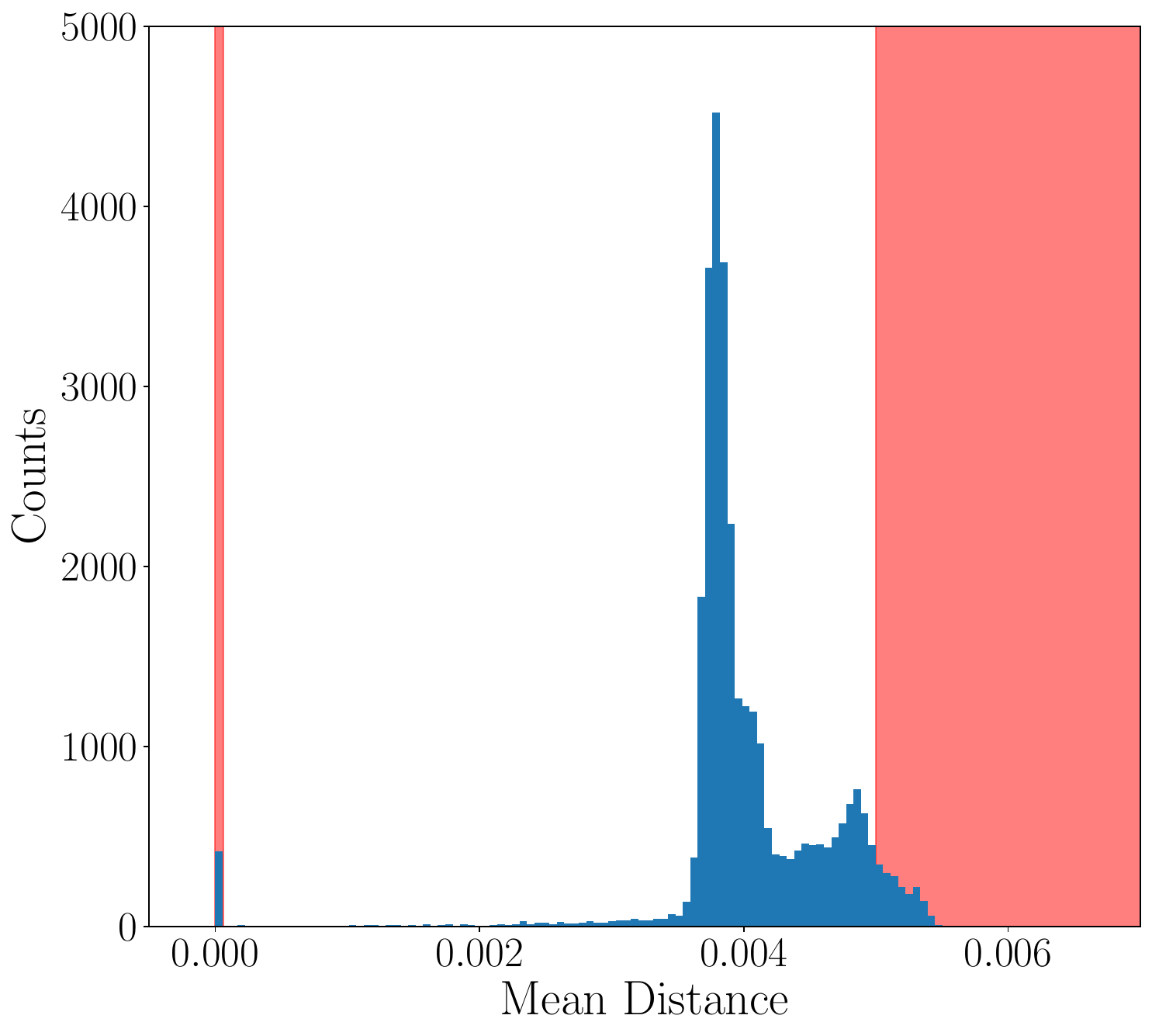}
}
\subfigure[\knnsj: In-degree Metric ]{
       \includegraphics[width=0.4\textwidth]{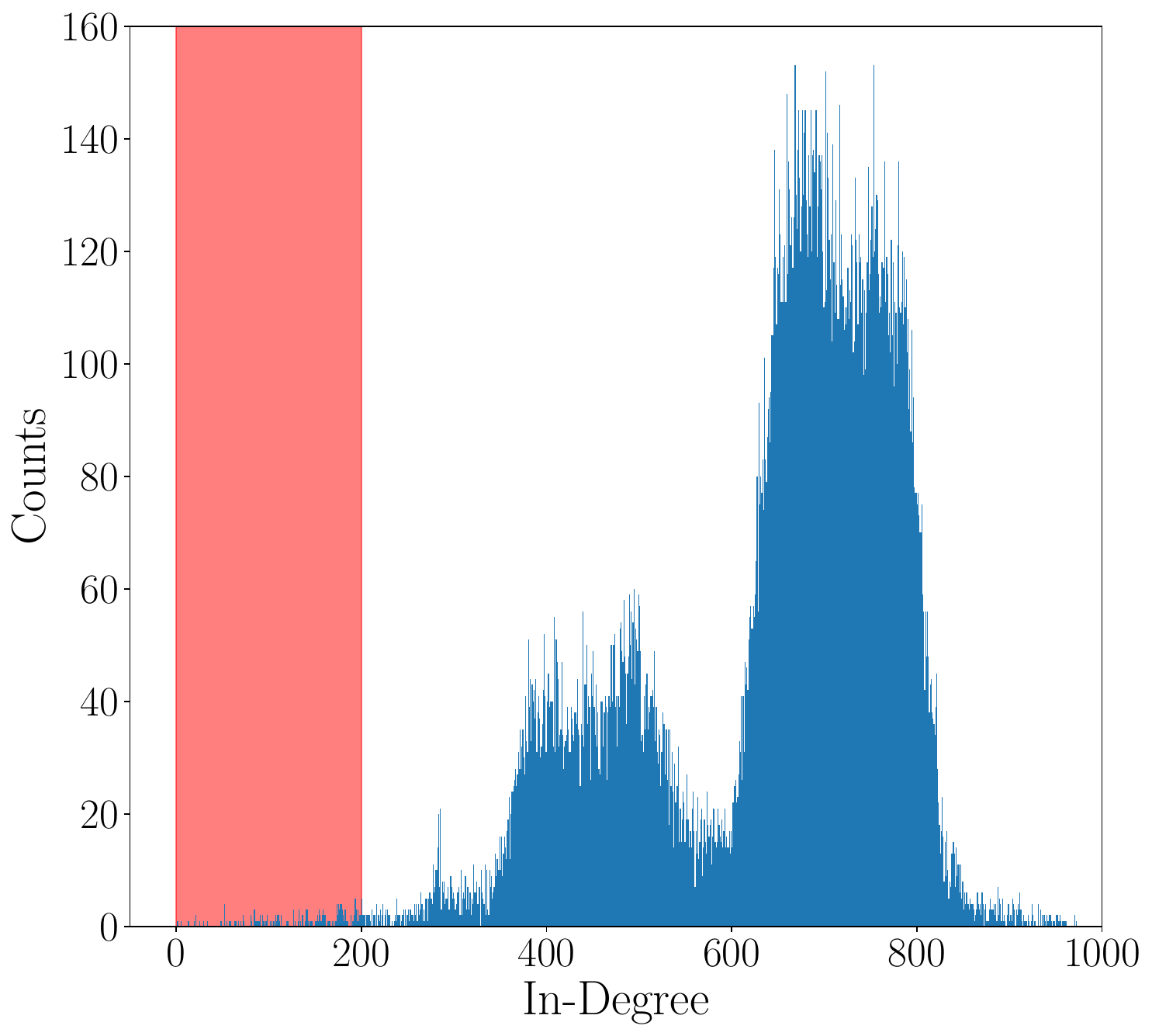}
}
\subfigure[\knnsj: Mean Distance to $k$ Neighbors Metric]{
       \includegraphics[width=0.4\textwidth]{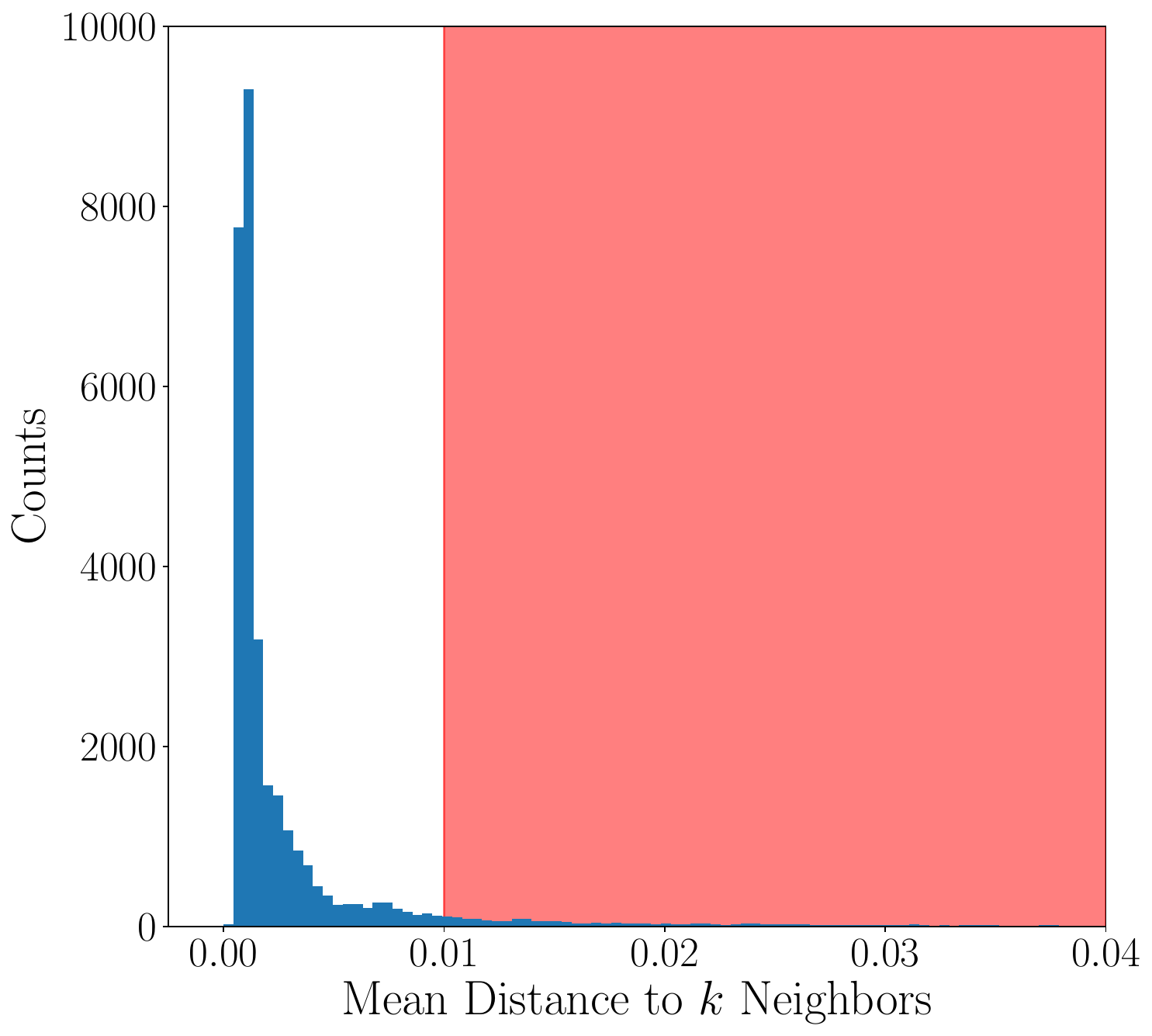}
}

    \caption{Histograms of the distribution of metric values for the corresponding plots in Figure~\ref{fig:DSSJ_KNNSJ_rotper_grcolor}. The red shaded regions are illustrative examples of where outliers are found in the respective distributions. The the mean distance metric for \dssj (b) yields two regions where outliers are found. Points with no neighbors within their search radius will yield a mean distance of 0 and so all points that do not have any neighbors are assigned a rank of 0. And those points that have large mean distances to their neighbors will also be outliers.}
   \label{fig:DSSJ_KNNSJ_rotper_grcolor_diagnostic}
\end{figure*}

Figure~\ref{fig:DSSJ_KNNSJ_rotper_grcolor} shows a $d=2$ feature space (rotation period vs. $g-r$ color) for the \dssj method in panels (a)--(b) and the \knnsj method in (c)--(d). In the figure, we arbitrarily denote 1\% of the points as outliers; in practice a user can select any percentage of points to be outliers. Furthermore, Figure~\ref{fig:DSSJ_KNNSJ_rotper_grcolor_diagnostic} shows histograms of the metric values that are used to compute the outlier rankings for the corresponding plots in Figure~\ref{fig:DSSJ_KNNSJ_rotper_grcolor}. This information is helpful for understanding why outliers are selected in each case. We make the following observations.

\begin{itemize}
\item Figure~\ref{fig:DSSJ_KNNSJ_rotper_grcolor}(a) (\dssj: number of neighbors metric) clearly finds points in the underdense regions of the feature space, but they are interspersed with other points that are considered inliers in this space. Had a larger outlier threshold been selected, more points in the sparse regions would be selected as outliers. This is because the method allows for two or more points to share the same rank. In this case, $>1$\% of the points have a rank of 0 (they have no neighbors), but only the first 1\% of these points are considered outliers. Figure~\ref{fig:DSSJ_KNNSJ_rotper_grcolor_diagnostic}(a) shows this more clearly, where there are a large number of points ($>1$\% of the total fraction of points) that do not have any neighbors, but not all of these points are classified as outliers. An example point of this type can be clearly observed at (x, y)$\approx$(0, 0) in Figure~\ref{fig:DSSJ_KNNSJ_rotper_grcolor}(a).
\item Figure~\ref{fig:DSSJ_KNNSJ_rotper_grcolor}(b)  (\dssj: mean distance metric) is very similar to that of Figure~\ref{fig:DSSJ_KNNSJ_rotper_grcolor}(a), except that some points tend to swap outlier/inlier classifications. 
\item Figure~\ref{fig:DSSJ_KNNSJ_rotper_grcolor}(c) (\knnsj: in-degree metric) clearly selects outliers that are clustered together as shown at a rotation period $\lesssim500$~h. This is expected as points located in a similar region of a feature space are likely to have a similar in-degree number (recall that the in-degree, or reverse nearest neighbors refers to the number of instances that a point is considered a neighbor of another point in its set of \knn). Counterintuitively, the outliers selected by this metric are clustered at a rotation period $\lesssim500$~h, and thus appear to be inliers. However, the reason they are not inliers is because the inlying points in this region have few of these outlying points in their set of $k$ nearest neighbors. Interestingly, this metric finds outliers that are largely different than both of the \dssj methods in Figure~\ref{fig:DSSJ_KNNSJ_rotper_grcolor}(a)--(b).
\item Figure~\ref{fig:DSSJ_KNNSJ_rotper_grcolor}(d) (\knnsj: mean distance to $k$ neighbors metric) is interesting to compare with the two \dssj methods in Figure~\ref{fig:DSSJ_KNNSJ_rotper_grcolor}(a)--(b). With this method, the outlying points at a rotation period $\gtrsim$2200~h and  $\lesssim$4500~h are all outliers, and are not interspersed with inliers as shown in Figure~\ref{fig:DSSJ_KNNSJ_rotper_grcolor}(a)--(b). The reason the outliers are all selected in this region is because the \knn algorithm requires each point to find $k$ neighbors which decreases the stochastic variance in this set of $k$ points. Therefore, the outliers will be selected in regions where there are similarly low densities. In contrast the \dssj methods do not require finding a minimum number of neighbors and so the set of neighbors found within $\epsilon$ of each point will have greater variation compared to each other, and therefore \dssj is more susceptible to stochastic variation in local density.
\end{itemize}

In summary, each of the outlier detection methods selects outliers based on different criteria. Thus, it may be useful for the user to select the outlier method that best suits their science case; for instance, from Figure~\ref{fig:DSSJ_KNNSJ_rotper_grcolor} a user may want to find asteroids that are fast rotators (short rotation periods) that have an unusual color. In this case, the \knn algorithm with the in-degree metric would be best. Another option is to limit outliers to those that have low rankings in multiple outlier detection methods which can be computed by averaging the outlier scores across methods.

\begin{figure*}[!t]
\centering
\subfigure[\dssj: Number of Neighbors Metric ]{
       \includegraphics[width=0.4\textwidth]{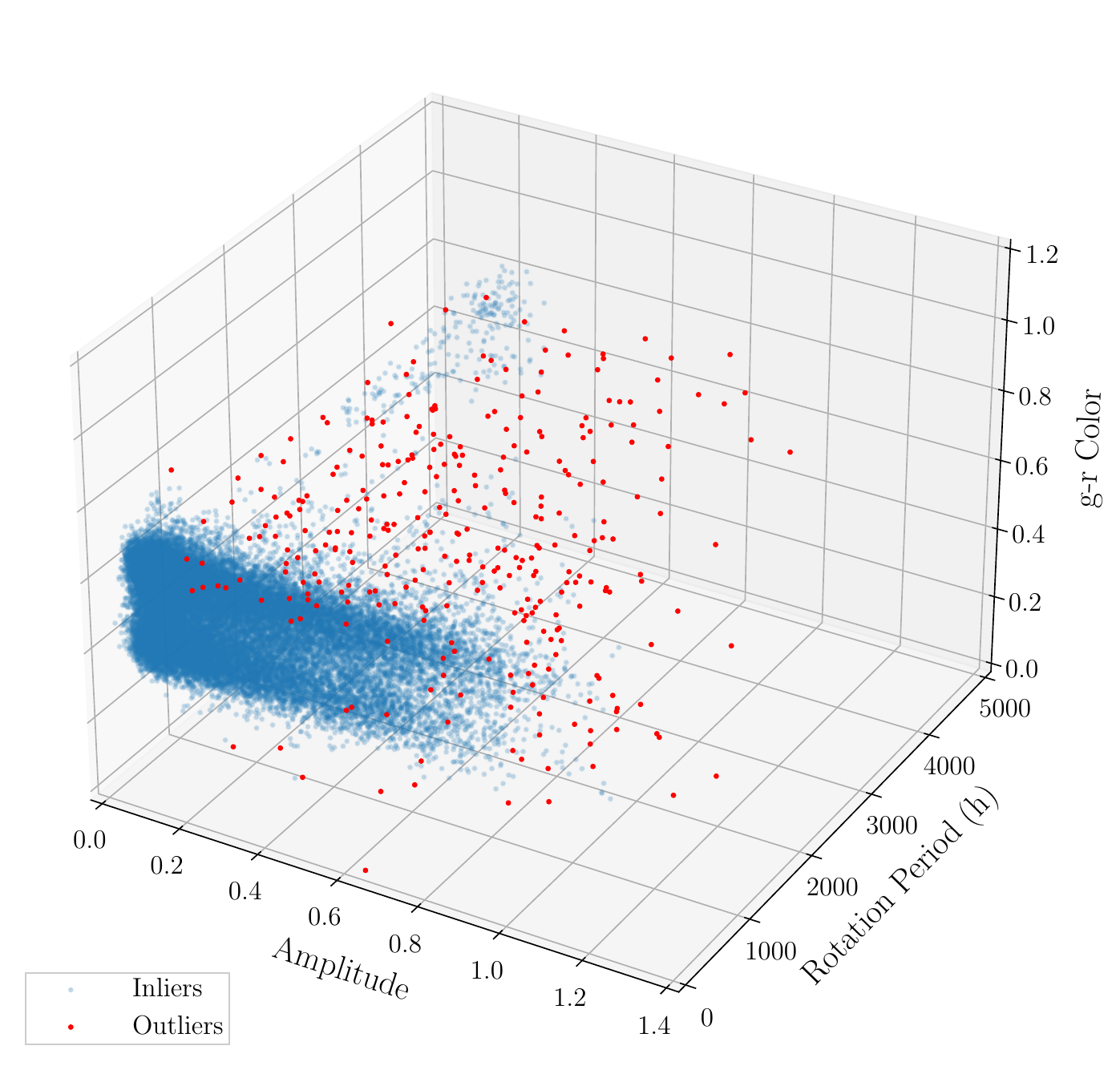}
}
\subfigure[\dssj: Mean Distance Metric]{
       \includegraphics[width=0.4\textwidth]{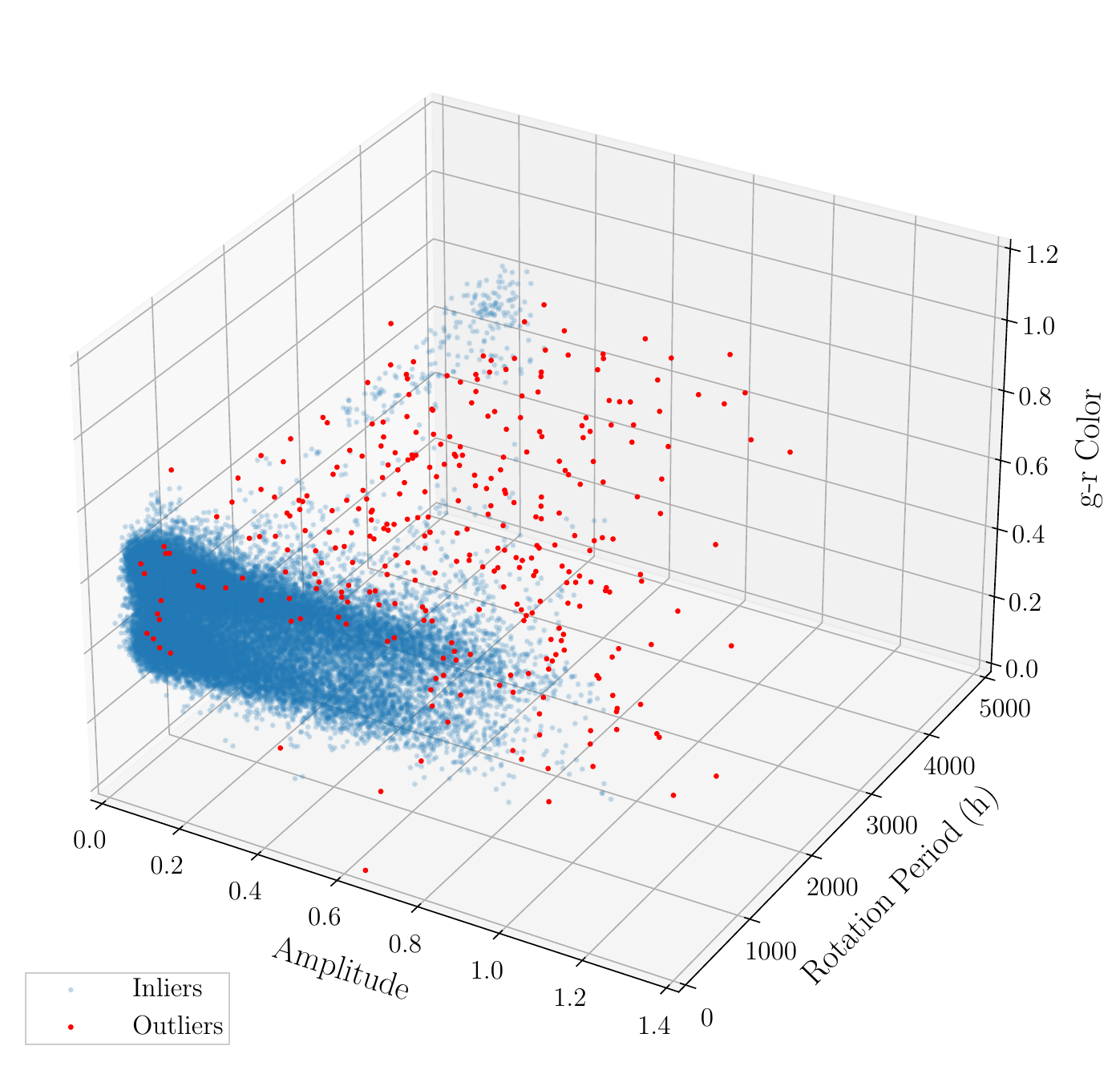}
}
\subfigure[\knnsj: In-Degree Metric ]{
       \includegraphics[width=0.4\textwidth]{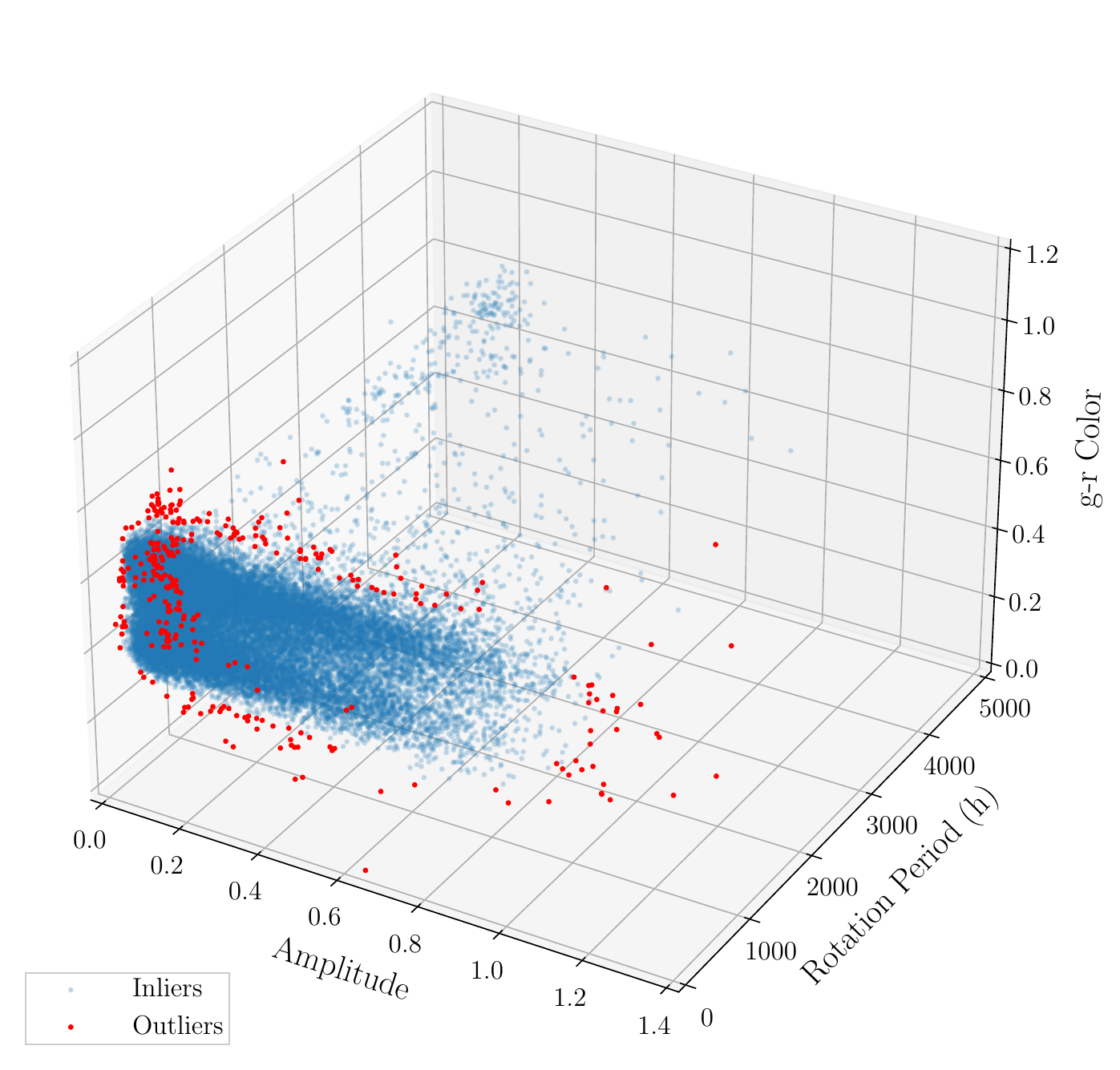}
}
\subfigure[\knnsj: Mean Distance to $k$ Neighbors Metric]{
       \includegraphics[width=0.4\textwidth]{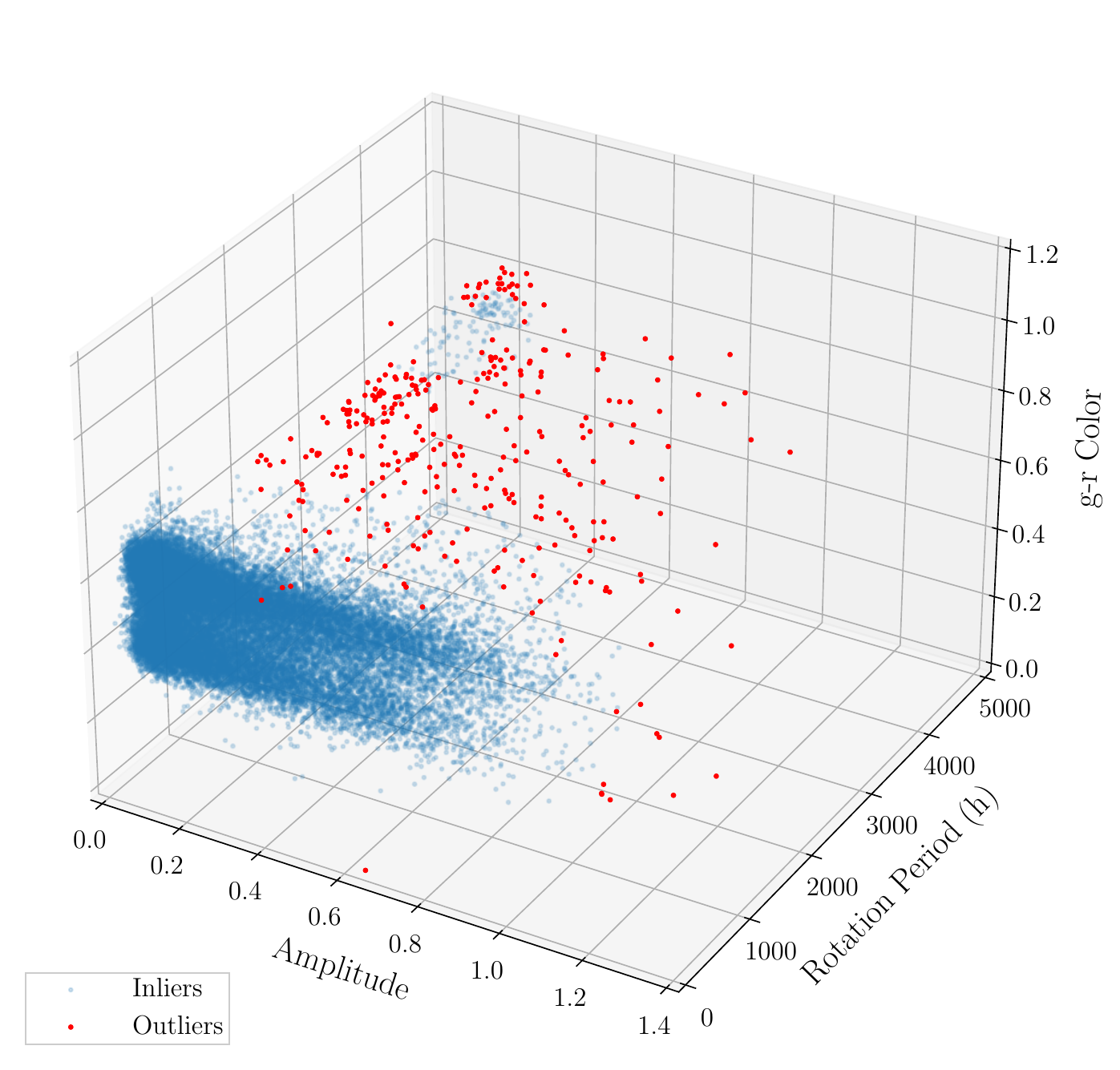}
}

    \caption{The same as Figure~\ref{fig:DSSJ_KNNSJ_rotper_grcolor}, but for a $d=3$ feature space (light curve amplitude, rotation period, and $g-r$ color).}
   \label{fig:DSSJ_KNNSJ_lcamp_rotper_grcolor}
\end{figure*}

Figure~\ref{fig:DSSJ_KNNSJ_lcamp_rotper_grcolor} shows a scatter plot of outliers detected in a $d=3$ feature space (light curve amplitude, rotation period, and $g-r$ color). Many observations are similar to that of Figure~\ref{fig:DSSJ_KNNSJ_rotper_grcolor} as described above.  To better understand the assigned outlier labels in Figure~\ref{fig:DSSJ_KNNSJ_lcamp_rotper_grcolor}, for the interested reader, Appendix~\ref{sec:appendix_comparison_outlier_methods} presents a comparison of rankings for this feature space. It is informative for understanding the nuances of the four outlier detection metrics, and illustrates why using an ensemble of methods/metrics are needed to identify outliers. With a single method, we may inadvertently omit detecting interesting Solar System objects, and our approach mitigates this potential drawback.

\subsection{Viability of Population Outlier Detection at LSST Scale}\label{sec:LSST_scale}
\begin{deluxetable*}{l|r|r|r}
\tablecaption{The response time and multi-GPU speedup for \dssj and \knnsj on \lsstsynt. The same permutations of the \snapshotA $d=15$ feature space are employed. The response times refer to computing 5\% of the 32,752 feature spaces for both 1 and 4 GPUs. The speedup of using 4 GPUs over 1 GPU is reported. The oversubscription factor for \dssj and \knnsj is $o=5$. }\label{tab:lsst_scale}
\tablewidth{\columnwidth}
\tabletypesize{\footnotesize}
\tablehead{
\colhead{Algorithm}&\colhead{Response Time (h) 1xGPU}&\colhead{Response Time (h) 4xGPUs}&\colhead{GPU Speedup}
}
\startdata
\dssj&10.83&2.94&3.68\\
\knnsj&45.58&11.55&3.95\\\hline 
Total Time &56.41&14.49&\\\hline
\enddata
\end{deluxetable*}

LSST will observe roughly 5 million asteroids over the 10 year survey; therefore, the response time and scalability measurements reported in the prior sections will not directly reflect the expected performance of the system when it is deployed on the LSST data stream. In this section, we report the performance of \snaps population outlier detection on the \lsstsynt dataset which is representative of the LSST dataset at the end of the 10 year survey (Section~\ref{sec:datasets_lsst_synthetic}).

In this experiment, for \dssj we set $s_{min}=32$ and $s_{max}=256$ and for \knnsj we set $k_{min}=2$ and $k_{max}=256$.  These maximum values are significantly lower than those used for \snapshotA; recall that in Section~\ref{sec:all_performance_results}, we selected large values of $s_{max}$ and $k_{max}$ using \snapshotA to report the conservative (worst-case) performance. \snapshotA has significant variations in the data densities of each feature space because it only contains 31,693 objects and therefore, the feature spaces are poorly sampled. The 5 million objects detected by LSST will reduce the degree of sparsity in the feature spaces which will reduce the variance in the feature space data densities. This implies that smaller  $s_{max}$ values for \dssj and $k_{max}$ neighbors for \knnsj will be needed to adequately sample the feature spaces when operating \snaps at LSST scale. Furthermore, as observed in Figures~\ref{fig:heatmap_selected_eps_values_num_neighbors}--\ref{fig:heatmap_selected_k_values_mean_distance}, $\epsilon^{select}$ and $k^{select}$ are generally closer to smaller values of $\epsilon$ and $k$ rather than larger values, so using smaller   $\epsilon^{select}$ and $k^{select}$ parameter values at LSST scale should still yield good sets of outlier rankings.


Due to the excessive execution times required of this experiment, we only report the time to compute the first 5\% of the 32,752 feature spaces and then extrapolate this time to the expected response time when computing all of the feature spaces. These results are reported in Table~\ref{tab:lsst_scale}. We achieve a multi-GPU speedup (using 4 GPUs over 1 GPU) of 3.68$\times$ and 3.95$\times$ on \dssj and \knnsj, respectively. Consequently, we achieve a very high $\geq92$\% parallel efficiency, where the parallel efficiency is the speedup divided by the number of GPUs.  Recall that on \snapshotA (Section~\ref{sec:all_performance_results}), multi-GPU performance was poor because the ratio of GPU to CPU (host-side) work was low. However, with 5 million data points, there is significantly more work to compute on the GPU and so the abovementioned ratio is much higher, leading to greater multi-GPU scalability. Extrapolating the results in Table~\ref{tab:lsst_scale} to all 32,752 feature spaces, we can compute population outliers for both \dssj and \knnsj at LSST scale in roughly 12~days using 4 GPUs. This will allow \snaps to provide users updated population outlier detection rankings once per month over the 10~year survey.


We highlight several caveats to this analysis. This level of performance is an estimate, as we only computed 5\% of the feature spaces, estimated selectivity and $k$ nearest neighbor parameter values, estimated the total number of asteroids detected by LSST at the end of the survey, and used a synthetic ZTF-like dataset which may not be representative of the LSST catalog. Furthermore, we anticipate that the GPUs SNAPS will use to compute population outlier scores will be upgraded over the next $\sim$10 years, which will decrease the computation time needed for this task. Despite these estimates, we believe that \snaps population outlier detection will scale to LSST data volumes even if some of these assumptions change during the 10~year survey. 

\section{Example Avenues of Scientific Investigation}\label{sec:science}

One of the science goals of SNAPS is to map ``edge cases'' within the asteroid population that reveal unusual physical properties that in turn suggest constraints on the origin and evolution of our Solar System. 
In this section, we suggest various scientific investigations that can be carried out using population outlier detection and multi-dimensional feature space exploration.

\subsection{Outliers in a Single Dimension}

The simplest outlier detection scheme is one that examines just a single dimension.
(1) There are five objects in \snapshotA with $H<5$, which corresponds to diameters larger than around 300~km. These rare objects --- some 0.02\% of \snapshotA --- turn out to be dwarf planets and large trans-Neptunian objects in the outer Solar System, and not main belt asteroids.
These objects have different compositions and histories and therefore require different analysis than the rest of the sample.
(2) Some objects in \snapshotA have a derived rotation period of exactly 5000~hours, which is the maximum period searched. These solutions are likely incorrect, and probably represent objects with very low lightcurve amplitudes where no true periodic signal can be identified. Leaving these likely-false solutions aside, there are 550~objects (1.7\% of SNAPShot1) with 
periods $>$1000~hours and $<$5000~hours.
This population was first identified in \citet{2021MNRAS.506.3872E},
who identified YORP, 
which alters the rotation state of an asymmetric object due to thermal
torques
\citep{2000Icar..148....2R,2006AREPS..34..157B}, 
as the most likely origin mechanism for these very long periods. This serves as an existence proof for YORP creating slow rotation rates, and puts strong constraints on the initial distribution of rotation rates in the asteroid belt, and the collision rate in the main belt. These constraints in turn constrain important details about the formation and evolution of the Solar System.

The amplitude of a lightcurve indicates the shape of the asteroid, through $\Delta {\rm mag} = 2.5 \log \frac{a}{b}$, where $a$ and $b$ are the major and minor axis lengths of the ellipsoidal asteroid shape.
There are 160~objects in \snapshotA that have lightcurve amplitudes $>$1~magnitude, which implies
very elongated asteroids where the long side is at least 2.5$\times$ longer than the small side. Elongated asteroids are interesting because they may require internal strength~\citep{McNeill_2018}. The existence of these very elongated objects --- present at the level of 0.5\% in \snapshotA ---
demands that asteroid evolution models explain the creation and persistence of such objects.

\subsection{Outliers in Multiple Dimensions}

Figure~\ref{fig:DSSJ_KNNSJ_rotper_grcolor} shows outlier sets for our four methods, in two dimensions. Most of the outlying points clearly have either the most extreme rotation periods or colors, which could be identified through  a one-dimensional search. However, there are some objects that are neither extreme in period nor color --- for example, objects near \{500,0.38\} --- that are identified in all four methods as outliers. Neither this period or color is extreme, but this combination of properties places these objects in the outermost envelope of the distribution of objects in period-color space. Thus, outliers can help define details of the multi-dimensional distributions.
The existence of objects near \{500, 0.38\} may indicate something about the internal structure and strength of those objects.
Figure~\ref{fig:DSSJ_KNNSJ_lcamp_rotper_grcolor}
adds amplitude to create a three-dimensional space, again helping to define the envelope of the distributions of these properties.

\begin{figure}[!t]
\centering
\includegraphics[width=1\columnwidth]{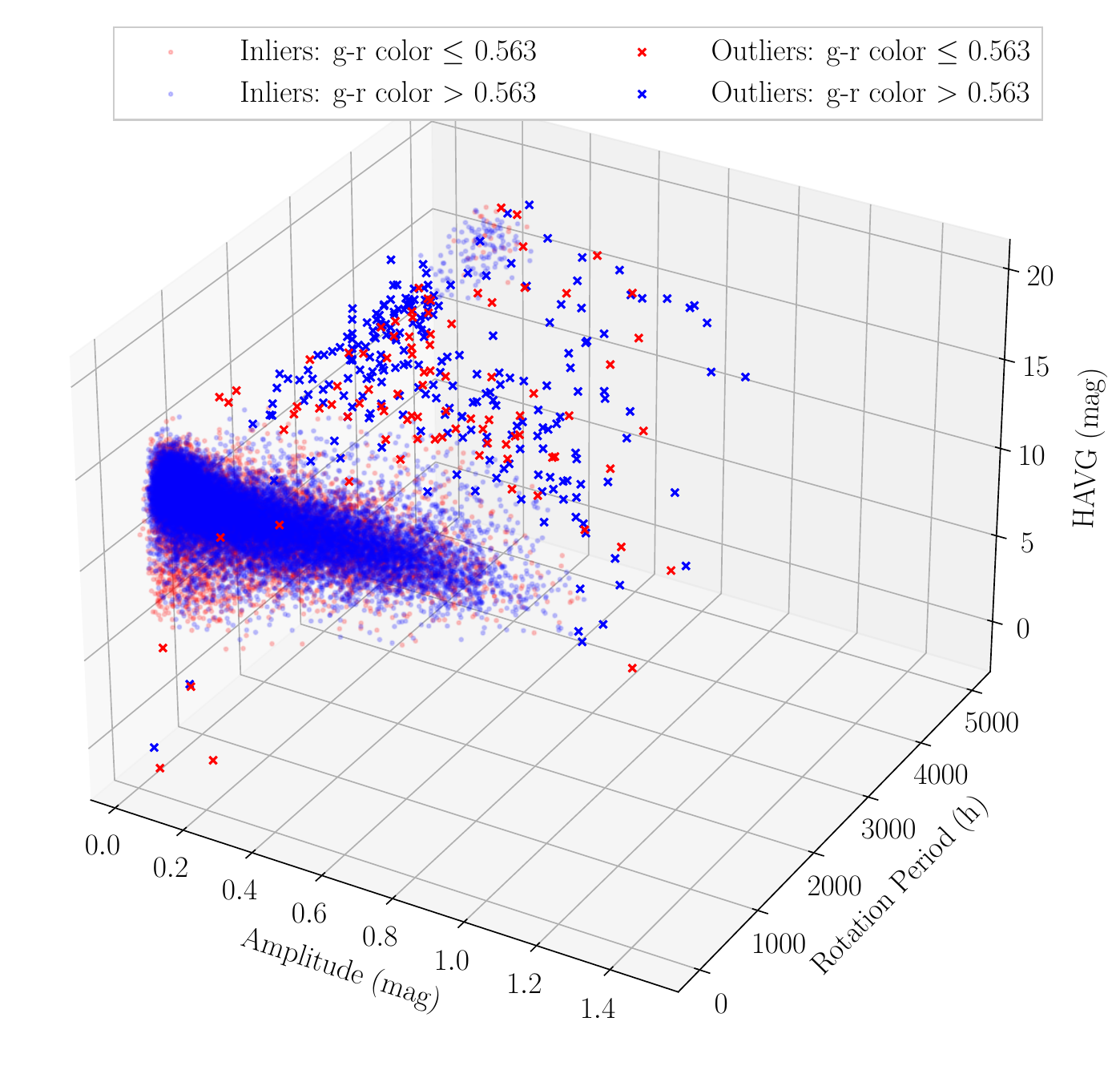}
    \caption{Outliers detected using the $k$NNSJ: Mean Distance to $k$ Neighbors Metric rankings in the light curve amplitude, rotation period, and HAVG feature space. The inliers (99\% of the objects) are plotted as translucent circles and the outliers with ``x'' markers (1\% of the objects). Red and blue markers denote C-type ($g-r$ color $\leq$ 0.563) and S-type ($g-r$ color $>$ 0.563) asteroids, respectively. Two dimensional slices through this space are shown in Figure~\ref{fig:science_KNNSJ_mean_distance_rotper_amp_havg_grColor_three_panel}.}
   \label{fig:science_KNNSJ_mean_distance_rotper_amp_havg_grColor}
\end{figure}

\begin{figure}[!t]
\centering
\includegraphics[width=1\columnwidth]{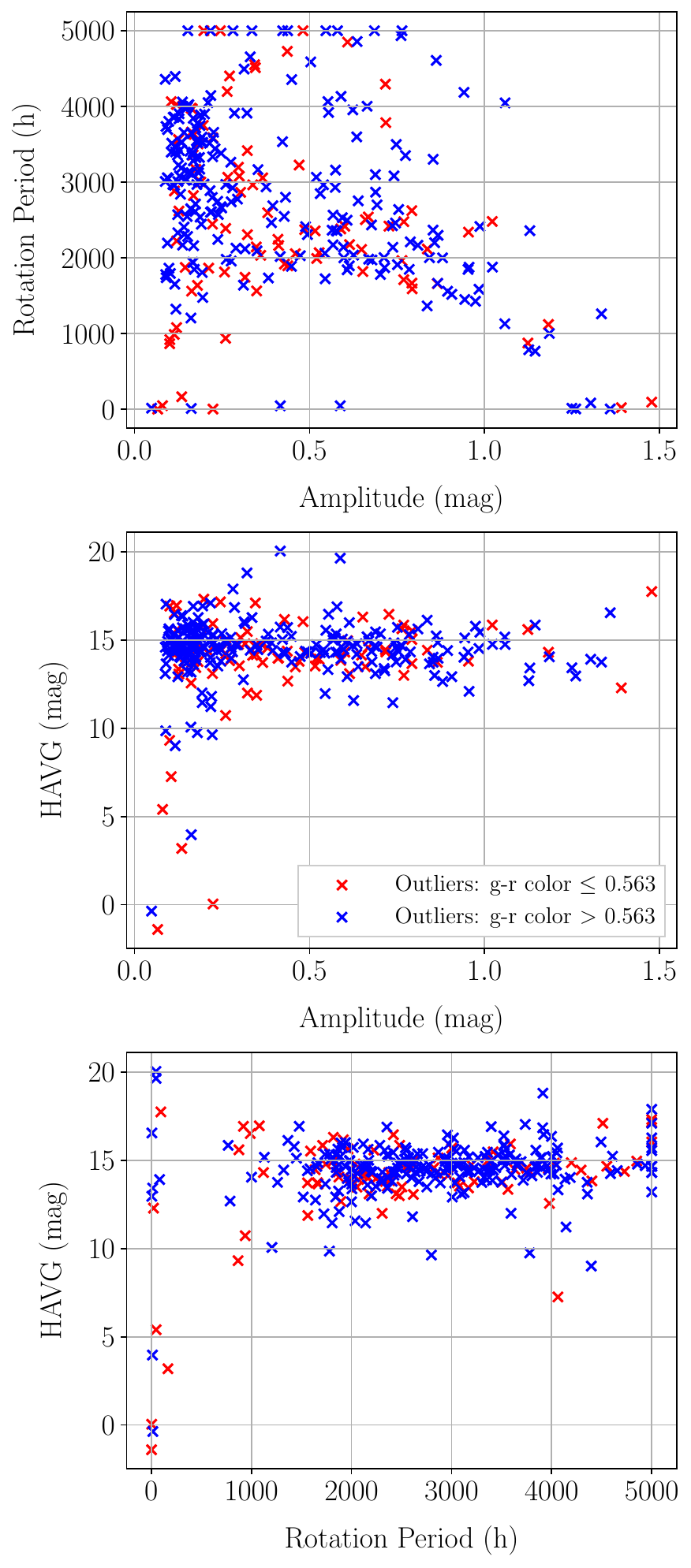}
    \caption{The same as Figure~\ref{fig:science_KNNSJ_mean_distance_rotper_amp_havg_grColor}, except that only the outliers are shown. The panels ordered from top to bottom are as follows: rotation period as a function of amplitude, HAVG as a function of amplitude, and HAVG as a function of rotation period.
    Outliers in this space are equally divided between C-type and S-type asteroids, indicating that composition is not a significant driver of unusual behaviors in these three dimensions.    }
   \label{fig:science_KNNSJ_mean_distance_rotper_amp_havg_grColor_three_panel}
\end{figure}

Figures~\ref{fig:science_KNNSJ_mean_distance_rotper_amp_havg_grColor} 
and~\ref{fig:science_KNNSJ_mean_distance_rotper_amp_havg_grColor_three_panel}
show outliers
in the parameter space of lightcurve amplitude, rotation period, and HAVG (a proxy for size). We
then separate S- and C-type asteroids using $g-r=0.563$ as the boundary between the two taxonomic types ~\citep[the boundary can be observed in Figure~4 in][]{trilling2023}.
The outliers are roughly equally divided between C- and S-type asteroids, suggesting that, to the degree that composition and internal properties correspond to asteroid color, unusual properties in this multi-dimensional space are not biased in favor of a particular asteroid composition.
These figures show a group of objects that are particularly unusual, with very long periods (2000--4000~h) and relatively small amplitudes (0.2~mag) and unremarkable H~magnitudes (around~15, so a few km in diameter). This three-dimensional outlier detection scheme reveals this population of very slowly rotating asteroids that are nearly round (low amplitude) and have sizes around 3~km, more or less. Asteroid dynamical models must include a mechanism for the existence of these bodies.

\begin{figure*}[!t]
\centering
\includegraphics[width=1\textwidth]{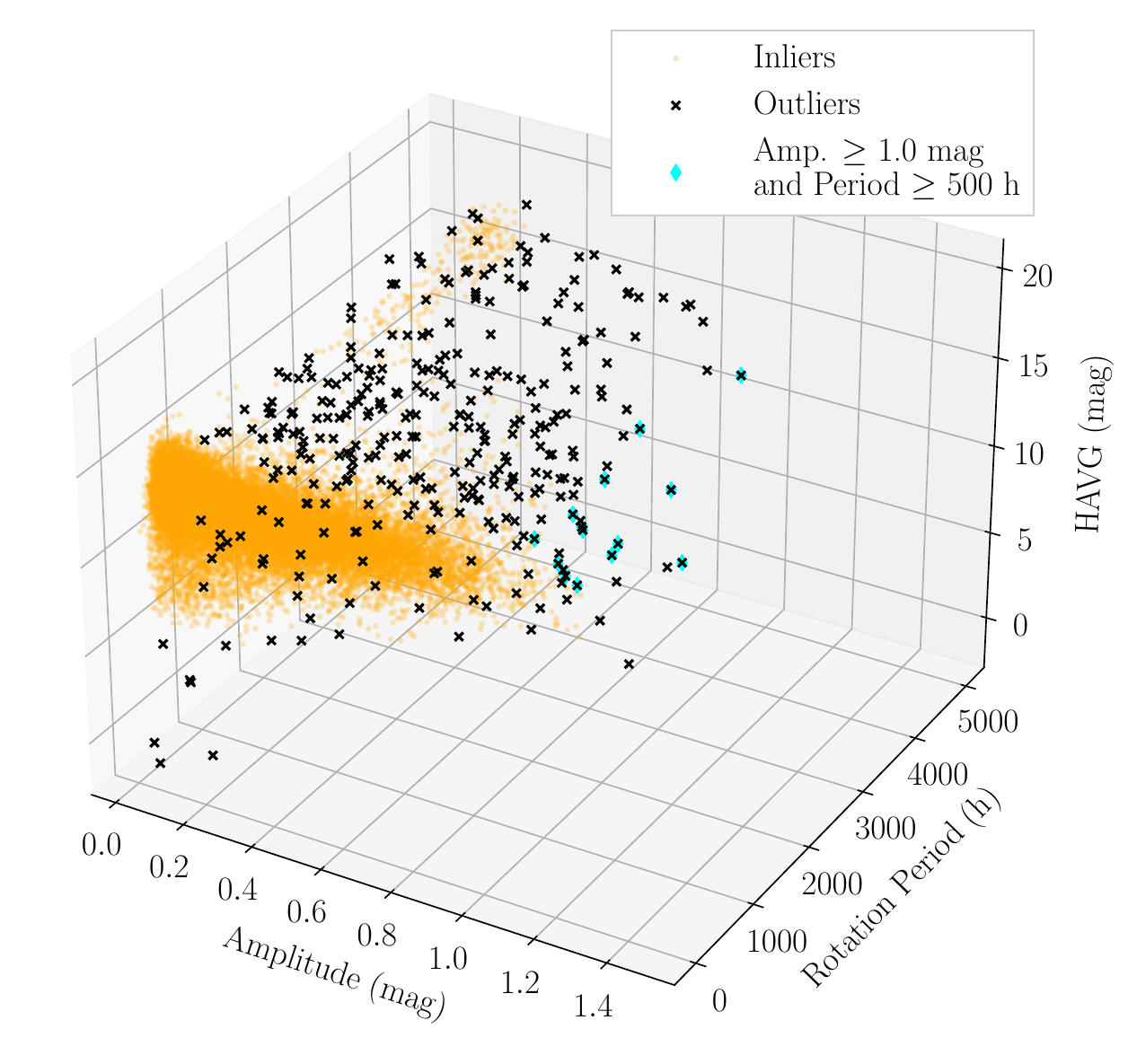}
    \caption{Outliers detected using the DSSJ: Number of Neighbors Metric in the light curve amplitude, rotation period, and HAVG feature space. The inliers (99\% of the objects) are plotted as translucent circles and the outliers are plotted with ``x'' markers (1\% of the objects). We highlight with cyan diamond markers objects with a lightcurve amplitude $\geq 1$~mag and rotation period of $\geq$ 500~h. There are only 14 objects with these properties and our method is able to detect all of these as outliers denoted by the 14 diamond markers imposed behind the ``x'' markers.}
   \label{fig:science_DSSJ_num_neighbors_rotper_amp_havg}
\end{figure*}

Figure~\ref{fig:science_DSSJ_num_neighbors_rotper_amp_havg} shows this same parameter space (amplitude, rotation period, HAVG), though with a different color coding to emphasize a particular subset of outliers: objects with amplitudes $\geq$1~mag and periods $\geq$500~hours. These are objects that are quite elongated and slowly rotating, in both cases showing unusual properties. In
general, the maximum amplitude that a single elongated body can have
is around 0.9~mag~\citep{2023arXiv230904034S}. Amplitudes greater than this probably
indicate binary or contact binary systems. Our population outlier
detection approaches may be detecting binary asteroids in the main
belt --- and, intriguingly, slowly rotating binaries, which implies a
relatively small amount of angular momentum. The origin and evolution
of these bodies is an interesting question and may provide further
clues to the overall evolutionary history of the asteroid belt. 
Furthermore, within this group, the ratio of C:S asteroids is 12:5, which may not be statistically significant but is at least suggestive that the mechanism of creating such unusual asteroids may depend on composition. The sizes are 1--10~km (H is estimated to be 12.5--17.5), potentially a key clue to the evolution of these bodies.

There are many other examples of multi-dimensional investigations that can probe the properties of asteroids. As an example of another extended science case, our team is identifying outliers in minimum required strength in a space that is
defined by a number of asteroid physical properties (Chernyavskaya et al., in prep.). In summary, the 
various techniques presented here to characterize large datasets are powerful tools that provide evidence that would not otherwise be available and allow
us to carry out experiments about the formation and evolution of the Solar System.

\section{Discussion \& Conclusions}\label{sec:conclusions}
This paper has presented the population outlier detection functionality of the \snaps alert broker.  We summarize our contributions as follows:

\begin{itemize}
\item We have constructed an asteroid outlier detection system that is demonstrated on the first SNAPS data release, \snapshotA, which employs state-of-the-art GPU algorithms. We showed that the system provides outlier detection capabilities in addition to feature space exploration and data prioritization which will help researchers rapidly identify objects of interest that may warrant further investigation. 
\item We utilize a feature space permutation approach that allows users to select objects of interest based on ZTF measurements, adjacent catalogs, and derived properties.  We show that this approach can assist in data exploration, prioritization, and visualization activities to accelerate the scientific discovery process.
\item We demonstrate that the outlier detection system is currently able to scale to the ZTF data rate. Through measurements on a synthetic dataset, we expect that the system will be able to derive population outliers at LSST scale and will be able to provide outlier detection rankings roughly once per month. The monthly data will be archived and made publicly available on our website to enable tracking how objects evolve over time relative to the population of objects.
\item We highlight several preliminary scientific results, and example avenues of scientific investigation that will be enabled by \snaps population outlier detection capabilities. 
\end{itemize}

We demonstrate the grasp of these approaches with a few example science cases, of the countless investigations that are possible. As described above, our calculated outlier lists will be published on our SNAPS web page and updated monthly, allowing all users to pursue their own experiments with our large and growing catalog.

It is clear that deriving outlier rankings across all feature spaces is computationally expensive. There are several methods that could be employed to reduce this computational cost. For example, while we derived outliers for 15 features in \snapshotA, we are aware that some of these features are redundant or correlated, and so we could reduce the total number of feature spaces that we provide to users by eliminating some of these feature spaces. However, the \snaps team will aim to provide outlier rankings for at least all permutations of $d=15$ features; however, these features will change slightly when we begin receiving data from the Rubin Observatory. At present, we believe that providing $d=15$ features should provide a reasonable level outlier detection and data space exploration for the community.

We showed that population outlier detection with \snaps is expected to scale to LSST data volumes assuming that the current functionality is sufficient for the LSST era. We predict that the community will desire additional functionality, such as providing binary classification labels (inlier vs. outlier) instead of only using ranked lists of objects. The \snaps team intends to add this functionality to the system in the future.

A distinct but complimentary service that SNAPS could provide to the community is clustering each of the feature spaces. This task is similar to the outlier detection methods described here, as clustering typically  relies on similarity searches (finding nearby points in a feature space). This would allow for the automated detection of groups of objects that may be of interest. A canonical example would be to derive asteroid families as a function of orbital elements~\citep{2008Icar..198..138P} while also examining additional features/dimensions that may be less obvious to explore.

We identified that one bottleneck in the system is using Python to invoke the GPU algorithms for each feature space. Since we process multiple feature spaces concurrently, this yields non-negligible memory pressure, which impacts the rate at which we can process a given feature space on a GPU. Thus, future work includes running a single process that executes all of the feature spaces on the GPU for \dssj and \knnsj. By undertaking this research direction, we anticipate that there will be interesting opportunities for data reuse, such as providing the set of $k$ nearest neighbors from \knnsj to \dssj for its computation, thus potentially eliminating a significant fraction of the total \dssj cost. Another direction that can be undertaken to improve performance is to leverage the tensor cores on many modern GPUs to perform Euclidean distance calculations~\citep{gallet2022leveraging}. These and other algorithmic advances are critical for reaching LSST scale while expanding the services offered by SNAPS, as many of the computational bottlenecks identified in this paper cannot be solved by hardware advances alone over the next $\gtrsim$10 years.

At present, other alert broker teams are preparing for LSST by developing software infrastructure to classify objects. We welcome collaboration with other teams that anticipate requiring the population outlier detection methods described here for their respective science domains.

\section*{Acknowledgments}
We thank the anonymous referee for a very thorough review of the manuscript which significantly strengthened the paper. MG is supported by the National Science Foundation under Grant No. 2042155. DET, MG, and MC are supported by the National Science Foundation under Grant No. 2206796. This work is supported by the Arizona Board of Regents and the Technology Research Initiative Research Fund (TRIF) Small Research Equipment Acquisition Program (SREAP). 

ZTF is a public-private partnership, with equal support from the ZTF Partnership and from the U.S. National Science Foundation through the Mid-Scale Innovations Program (MSIP).
The ZTF partnership is a consortium of the following universities and institutions (listed in descending longitude): TANGO Consortium of Taiwan; Weizmann Institute of Sciences, Israel; Oskar Klein Center, Stockholm University, Sweden; Deutsches Elektronen-Synchrotron \& Humboldt University, Germany; Ruhr University, Germany; Institut national de physique nucl\'eaire et de physique des particules, France; University of Warwick, UK; Trinity College, Dublin, Ireland; University of Maryland, College Park, USA; Northwestern University, Evanston, USA; University of Wisconsin, Milwaukee, USA; Lawrence Livermore National Laboratory, USA; IPAC, Caltech, USA; Caltech, USA.

\software{NumPy~\citep{harris2020array} and pandas~\citep{reback2020pandas}.}

\bibliographystyle{aasjournal}

\begin{thebibliography}{}
\expandafter\ifx\csname natexlab\endcsname\relax\def\natexlab#1{#1}\fi
\providecommand{\url}[1]{\href{#1}{#1}}
\providecommand{\dodoi}[1]{doi:~\href{http://doi.org/#1}{\nolinkurl{#1}}}
\providecommand{\doeprint}[1]{\href{http://ascl.net/#1}{\nolinkurl{http://ascl.net/#1}}}
\providecommand{\doarXiv}[1]{\href{https://arxiv.org/abs/#1}{\nolinkurl{https://arxiv.org/abs/#1}}}

\bibitem[{Bellm {et~al.}(2018)Bellm, Kulkarni, Graham, Dekany, Smith, Riddle,
  Masci, Helou, Prince, Adams, {et~al.}}]{bellm2018zwicky}
Bellm, E.~C., Kulkarni, S.~R., Graham, M.~J., {et~al.} 2018, Publications of
  the Astronomical Society of the Pacific, 131, 018002,
  \dodoi{10.1088/1538-3873/aaecbe}

\bibitem[{{Bottke} {et~al.}(2006){Bottke}, {Vokrouhlick{\'y}}, {Rubincam}, \&
  {Nesvorn{\'y}}}]{2006AREPS..34..157B}
{Bottke}, William~F., J., {Vokrouhlick{\'y}}, D., {Rubincam}, D.~P., \&
  {Nesvorn{\'y}}, D. 2006, Annual Review of Earth and Planetary Sciences, 34,
  157, \dodoi{10.1146/annurev.earth.34.031405.125154}

\bibitem[{Bowell {et~al.}(1989)Bowell, Hapke, Domingue, Lumme, Peltoniemi, \&
  Harris}]{bowellApplicationPhotometricModels1989}
Bowell, E., Hapke, B., Domingue, D., {et~al.} 1989, Asteroids II, 524

\bibitem[{Campos {et~al.}(2016)Campos, Zimek, Sander, Campello, Micenkov{\'a},
  Schubert, Assent, \& Houle}]{campos2016evaluation}
Campos, G.~O., Zimek, A., Sander, J., {et~al.} 2016, Data mining and knowledge
  discovery, 30, 891, \dodoi{10.1007/s10618-015-0444-8}

\bibitem[{Capodieci {et~al.}(2017)Capodieci, Cavicchioli, Valente, \&
  Bertogna}]{capodieci2017sigamma}
Capodieci, N., Cavicchioli, R., Valente, P., \& Bertogna, M. 2017, in
  Proceedings of the 25th International Conference on Real-Time Networks and
  Systems, 48--57, \dodoi{10.1145/3139258.3139270}

\bibitem[Carry et al.(2024)]{2024A&A...687A..38C} Carry, B., Peloton, J., Le Montagner, R., et al.\ 2024, Astronomy \& Astrophysics, 687, A38. \dodoi{10.1051/0004-6361/202449789}

\bibitem[{Coughlin {et~al.}(2021)Coughlin, Burdge, Duev, Katz, Van~Roestel,
  Drake, Graham, Hillenbrand, Mahabal, Masci, {et~al.}}]{coughlin2021ztf}
Coughlin, M.~W., Burdge, K., Duev, D.~A., {et~al.} 2021, Monthly Notices of the
  Royal Astronomical Society, 505, 2954, \dodoi{10.1093/mnras/stab1502}

\bibitem[{Drake {et~al.}(2014)Drake, Graham, Djorgovski, Catelan, Mahabal,
  Torrealba, García-Álvarez, Donalek, Prieto, Williams, Larson, sen,
  Belokurov, Koposov, Beshore, Boattini, Gibbs, Hill, Kowalski, Johnson, \&
  Shelly}]{CRTS-survey}
Drake, A.~J., Graham, M.~J., Djorgovski, S.~G., {et~al.} 2014, The
  Astrophysical Journal Supplement Series, 213, 9,
  \dodoi{10.1088/0067-0049/213/1/9}

\bibitem[{{Erasmus} {et~al.}(2021){Erasmus}, {Kramer}, {McNeill}, {Trilling},
  {Janse van Rensburg}, {van Belle}, {Tonry}, {Denneau}, {Heinze}, \&
  {Weiland}}]{2021MNRAS.506.3872E}
{Erasmus}, N., {Kramer}, D., {McNeill}, A., {et~al.} 2021, \mnras, 506, 3872,
  \dodoi{10.1093/mnras/stab1888}

\bibitem[{Förster {et~al.}(2021)Förster, Cabrera-Vives, Castillo-Navarrete,
  Estévez, Sánchez-Sáez, Arredondo, Bauer, Carrasco-Davis, Catelan,
  Elorrieta, Eyheramendy, Huijse, Pignata, Reyes, Reyes, Rodríguez-Mancini,
  Ruz-Mieres, Valenzuela, Álvarez Maldonado, Astorga, Borissova, Clocchiatti,
  Cicco, Donoso-Oliva, Hernández-García, Graham, Jordán, Kurtev, Mahabal,
  Maureira, Muñoz-Arancibia, Molina-Ferreiro, Moya, Palma, Pérez-Carrasco,
  Protopapas, Romero, Sabatini-Gacitua, Sánchez, Martín, Sepúlveda-Cobo,
  Vera, \& Vergara}]{Forster2021}
Förster, F., Cabrera-Vives, G., Castillo-Navarrete, E., {et~al.} 2021, The
  Astronomical Journal, 161, 242, \dodoi{10.3847/1538-3881/abe9bc}

\bibitem[{Gallet \& Gowanlock(2021)}]{gallet2021heterogeneous}
Gallet, B., \& Gowanlock, M. 2021, Data Science and Engineering, 6, 39,
  \dodoi{https://doi.org/10.1007/s41019-020-00145-x}

\bibitem[{Gallet \& Gowanlock(2022)}]{gallet2022leveraging}
Gallet, B., \& Gowanlock, M. 2022, in 2022 IEEE 29th International Conference
  on High Performance Computing, Data, and Analytics (HiPC), IEEE, 135--144,
  \dodoi{10.1109/HiPC56025.2022.00029}

\bibitem[{Gowanlock(2021)}]{Gowanlock2021KNNJPDC}
Gowanlock, M. 2021, Journal of Parallel and Distributed Computing, 149, 119,
  \dodoi{https://doi.org/10.1016/j.jpdc.2020.11.004}

\bibitem[{Gowanlock {et~al.}(2023)Gowanlock, Gallet, \&
  Donnelly}]{Gowanlock2023}
Gowanlock, M., Gallet, B., \& Donnelly, B. 2023, in Computational Science --
  ICCS 2023, ed. J.~Miky{\v{s}}ka, C.~de~Mulatier, M.~Paszynski, V.~V.
  Krzhizhanovskaya, J.~J. Dongarra, \& P.~M. Sloot (Cham: Springer Nature
  Switzerland), 357--364

\bibitem[{Gowanlock \& Karsin(2019)}]{Gowanlock2019DaMoN}
Gowanlock, M., \& Karsin, B. 2019, in Proceedings of the 15th International
  Workshop on Data Management on New Hardware, DaMoN'19 (Association for
  Computing Machinery), \dodoi{10.1145/3329785.3329920}

\bibitem[{Harris {et~al.}(2020)Harris, Millman, van~der Walt, Gommers,
  Virtanen, Cournapeau, Wieser, Taylor, Berg, Smith, Kern, Picus, Hoyer, van
  Kerkwijk, Brett, Haldane, del R{\'{i}}o, Wiebe, Peterson,
  G{\'{e}}rard-Marchant, Sheppard, Reddy, Weckesser, Abbasi, Gohlke, \&
  Oliphant}]{harris2020array}
Harris, C.~R., Millman, K.~J., van~der Walt, S.~J., {et~al.} 2020, Nature, 585,
  357, \dodoi{10.1038/s41586-020-2649-2}

\bibitem[{Hautamaki {et~al.}(2004)Hautamaki, Karkkainen, \&
  Franti}]{hautamaki2004outlier}
Hautamaki, V., Karkkainen, I., \& Franti, P. 2004, in Proceedings of the 17th
  International Conference on Pattern Recognition, 2004. ICPR 2004., Vol.~3,
  IEEE, 430--433, \dodoi{10.1109/ICPR.2004.1334558}

\bibitem[{Ivezi{\'c} {et~al.}(2019)Ivezi{\'c}, Kahn, Tyson, Abel, Acosta,
  Allsman, Alonso, AlSayyad, Anderson, Andrew, {et~al.}}]{LSST}
Ivezi{\'c}, {\v{Z}}., Kahn, S.~M., Tyson, J.~A., {et~al.} 2019, The
  Astrophysical Journal, 873, 111, \dodoi{10.3847/1538-4357/ab042c}

\bibitem[{Knorr \& Ng(1997)}]{knorr1997unified}
Knorr, E.~M., \& Ng, R.~T. 1997, in KDD, Vol.~97, 219--222

\bibitem[Le Montagner et~al.(2023)]{2023A&A...680A..17L} Le Montagner, R., Peloton, J., Carry, B., et al.\ 2023, Astronomy \& Astrophysics, 680, A17. \dodoi{10.1051/0004-6361/202346905}

\bibitem[{Lu {et~al.}(2022)Lu, Curtis, Angus, David, \&
  Hattori}]{lu2022bridging}
Lu, Y.~L., Curtis, J.~L., Angus, R., David, T.~J., \& Hattori, S. 2022, The
  Astronomical Journal, 164, 251, \dodoi{10.3847/1538-3881/ac9bee}

\bibitem[{Mainzer {et~al.}(2019)Mainzer, Bauer, Cutri, Grav, Kramer, Masiero,
  Sonnett, , \& Wright}]{neowise}
Mainzer, A., Bauer, J., Cutri, R., {et~al.} 2019, NEOWISE Diameters and Albedos
  V2.0, Tech. rep., \dodoi{10.26033/18S3-2Z54}

\bibitem[{Matheson {et~al.}(2021)Matheson, Stubens, Wolf, Lee, Narayan, Saha,
  Scott, Soraisam, Bolton, Hauger, {et~al.}}]{matheson2021antares}
Matheson, T., Stubens, C., Wolf, N., {et~al.} 2021, The Astronomical Journal,
  161, 107, \dodoi{10.3847/1538-3881/abd703}

\bibitem[{McNeill {et~al.}(2018)McNeill, Fitzsimmons, Jedicke, Lacerda, Lilly,
  Thompson, Trilling, DeMooij, Hooton, \& Watson}]{McNeill_2018}
McNeill, A., Fitzsimmons, A., Jedicke, R., {et~al.} 2018, The Astronomical
  Journal, 156, 282, \dodoi{10.3847/1538-3881/aaeb8c}

\bibitem[{Möller {et~al.}(2020)Möller, Peloton, Ishida, Arnault, Bachelet,
  Blaineau, Boutigny, Chauhan, Gangler, Hernandez, Hrivnac, Leoni, Leroy,
  Moniez, Pateyron, Ramparison, Turpin, Ansari, Allam Jr, Bajat, Biswas,
  Boucaud, Bregeon, Campagne, Cohen-Tanugi, Coleiro, Dornic, Fouchez, Godet,
  Gris, Karpov, Nebot Gomez-Moran, Neveu, Plaszczynski, Savchenko, \&
  Webb}]{Moller2020}
Möller, A., Peloton, J., Ishida, E. E.~O., {et~al.} 2020, Monthly Notices of
  the Royal Astronomical Society, 501, 3272, \dodoi{10.1093/mnras/staa3602}

\bibitem[{pandas~development team(2020)}]{reback2020pandas}
pandas~development team, T. 2020, pandas-dev/pandas: Pandas, latest,  Zenodo,
  \dodoi{10.5281/zenodo.3509134}

\bibitem[{Pankratius {et~al.}(2016)Pankratius, Li, Gowanlock, Blair, Rude,
  Herring, Lind, Erickson, \& Lonsdale}]{ComputerAidedDiscovery}
Pankratius, V., Li, J., Gowanlock, M., {et~al.} 2016, IEEE Intelligent Systems,
  31, 3, \dodoi{10.1109/MIS.2016.60}

\bibitem[{{Parker} {et~al.}(2008){Parker}, {Ivezi{\'c}}, {Juri{\'c}}, {Lupton},
  {Sekora}, \& {Kowalski}}]{2008Icar..198..138P}
{Parker}, A., {Ivezi{\'c}}, {\v{Z}}., {Juri{\'c}}, M., {et~al.} 2008, \icarus,
  198, 138, \dodoi{10.1016/j.icarus.2008.07.002}

\bibitem[{{Rubincam}(2000)}]{2000Icar..148....2R}
{Rubincam}, D.~P. 2000, \icarus, 148, 2, \dodoi{10.1006/icar.2000.6485}

\bibitem[{{S{\'a}nchez-S{\'a}ez} {et~al.}(2021){S{\'a}nchez-S{\'a}ez}, {Reyes},
  {Valenzuela}, {F{\"o}rster}, {Eyheramendy}, {Elorrieta}, {Bauer},
  {Cabrera-Vives}, {Est{\'e}vez}, {Catelan}, {Pignata}, {Huijse}, {De Cicco},
  {Ar{\'e}valo}, {Carrasco-Davis}, {Abril}, {Kurtev}, {Borissova}, {Arredondo},
  {Castillo-Navarrete}, {Rodriguez}, {Ruz-Mieres}, {Moya},
  {Sabatini-Gacit{\'u}a}, {Sep{\'u}lveda-Cobo}, \&
  {Camacho-I{\~n}iguez}}]{Sanchez2021}
{S{\'a}nchez-S{\'a}ez}, P., {Reyes}, I., {Valenzuela}, C., {et~al.} 2021, \aj,
  161, 141, \dodoi{10.3847/1538-3881/abd5c1}

\bibitem[{Shappee {et~al.}(2014)Shappee, Prieto, Grupe, Kochanek, Stanek, Rosa,
  Mathur, Zu, Peterson, Pogge, Komossa, Im, Jencson, Holoien, Basu, Beacom,
  Szczygieł, Brimacombe, Adams, Campillay, Choi, Contreras, Dietrich,
  Dubberley, Elphick, Foale, Giustini, Gonzalez, Hawkins, Howell, Hsiao, Koss,
  Leighly, Morrell, Mudd, Mullins, Nugent, Parrent, Phillips, Pojmanski,
  Rosing, Ross, Sand, Terndrup, Valenti, Walker, \& Yoon}]{ASAS-SN-survey}
Shappee, B.~J., Prieto, J.~L., Grupe, D., {et~al.} 2014, The Astrophysical
  Journal, 788, 48, \dodoi{10.1088/0004-637X/788/1/48}

\bibitem[{Smith {et~al.}(2019)Smith, Williams, Young, Ibsen, Smartt, Lawrence,
  Morris, Voutsinas, \& Nicholl}]{Smith2019}
Smith, K.~W., Williams, R.~D., Young, D.~R., {et~al.} 2019, Research Notes of
  the AAS, 3, 26, \dodoi{10.3847/2515-5172/ab020f}

\bibitem[{Soraisam {et~al.}(2020)Soraisam, Saha, Matheson, Lee, Narayan, Vivas,
  Scheidegger, Oppermann, Olszewski, Sinha,
  {et~al.}}]{soraisam2020classification}
Soraisam, M.~D., Saha, A., Matheson, T., {et~al.} 2020, The Astrophysical
  Journal, 892, 112, \dodoi{10.3847/1538-4357/ab7b61}

\bibitem[{{Strauss} {et~al.}(2023){Strauss}, {Trilling}, {Bernardinelli},
  {Beach}, {Oldroyd}, {Sheppard}, {Schlichting}, {Gerdes}, {Adams}, {Chandler},
  {Fuentes}, {Holman}, {Juri{\'c}}, {Lin}, {Markwardt}, {McNeill}, {Mommert},
  {Napier}, {Payne}, {Ragozzine}, {Rivkin}, {Smotherman}, \&
  {Trujillo}}]{2023arXiv230904034S}
{Strauss}, R., {Trilling}, D.~E., {Bernardinelli}, P.~H., {et~al.} 2023, arXiv
  e-prints, arXiv:2309.04034, \dodoi{10.48550/arXiv.2309.04034}

\bibitem[{Tonry {et~al.}(2018)Tonry, Denneau, Heinze, Stalder, Smith, Smartt,
  Stubbs, Weiland, \& Rest}]{ATLAS-survey}
Tonry, J.~L., Denneau, L., Heinze, A.~N., {et~al.} 2018, Publications of the
  Astronomical Society of the Pacific, 130, 064505,
  \dodoi{10.1088/1538-3873/aabadf}

\bibitem[{Trilling {et~al.}(2023)Trilling, Gowanlock, Kramer, McNeill,
  Donnelly, Butler, \& Kececioglu}]{trilling2023}
Trilling, D.~E., Gowanlock, M., Kramer, D., {et~al.} 2023, The Astronomical
  Journal, 165, 111, \dodoi{10.3847/1538-3881/acac7f}

\bibitem[{van Roestel {et~al.}(2021)van Roestel, Duev, Mahabal, Coughlin,
  Mr{\'o}z, Burdge, Drake, Graham, Hillenbrand, Bellm, {et~al.}}]{van2021ztf}
van Roestel, J., Duev, D.~A., Mahabal, A.~A., {et~al.} 2021, The Astronomical
  Journal, 161, 267, \dodoi{10.3847/1538-3881/abe853}

\bibitem[{Wagstaff {et~al.}(2013)Wagstaff, Lanza, Thompson, Dietterich, \&
  Gilmore}]{wagstaff2013guiding}
Wagstaff, K.~L., Lanza, N.~L., Thompson, D.~R., Dietterich, T.~G., \& Gilmore,
  M.~S. 2013, in Twenty-Seventh AAAI Conference on Artificial Intelligence,
  \dodoi{10.1609/aaai.v27i1.8561}

\bibitem[{Zhou {et~al.}(2017)Zhou, Pan, Wang, \& Vasilakos}]{ZHOU2017350}
Zhou, L., Pan, S., Wang, J., \& Vasilakos, A.~V. 2017, Neurocomputing, 237,
  350, \dodoi{10.1016/j.neucom.2017.01.026}

\bibitem[{Zimek {et~al.}(2012)Zimek, Schubert, \& Kriegel}]{zimek2012survey}
Zimek, A., Schubert, E., \& Kriegel, H.-P. 2012, Statistical Analysis and Data
  Mining: The ASA Data Science Journal, 5, 363, \dodoi{10.1002/sam.11161}

\end{thebibliography}

\appendix
\section{Comparison of Rankings Between Outlier Detection Methods}\label{sec:appendix_comparison_outlier_methods} 
\begin{figure*}[!t]
\centering
\subfigure[\dssj: Ranking Comparison]{
       \includegraphics[width=0.4\textwidth]{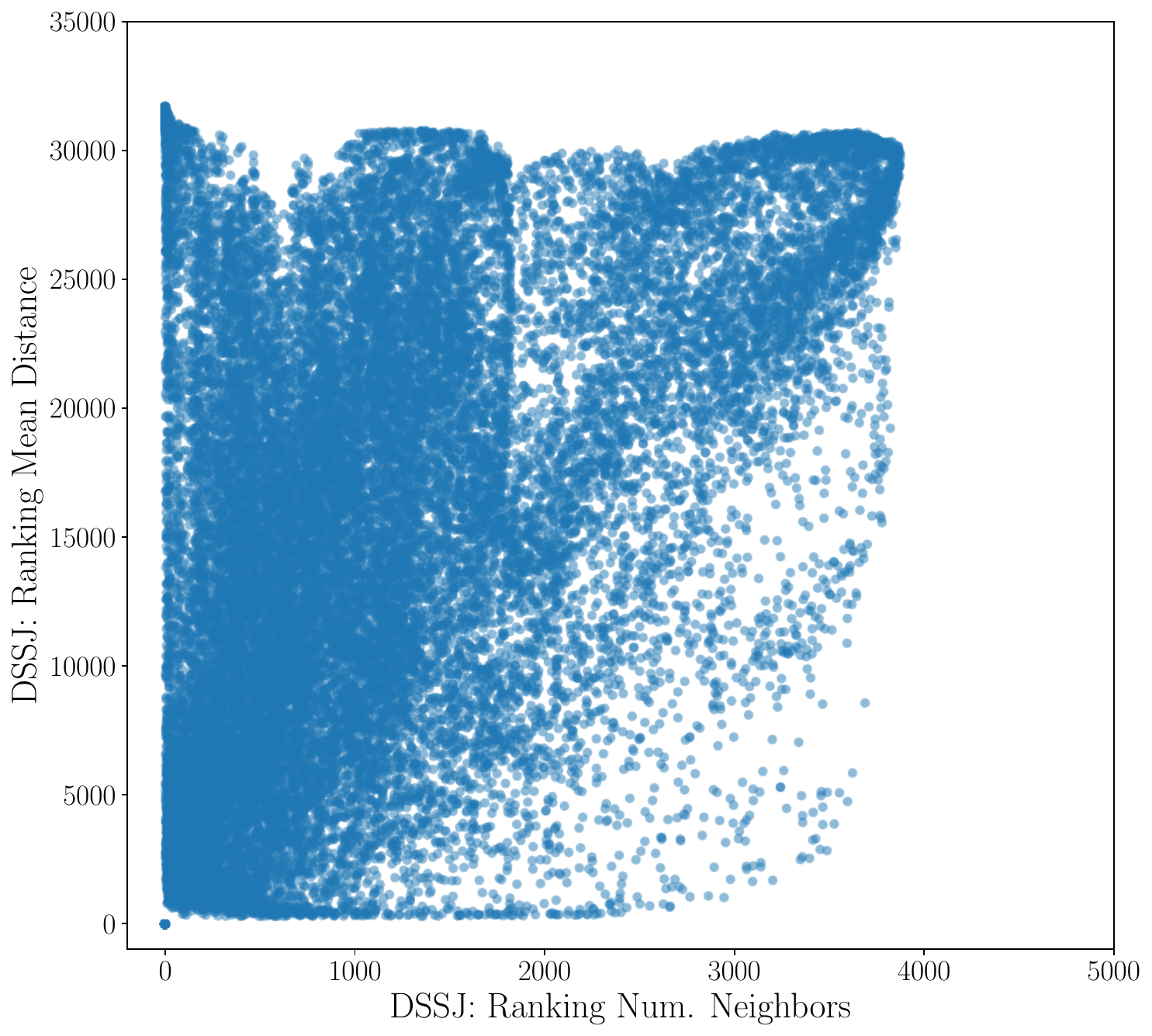}
}
\subfigure[\knnsj: Ranking Comparison]{
       \includegraphics[width=0.4\textwidth]{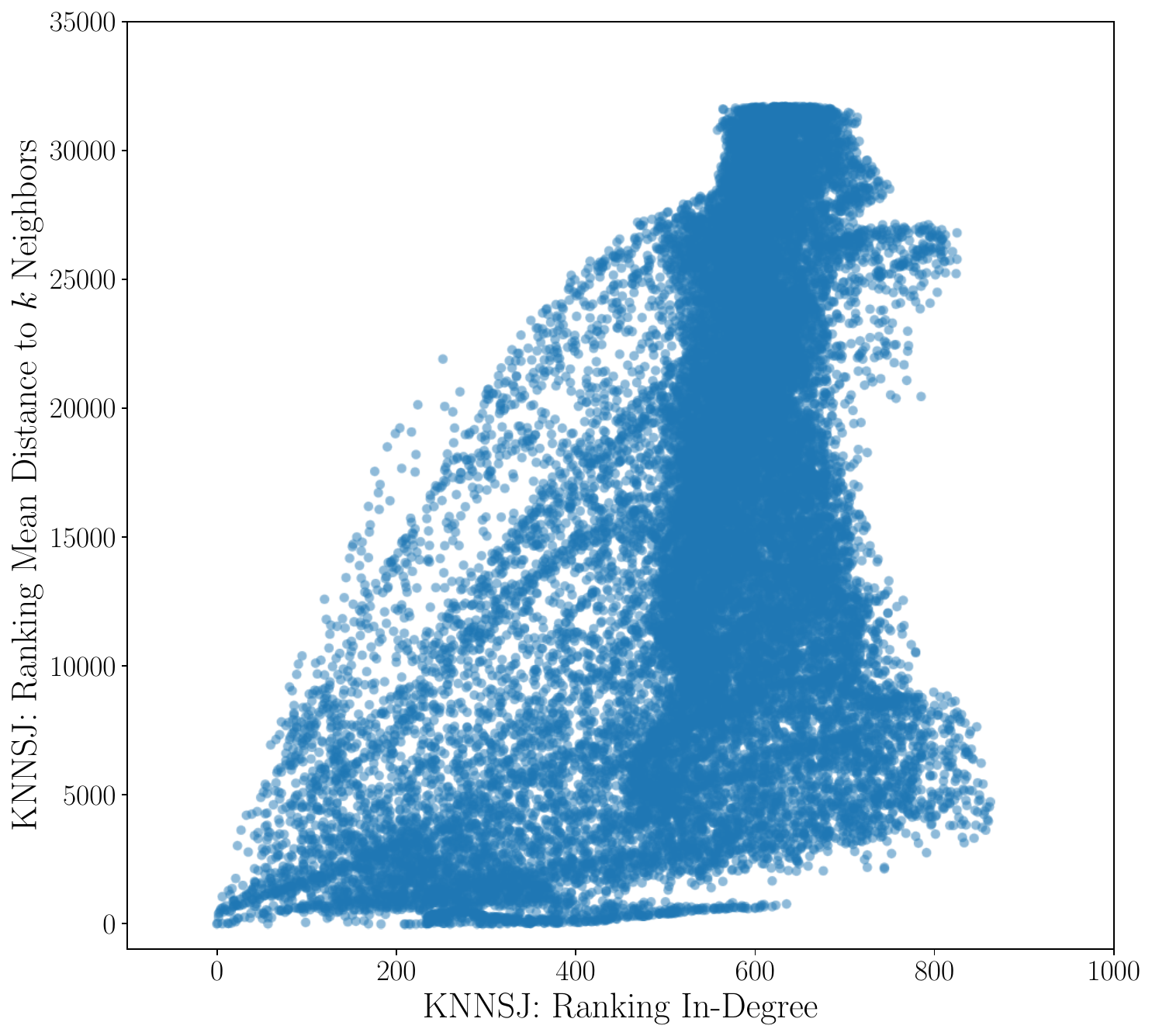}
}
\subfigure[\dssj vs. \knnsj: Distance-Oblivious Comparison]{
       \includegraphics[width=0.4\textwidth]{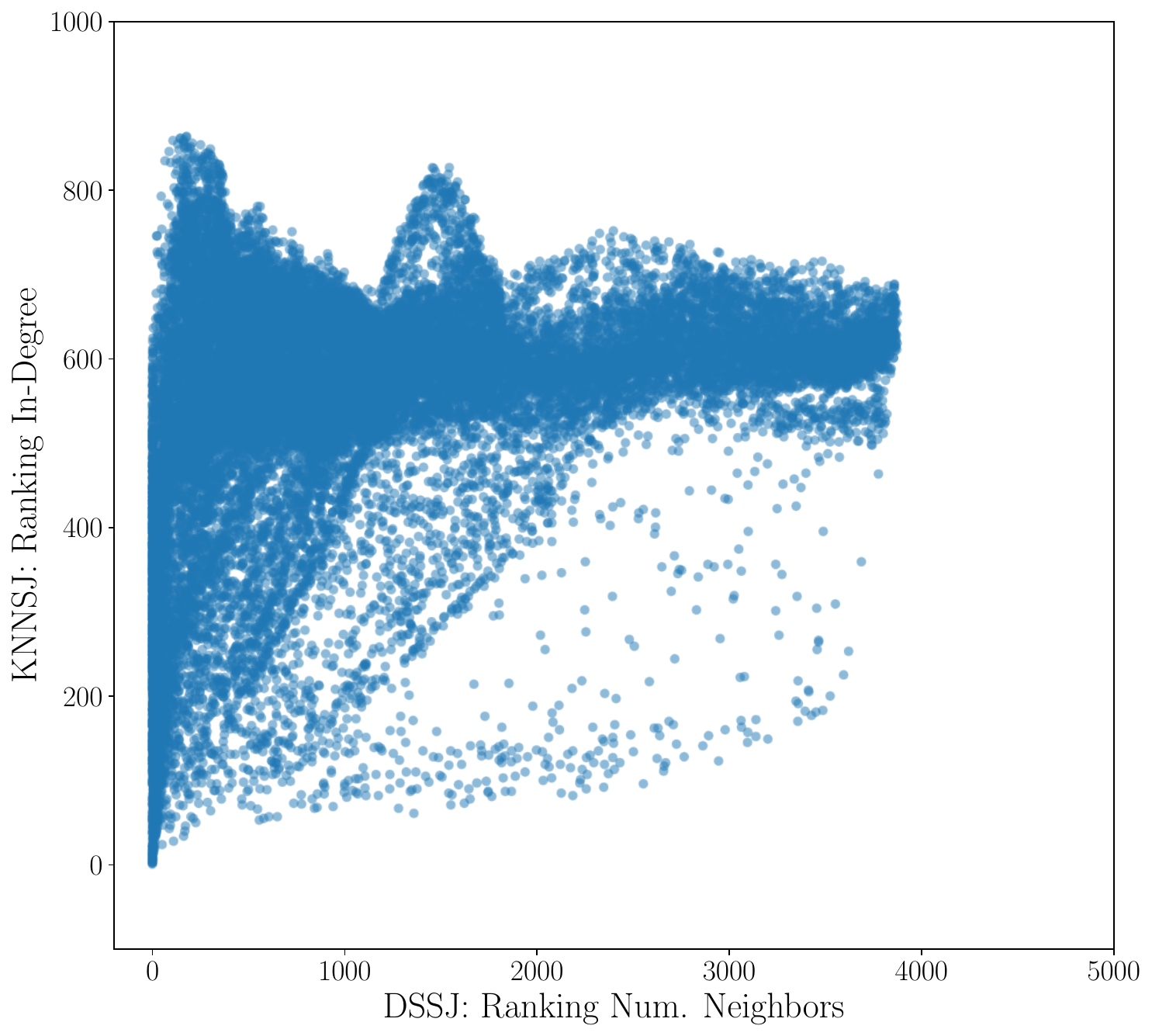}
}
\subfigure[\dssj vs. \knnsj: Distance-Aware Comparison]{
       \includegraphics[width=0.4\textwidth]{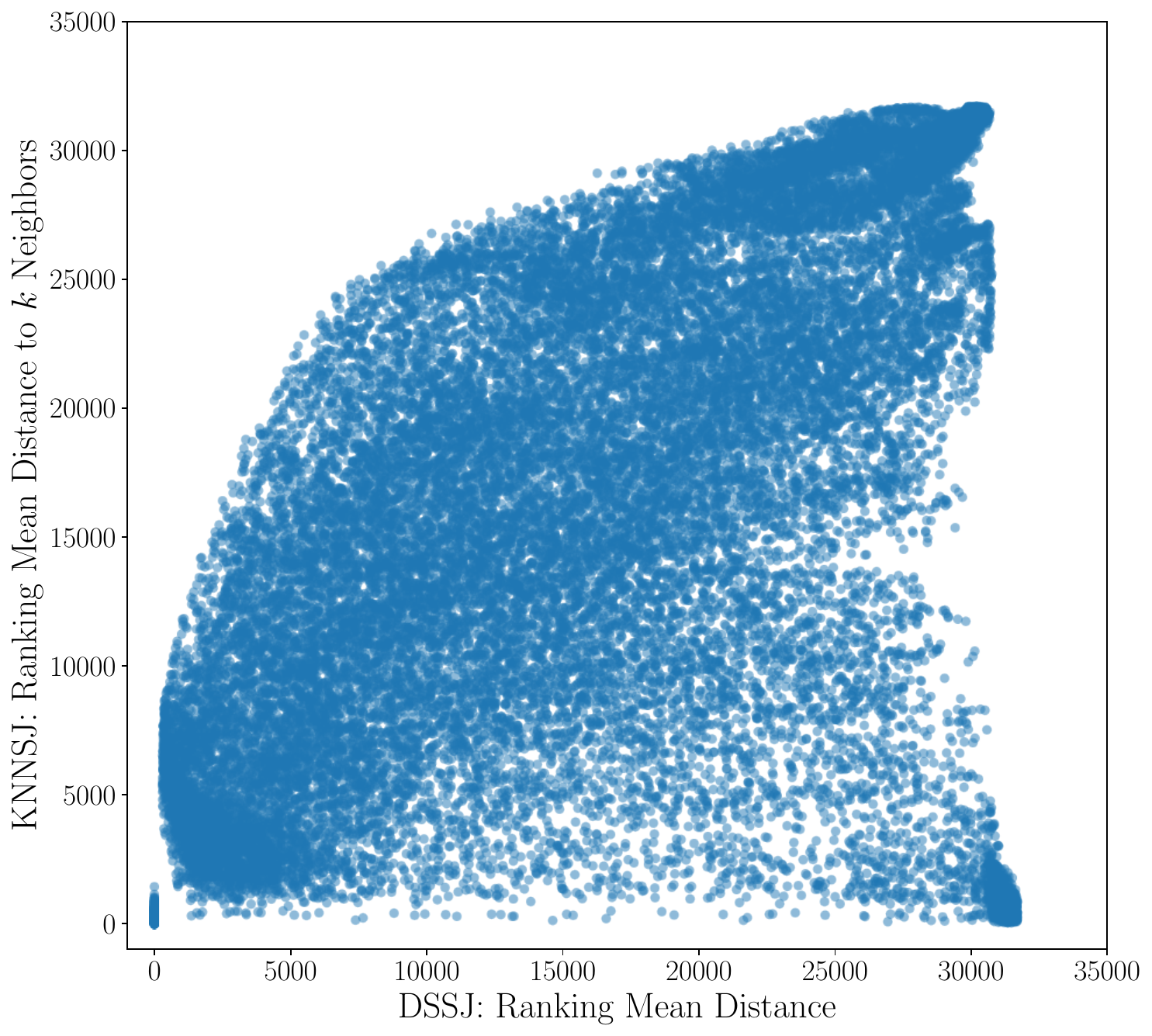}
}

    \caption{Scatterplots showing a comparison of rankings for the same $d=3$ feature space shown in Figure~\ref{fig:DSSJ_KNNSJ_lcamp_rotper_grcolor}. (a) Ranking comparison for the two \dssj metrics. (b) Ranking comparison for the two \knnsj metrics. (c) Comparison of the two distance-oblivious metrics. (d) Comparison of the two distance-aware metrics. Note that across all figures, the distance-oblivious metrics (\dssj: Number of Neighbors and \knnsj: In-degree) have numerous points that share the same rank, and so they have a smaller range of possible rank values compared to the distance-aware metrics (\dssj: Mean Distance and \knnsj: Mean Distance to $k$ Neighbors).}
   \label{fig:DSSJ_KNNSJ_lcamp_rotper_grcolor_ranking_scatter}
\end{figure*}

From Figure~\ref{fig:DSSJ_KNNSJ_lcamp_rotper_grcolor} we observed that there may be minimal overlap between those points that are selected as outliers between the outlier detection metrics. Figure~\ref{fig:DSSJ_KNNSJ_lcamp_rotper_grcolor_ranking_scatter} shows another perspective on the comparison of rankings that were shown in Figure~\ref{fig:DSSJ_KNNSJ_lcamp_rotper_grcolor}, where we plot the rankings for each object as a scatterplot.  Figure~\ref{fig:DSSJ_KNNSJ_lcamp_rotper_grcolor_ranking_scatter} shows the following subplots: (a) shows a ranking comparison for the two \dssj metrics, (b) shows a comparison for the two \knnsj metrics, (c) shows a comparison of the two distance-oblivious metrics, and (d) shows a comparison between the two distance-aware metrics. Recall that in all figures, the outliers are those with a low rank value. 

Figure~\ref{fig:DSSJ_KNNSJ_lcamp_rotper_grcolor_ranking_scatter}(a) shows a comparison of \dssj metrics where we observe that many of the points that have few neighbors (with a rank $\gtrsim0$) have a large range of possible ranks using the mean distance metric. This shows that a large mean distance to neighboring points may be achieved when there are few or many points within the search radius $\epsilon$ of a given point. 

Figure~\ref{fig:DSSJ_KNNSJ_lcamp_rotper_grcolor_ranking_scatter}(b) shows a comparison of \knnsj metrics. Here we find that there are consistencies between outlier scores for both metrics when the rankings are $\gtrsim0$. Interestingly, there is an overabundance of in-degree rankings with values between $\sim 500-700$, although these points will not be considered outliers in this feature space.  

Figure~\ref{fig:DSSJ_KNNSJ_lcamp_rotper_grcolor_ranking_scatter}(c) compares the rankings derived by the distance-oblivious metrics for \dssj and \knnsj. We find that there is little correlation between these the rankings of these two methods. This is because it is possible to have a low in-degree value in the \knn graph, but yet have several neighbors found within the search radius $\epsilon$. Despite this, there are still numerous points that have ranks near 0 in both metrics which would indicate that they are outliers using both metrics.

Figure~\ref{fig:DSSJ_KNNSJ_lcamp_rotper_grcolor_ranking_scatter}(d) compares the rankings derived by the distance-aware metrics for \dssj and \knnsj. We clearly observe that there is a deficit of points assigned a low mean distance rank and a high ($\gtrsim 10000$) mean distance to $k$ neighbors rank. Compared to the other plots, the rankings appear to be the most correlated, although there are still major differences in the assigned rankings.

\end{document}